\documentclass[seceqn,draft]{elsart1p}

\usepackage{epsf}

\usepackage{amssymb}

\def\sect#1{section~{\ref{#1}}}
\def\eqn#1{eq.~(\ref{#1})}
\def\Eqn#1{Equation~(\ref{#1})}
\def\eqns#1#2{eqs.~(\ref{#1}) and~(\ref{#2})}

\def\Shift#1#2{{[#1,#2\rangle}}
\def\spa#1.#2{\left\langle#1\,#2\right\rangle}
\def\spb#1.#2{\left[#1\,#2\right]}
\def\spash#1.#2{\spa{\smash{#1}}.{\smash{#2}}}
\def\spbsh#1.#2{\spb{\smash{#1}}.{\smash{#2}}}
\def\lor#1.#2{\left(#1\,#2\right)}
\def\sand#1.#2.#3{%
\left\langle\smash{#1}{\vphantom1}^{-}\right|{#2}%
\left|\smash{#3}{\vphantom1}^{-}\right\rangle}
\def\sandpp#1.#2.#3{%
\left\langle\smash{#1}{\vphantom1}^{+}\right|{#2}%
\left|\smash{#3}{\vphantom1}^{+}\right\rangle}
\def\sandpm#1.#2.#3{%
\left\langle\smash{#1}{\vphantom1}^{+}\right|{#2}%
\left|\smash{#3}{\vphantom1}^{-}\right\rangle}
\def\sandmp#1.#2.#3{%
\left\langle\smash{#1}{\vphantom1}^{-}\right|{#2}%
\left|\smash{#3}{\vphantom1}^{+}\right\rangle}
\def\vK{\vphantom{\hat K}}
\def\vp{\vphantom{\dot A}}

\def\tr{\mathop{\rm tr}\nolimits}
\def\Tr{\mathop{\rm Tr}\nolimits}
\def\Ord{{\cal O}}
\def\e{\epsilon}
\def\eps{\epsilon}
\def\Res{\mathop{\rm Res}}
\def\Im{\mathop{\rm Im}}
\def\I{{\cal I}}
\def\tlambda{{\tilde\lambda}}
\def\teta{{\tilde\eta}}

\def\Ll{\mathop{\rm L{}}\nolimits}
\def\Kh{{\hat K}}
\def\Ksl{{\s K}}
\def\ksl{\s{k}}

\def\Kinsl{K}
\def\Ellin{\ell}
\def\ve{\varepsilon}
\def\pol{\varepsilon}
\def\fig#1{fig.~{\ref{#1}}}
\def\Fig#1{Figure~{\ref{#1}}}
\def\figs#1#2{figs.~{\ref{#1}} and {\ref{#2}}}

\def\tree{{\rm tree}}

\def\nn{\nonumber}
\newbox\charbox
\newbox\slabox
\def\s#1{{      % Feynman slash
        \setbox\charbox=\hbox{$#1$}
        \setbox\slabox=\hbox{$/$}
        \dimen\charbox=\ht\slabox
        \advance\dimen\charbox by -\dp\slabox
        \advance\dimen\charbox by -\ht\charbox
        \advance\dimen\charbox by \dp\charbox
        \divide\dimen\charbox by 2
        \raise-\dimen\charbox\hbox to \wd\charbox{\hss/\hss}
        \llap{$#1$}
}}

\def\Cuth{{\widehat {C}}}
\def\CuthRat{{\widehat {CR}}}
\def\Remaining{{\widehat {R}}}
\def\Vertex{R}
\def\Remaining{{\widehat {R}}}
\def\cg{c_\Gamma}
\def\PureCut{C}
\def\Li{\mathop{\rm Li}\nolimits}
\def\ellsl{\s{\ell}}
\def\NeqFour{{\cal N}=4}
\def\onehalf{{\textstyle\frac12}}
\def\oneloop{{1 \mbox{-} \rm loop}}
\def\Fact{{\cal F}}

\def\Overlap{O}
\def\Inf{\mathop{\rm Inf}}

\begin{document}

\begin{frontmatter}



\noindent
UCLA/07/TEP/11
\hfill SLAC--PUB--12447
\hfill SPhT--T07/039

 \title{On-Shell Methods in Perturbative QCD}
 \author[author1]{Zvi Bern,\thanksref{thanks1}}
 \author[author2]{Lance J. Dixon\thanksref{thanks2}}
 \author[author3]{and David A. Kosower\thanksref{thanks3}}
\thanks[thanks1]{Supported by the US Department of Energy under
contract DE--FG03--91ER40662}
\thanks[thanks2]{Supported by the US Department of Energy under
contract DE--AC02--76SF00515}
\thanks[thanks3]{Laboratory
   of the {\it Direction des Sciences de la Mati\`ere\/}
   of the {\it Commissariat \`a l'Energie Atomique\/} of France}
%\corauth[cor1]{cor1test}
\address[author1]{Department of Physics and Astronomy, UCLA,
Los Angeles, CA 90095--1547, USA}
\address[author2]{Stanford Linear Accelerator Center,  
Stanford University, Stanford, CA 94309, USA}
\address[author3]{Service de Physique Th\'eorique,
CEA--Saclay, F--91191 Gif-sur-Yvette cedex, France}


%\author{}

%\address{}

\begin{abstract}
We review on-shell methods for computing multi-parton scattering 
amplitudes in perturbative QCD, utilizing their unitarity and 
factorization properties.  We focus on aspects which are useful
for the construction of one-loop amplitudes needed for phenomenological 
studies at the Large Hadron Collider.
\end{abstract}

\end{frontmatter}

% main text
%%%%%%%%%%%%%%%%%%%%%%%%%%%%%%%%%%%%%%%%%%%%%%%%%%%%%%%
\section{Introduction}
\label{IntroSection}

The Large Hadron Collider (LHC) is poised to begin exploration of 
the multi-TeV energy frontier within the next year.  
It is widely anticipated that physics beyond the Standard Model
will emerge at this scale, most likely via the production
of new, heavy particles which may be associated with the mechanism
of electroweak symmetry breaking.  At the very least, the scalar 
Higgs boson of the Standard Model awaits discovery.   New, heavy
particles typically will decay rapidly into the known leptons, 
neutrinos, quarks and gluons.   When the new physics model
includes a dark matter candidate (as in supersymmetric models
 with conserved $R$ parity), this particle can terminate
the decay chain.  In this case, sharp peaks in invariant-mass 
distributions may be scarce; they can be replaced by missing-energy 
signals.  Quite often then, the signals for new physics have
to be assessed against a significant background of Standard Model physics.
The better the background is understood, the better the prospects 
for discovery.   Once new physics is discovered, we will want to 
measure its properties precisely, via production cross sections,
branching ratios and masses.   A numerically precise understanding 
of the backgrounds and of theoretical aspects of luminosity measurement 
will be essential to this endeavor.

In some cases backgrounds can be understood without much theoretical
input.  For example, in the decay of a light Higgs boson to a pair of
photons, the signal is a very narrow peak in the di-photon invariant
mass.  The QCD background, in contrast, is a smooth distribution which
can be interpolated easily under a candidate peak.  However, many
signals involve much broader kinematic distributions, with final
states including jets and missing energy in addition to charged
leptons and photons. A classic example is the production of a 
Higgs boson in association with a $W$ boson at the Tevatron,
with the Higgs decaying to a $b\bar{b}$ pair, and the $W$ decaying
to a charged lepton plus a neutrino.  For such signals, a much more detailed
understanding of the backgrounds is typically required.

Many methods are available for computing Standard Model backgrounds at
the leading order (LO) in QCD perturbation theory.  For example, {\tt
MADGRAPH}~\cite{MADGRAPH}, {\tt CompHEP}~\cite{CompHEP} and {\tt
AMEGIC++}~\cite{AMEGIC} automatically sum tree-level Feynman graphs for
helicity amplitudes.  Other programs, such as {\tt ALPGEN}~\cite{ALPGEN} 
and {\tt HELAC}~\cite{HELAC} are based on `off-shell' recursive 
algorithms~\cite{BGRecursion,LaterRecursive}.   For these algorithms,
the building blocks are quantities in which at least one external leg is
off shell (in contrast to the on-shell recursion 
relations~\cite{BCFRecursion,BCFW} described later in this article).   
These recursion relations were first constructed in the QCD
context by Berends and Giele~\cite{BGRecursion}, and applied
early on to matrix elements for backgrounds to top quark 
production~\cite{VECBOS}.

For quite a while, simpler processes have been incorporated
into parton-shower Monte Carlo programs that provide realistic event
simulation at the hadron level.
These programs, including {\tt PYTHIA}~\cite{PYTHIA} and
{\tt HERWIG}~\cite{HERWIG}, perform parton showering and
hadronization.  They implement an approximation to the perturbative
expansion that resums leading logarithms.
More complex processes are now being included in this
framework, using leading-order parton-level matrix elements provided
by {\tt MADGRAPH} or {\tt ALPGEN}, for example, combined with a
matching scheme~\cite{Matching} that avoids double-counting between the
leading-order matrix elements and the parton shower.  Yet these
leading-order results often have a strong sensitivity to higher-order
corrections.  Gluon-initiated processes are particularly sensitive.
For example, the cross section for production of the Standard Model
Higgs boson via gluon-fusion at the LHC is boosted by roughly a factor
of two as one goes from leading order to next-to-next-to-leading order
(NNLO) in the perturbative QCD expansion~\cite{NNLOHiggs}.

For this reason, quantitative estimates for most processes require a
calculation at next-to-leading order (NLO).  Ideally, one
would also match such results to a parton-shower Monte Carlo as
well~\cite{MCNLO,NLOPartonShowers}.  For a growing list of processes,
this has been achieved, particularly within the program
{\tt MC@NLO}~\cite{MCNLO}.

NLO calculations require knowledge of both virtual and real-emission
corrections to the basic process.  The real-emission corrections are
constructed from tree-level amplitudes with one additional parton present,
either an additional gluon, or a quark--antiquark pair replacing
a gluon in the LO process.  They can be computed using the same
tree-level techniques used for the basic process.  In addition, we need a
method for extracting the infrared singularities arising from integration
of the real-emission contributions over unresolved regions of phase space.
These singularities cancel against those in the virtual corrections
or against counterterms from the evolution of parton distributions.  
Several such methods have been developed for use in generic NLO
processes~\cite{FKS,GGK,CS,CDST}.
Subtraction methods based on dipole subtraction~\cite{CS,CDST}
are the most widely used today.  A related subtraction method,
based on so-called antenna factorization~\cite{Antenna}, 
has even been extended to NNLO~\cite{Gehrmann}.
In this review, we will focus on techniques
for computing the virtual one-loop corrections to processes.  
Once the virtual corrections are known, their incorporation into
a numerical program for the full NLO result is straightforward, 
because they do not need to be integrated over unresolved regions of
phase space, and all of their infrared singularities will be manifest.

The computation of one-loop virtual corrections for processes with
multiple partons, plus electroweak particles, is the bottleneck
currently limiting availability of NLO results.  The state of the art
for complete computations is processes with up to three objects ---
jets, vector bosons, or scalars --- in the final state (see {\it e.g.}
refs.~\cite{ttH,NLOThreeJet,MCFM,HbbWbb,EGZHjj,NLOttjet,LMPTriboson}).
A number of new approaches aimed at one-loop multi-parton amplitudes
are currently under development. Some are already producing results
applicable to processes with four final-state objects.  These
approaches fall into three basic categories: improved traditional
(including semi-numerical)~\cite{Davydychev,Passarino,GRACEQCD,DDReduction,
GieleGloverNumerical,AguilaPittau,EGZH4p, BGHPS, EGZSemiNum, 
EGZ6g, XYZ, BinothRat, 
OPP}, purely numerical~\cite{NSSubtract,LMPTriboson}, and on-shell
analytic~\cite{Neq4Oneloop,Neq1Oneloop,DDimU, UnitarityMachinery,
NeqOneNMHVSixPt, BCFGeneralized, BBDPSQCD, BBCF, OnShellOneLoop, Qpap,
Bootstrap, Genhel,LoopMHV, BFM,  ABFKM,MastroliaTriple,
BFMassive}.

The semi-numerical approach of
Ellis, Giele and Zanderighi~\cite{EGZSemiNum} has already been
used to compute loop corrections to the amplitudes involving a Higgs
boson and four external partons~\cite{EGZH4p}, and to the six-gluon
amplitude~\cite{EGZ6g}.  Nagy and Soper~\cite{NSSubtract} have proposed 
a subtraction method for use in a purely numerical evaluation of one-loop
amplitudes, and have evaluated the six-photon
helicity amplitudes purely numerically~\cite{NS6ph}.  The six-photon
result has been reproduced very recently using both on-shell analytical
and semi-numerical methods~\cite{BinothPhotons,OPPPhotons}.
Binoth {\it et al.}~\cite{BGHPS} have developed a combined
algebraic/numerical algorithm for multi-parton amplitudes; 
the algebraic part has been used to determine
the rational parts of several types of amplitudes, including the
six-photon amplitude (for which the rational parts vanish).  The full
algorithm is also being incorporated into a NLO QCD framework by the
GRACE collaboration~\cite{GRACEQCD}.  A purely numerical approach
combining sector decomposition and contour deformations has recently
been used to calculate tri-vector boson production~\cite{LMPTriboson}.
Denner and Dittmaier have developed a numerically stable method for
reducing one-loop tensor integrals~\cite{DDReduction}, which has been
used in various electroweak processes, but also in NLO QCD computations,
including the production of a top quark pair in association with a jet at 
hadron colliders~\cite{NLOttjet}.

In this review we describe on-shell analytic methods 
for one-loop computations.  The `on-shell' terminology means that
essentially all information is extracted from simpler 
(lower-loop and lower-point) amplitudes for physical states.  
In contrast, the conventional Feynman-diagram
approach requires building blocks with off-shell states.
The on-shell approach effectively restricts the states used in
a calculation to physical ones.
The restriction falls on some internal as well as external states.  
The approach relies on three
general properties of perturbative amplitudes in any field theory:
factorization, unitarity, and the existence of a representation in terms
of Feynman integrals.   An earlier form
of on-shell methods was used to compute the one-loop
amplitudes for $e^+e^- \rightarrow Z \rightarrow 4$ partons and (by
crossing) $pp \rightarrow W,Z$ + 2 jets~\cite{Zqqgg}.  The latter have
been implemented in the MCFM program~\cite{MCFM}.  More recently, all
six-gluon helicity amplitudes have been
computed (primarily) with these methods~\cite{Neq4Oneloop,Neq1Oneloop,
NeqOneNMHVSixPt,BBDPSQCD,BBCF,BFM,Genhel,LoopMHV,XYZ}. 

On-shell methods provide a means for determining scattering amplitudes
directly from their poles and cuts.  Perturbative unitarity, applied
to a one-loop amplitude, determines its branch cuts in terms of
products of tree amplitudes~\cite{OldUnitarity}.  
The unitarity method~\cite{Neq4Oneloop,Neq1Oneloop} provides a 
technique for producing functions with the correct branch cuts in
all channels.  Its power is enhanced by relying on the
decomposition of loop amplitudes into a basis of loop-integral
functions.  Matching the cuts with the cuts of basis integrals
in this decomposition provides
a direct and efficient means for producing expressions 
for amplitudes.  It is most convenient to use dimensional regularization
to handle both infrared and ultraviolet
divergences in massless gauge-theory scattering amplitudes.
  We take the number of
space-time dimensions to be $D=4-2\e$.  We can
 use fully $D$-dimensional states and momenta in the unitarity method
to obtain complete amplitudes, at least when all internal particles
are massless~\cite{DDimU}.  (In the older language of dispersion
relations~\cite{OldUnitarity}, amplitudes can be reconstructed fully
from cuts in $D$ dimensions because of the convergence of dispersion
integrals in dimensional regularization~\cite{VanNeervenUnitarity}.)
In most cases, however, it is much simpler to use four-dimensional
states and momenta in the cuts.  This procedure correctly yields all
terms in the amplitudes with logarithms and
polylogarithms~\cite{Neq1Oneloop}, but it generically drops rational
terms, which have to be recovered via another method.

The basic framework of the unitarity method was set up in
refs.~\cite{Neq4Oneloop,Neq1Oneloop}.  In ref.~\cite{Zqqgg},
generalized unitarity was introduced as a means for simplifying
cut calculations, by limiting the number of integral functions
contributing to a cut.  These techniques were applied to a 
variety of calculations in QCD
and supersymmetric gauge theories~\cite{TwoLoopSplit,Neq47pt}.  More
recent improvements to the unitarity method, by Britto, Cachazo and
Feng~\cite{BCFGeneralized}, use complex momenta within generalized
unitarity, allowing for a simple and purely algebraic determination of
all box integral coefficients.  Britto, Buchbinder, Cachazo and 
Feng~\cite{BBCF} have shown how triangle and bubble integral coefficients
may be evaluated by extracting residues in contour integrals.  
This approach has been used to compute various contributions to 
the six-gluon amplitude~\cite{BBCF,BFM}.  Very recently, Forde
has proposed an efficient new approach to computing these 
coefficients~\cite{FordeTriBub}.

The four-dimensional version of the unitarity method leaves undetermined
additive rational-function terms in the amplitudes.  These rational
functions can be characterized by their kinematic poles.  An efficient,
systematic means for constructing these terms from
their poles and residues is to use on-shell recursion 
relations~\cite{OnShellOneLoop,Qpap,Bootstrap,Genhel,LoopMHV}, which
were first devised by Britto, Cachazo, Feng and Witten,
as a means for constructing tree-level amplitudes~\cite{BCFRecursion,BCFW}.
In a related development,
Brandhuber {\it et al.}~\cite{BMST} and
Anastasiou {\it et al.}~\cite{ABFKM} have investigated how to determine
the rational parts of amplitudes from $D$-dimensional unitarity, 
extending earlier work~\cite{DDimU,OneLoopReview}.
Another approach to computing rational terms uses a clever organization
of Feynman diagrams, together with the observation that only a limited 
set of tensor integrals can contribute to the rational
terms~\cite{XYZ,BinothRat}.  Brandhuber {\it et al.}~\cite{BSTZ}
have argued that rational terms can
be obtained using a set of Lorentz-violating counterterms.  Finally,
Ossola, Papadopoulos and Pittau~\cite{OPP,OPPPhotons} 
have developed a new loop-integral decomposition method,
which we expect will mesh well with generalized unitarity techniques
reviewed here. 

Most of the recent development of on-shell techniques at one loop
has focused on QCD amplitudes in which all of the external particles
are gluons.  Amplitudes containing massless external quarks
pose no inherent difficulty.  The same basis of integrals can be
used, and individual amplitudes should have comparable analytic complexity.  
However, they have fewer discrete symmetries, so the number of 
amplitudes that need to be computed is larger for the same total number 
of partons.  Amplitudes with
external electroweak particles (vector bosons or Higgs bosons)
in addition to QCD partons are of even greater phenomenological interest.  
Such amplitudes can also be built from the same basis of integrals.
(For amplitudes with massive internal lines, such as in top-quark 
production, additional integrals are required.)  In this review, 
in order to simplify the discussion,
we focus primarily on amplitudes with external gluons only.

Recent years have also seen
the emergence of on-shell methods at tree level, which 
have provided a basis for some of the recent developments at one loop.
We will give some tree-level examples, but refer the reader to 
refs.~\cite{CSLectures,TwistorReview,SW} for a more extensive exposition.

It is also worth noting that the improved understanding of the structure of
scattering amplitudes, stemming from on-shell methods, has also led to
advances in more theoretical issues related to the AdS/CFT
conjecture and in quantum gravity -- see {\it e.g.}
refs.~\cite{TheoreticalAdvances}.  

This review is organized as follows. In \sect{KinematicsSection} we
describe how the use of complex momenta is intrinsic to on-shell
methods.  In \sect{TreeRecursionSection} we review the BCFW on-shell
recursion relations for tree amplitudes, giving a few examples.  We turn to
one-loop amplitudes in \sect{UnitaritySection}, where we describe the
unitarity-based method and some recent improvements.  In
\sect{LoopRecursionSection}, we show how on-shell recursion is
an efficient way to construct rational terms in one-loop amplitudes.
We give our conclusions in \sect{ConclusionsSection}, and comment on 
the prospects for evaluating new amplitudes of phenomenological interest
at the LHC with these methods.

%%%%%%%%%%%%%%%%%%%%%%%%%%%%%%%%%%%%%%%%%%%%%%%%%%%%%%%

\section{Uses of Complex Kinematics}
\label{KinematicsSection}

Multi-parton amplitudes in QCD are functions of a large
number of variables, describing polarization states,
color quantum numbers, and kinematics.  It is important to disentangle
the dependence on all these variables to as great an extent
as possible.  In this section we focus on techniques for
separating and simplifying the kinematic behavior.  
The polarization-state dependence may be handled efficiently
by computing helicity amplitudes, using the spinor-helicity formalism 
for external gluons~\cite{SpinorHelicity,TreeReview}.  

The color dependence can be organized using the notion 
of color-ordered subamplitudes, or primitive amplitudes.
Primitive amplitudes are defined for specific cyclic orderings
of the external partons, but carry no explicit color indices.
They can be computed using `color-ordered' Feynman diagrams.
Such diagrams can be drawn on the plane for the specified external
ordering of the partons.
(There are also restrictions on the direction of fermion 
flow through the diagrams in loop amplitudes.)
The color indices have been stripped from all the 
vertices and propagators for these diagrams. 
The full amplitude can be assembled from the primitive
amplitudes by dressing them with appropriate color factors.

For example, for $n$-gluon amplitudes, the tree-level color factors
have the form of single traces, $\tr(T^{a_1}T^{a_2}\cdots T^{a_n})$,
where $T^{a_i}$ is an $SU(N_c)$ generator matrix in the fundamental
representation, for gluon $i$~\cite{TreeColor,TreeReview}.
The coefficients of these color structures define the tree-level
primitive amplitudes $A_n^\tree(1,2,3,\ldots,n)$.
At one loop, the $n$-gluon color factors can be either single traces
--- with an additional factor of $N_c$ present ---
or double traces.  The coefficients of the single traces, which
would dominate in the large-$N_c$ limit, are given
directly by one-loop primitive amplitudes.   The coefficients
of the double traces are given by sums over permutations of
primitive amplitudes~\cite{BKColor,Neq4Oneloop,OneLoopReview}.
Therefore we can focus on the coefficients of 
$N_c \, \tr(T^{a_1}T^{a_2}\cdots T^{a_n})$, which we denote here
by $A_{n}^\oneloop(1,2,3,\ldots,n)$.  (Elsewhere in the literature, they
are often denoted $A_{n;1}$, to distinguish them from the double-trace
coefficients $A_{n;c}$ for $c>1$.)
The one-loop primitive amplitudes $A_n^\oneloop$ have the same symmetry under
cyclic permutations of the external legs as the tree amplitudes $A_n^\tree$.
Similar results hold when external quarks --- fermions in the fundamental
representation --- are present as well~\cite{TwoQuarkThreeGluon}.
We refer the reader to previous reviews~\cite{TreeReview,OneLoopReview}
for a more complete description of helicity and color
decompositions.  

In this review, we concentrate on methods for computing
the building blocks, primitive helicity amplitudes,
which are functions only of the kinematic variables.
A key property of primitive amplitudes is that their
singularities, branch cuts (at loop level) 
as well as factorization poles, depend only on a restricted set of 
kinematic variables.  This set consists of
the squares of sums of cyclically adjacent momenta,
$K_{i\cdots j} = k_i + k_{i+1} + \cdots + k_{j-1} + k_j$, where
all indices are taken modulo $n$.
Furthermore, the singularities are determined by lower-loop
primitive amplitudes in the case of cuts, and by lower-point
primitive amplitudes in the case of poles.   This reducibility
suggests the possibility of developing a recursive computational
framework.

The singularity information can be accessed more easily
if we define primitive amplitudes for suitable complex, yet
on-shell, values of the external momenta.  This simple observation
came out of Witten's development of twistor string 
theory~\cite{WittenTopologicalString}, and many subsequent papers 
sparked by that article.  Witten's work demonstrated a remarkable
simplicity for tree amplitudes, and suitable parts of loop
amplitudes, when they were mapped into the twistor space of
Penrose~\cite{Penrose} in which twistor strings
propagate.   This space can be defined as a `half Fourier transform'
of the space of left- and right-handed two-component (Weyl) spinors 
$u_{\pm}(k_i)$ associated with a massless vector, $k_i^2 = 0$.
The left-handed spinors are Fourier transformed, while the
right-handed spinors remain as coordinates.

Here we will not rely directly on any concrete properties of
twistor space.  However, two conceptual ideas from that work
underpin our approach:
\begin{enumerate}  
\item Use two-component spinor variables as the independent
variables for scattering processes, rather than four-component
momenta.
\item
Treat opposite-helicity spinors as independent variables.
For real Minkowski momenta, there is a complex-conjugation 
relation between the two. This treatment requires 
momenta to be generically complex. 
\end{enumerate}

Complex momenta are certainly not a new notion.  Wick rotation allows
the analytic properties of amplitudes for real Minkowski momenta
to be deduced by continuation from the Euclidean region, in 
which the time components of Minkowski momenta are imaginary.
Similarly, the complex analyticity of the $S$ matrix, expressed as a 
function of the Lorentz-invariant products $s_{ij} = 2 k_i\cdot k_j$,
depends implicitly on having complex momenta $k_i$.
On the other hand, arriving at complex momenta by thinking
of the spinor variables as fundamental, leads to new ways
of organizing the kinematic properties of helicity amplitudes.

%%%%%%%%%%%%%%%%%%%%%%%%%%%%%%%%%%%%%%

\subsection{Complex Kinematics and the Three-point Amplitude}
\label{KinThreeSubsection}

\begin{figure}[t]
\centerline{\epsfxsize 3.0truein\epsfbox{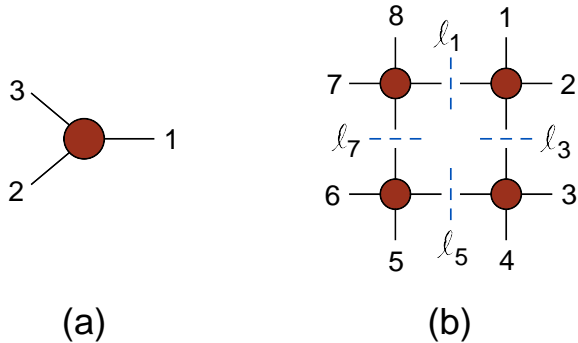}}
\caption{Two types of processes involving massless particles
for which complex momenta are very useful:  
(a) three-point amplitudes, and (b) generalized unitarity cuts. }
\label{complexkinFigure}
\end{figure}

The first hint that this approach might be useful comes from 
investigating three-point helicity 
amplitudes~\cite{WittenTopologicalString}, displayed in 
\fig{complexkinFigure}(a).
For real momenta, an on-shell process with three external massless
legs $i,j,k$ is always singular, because $s_{ij} = 0$ for all three
pairs of legs. These conditions force the three momenta to be 
collinear with each other --- if they are real --- which in turn 
makes all kinematic quantities vanish.  Using complex momenta 
to define the amplitude for three massless particles 
is also not a new idea.  For example, Goroff and Sagnotti~\cite{GoroffSagnotti}
used complex momenta to define non-singular three-graviton kinematics, 
in their on-shell computation of the two-loop divergence
in pure Einstein gravity.
However, there is a natural way to take the kinematics
to be complex using spinor variables, which meshes neatly
with the structure of helicity amplitudes.  

%%%%%% FIGURE %%%%%%%%%%%
%%%%%%%%%%%%%%%%%

First we introduce a shorthand notation for the two-component
(Weyl) spinors associated with an $n$-parton 
process~\cite{WittenTopologicalString}:
\begin{equation}
(\lambda_i)_\alpha \equiv [u_+(k_i)]_{\alpha} \,,
\qquad
(\tlambda_i)_{\dot\alpha} \equiv [u_-(k_i)]_{\dot\alpha} \,,
\qquad i = 1,2,\ldots,n.
\label{lambdadef}
\end{equation}
It is also convenient to describe the spinors with a `bra'
and `ket' notation,
\begin{equation}
\lambda_i\ =\ |i^+\rangle\ =\ \langle i^-| \,,
\qquad
\tlambda_i\ =\ |i^-\rangle\ =\ \langle i^+| \,.
\label{braketdef}
\end{equation}
One can always reconstruct the momenta from the spinors,
using the positive-energy projector for massless spinors, 
$u(k)\bar{u}(k) = \ksl$, or
\begin{equation}
k_i^\mu (\sigma_\mu)_{\alpha\dot\alpha}
= (\ksl_i)_{\alpha\dot\alpha}
= (\lambda_i)_\alpha (\tlambda_i)_{\dot\alpha} \,.
\label{kfact}
\end{equation}
\Eqn{kfact} shows that a massless momentum vector, written
as a bi-spinor, is simply the product of a left-handed spinor 
with a right-handed one.  This result is also valid for complex
momenta.

Lorentz-invariant spinor products can be defined using
the antisymmetric tensors $\pol^{\alpha\beta}$ 
and $\pol^{\dot\alpha\dot\beta}$ for the $SU(2)$ factors
in the Lorentz algebra, $SL(2,R) \sim SU(2)_L \times SU(2)_R$:
\begin{eqnarray}
\spa{j}.{l}
&=& \ve^{\alpha\beta} (\lambda_j)_\alpha (\lambda_l)_\beta
= \bar{u}_-(k_j) u_+(k_l)\,,
\label{spinorproddefa}\\
\spb{j}.{l}
&=& \ve^{\dot\alpha\dot\beta} 
(\tlambda_j)_{\dot\alpha} (\tlambda_l)_{\dot\beta}
= \bar{u}_+(k_j) u_-(k_l)\,.
\label{spinorproddefb}
\end{eqnarray}
(The sign of $\spb{j}.{l}$ in \eqn{spinorproddefb} matches that
in most of the QCD literature; in much of the `twistor' literature,
{\it e.g.} refs.~\cite{WittenTopologicalString,BCFGeneralized,
BCFRecursion,BCFW}, the opposite sign is used.)
These products are antisymmetric, $\spa{j}.{l} = - \spa{l}.{j}$, 
$\spb{j}.{l} = - \spb{l}.{j}$.  The usual momentum dot products
can be constructed from the spinor products using the relation,
\begin{equation}
\spa{l}.{j} \spb{j}.{l} 
= {1\over2} \Tr[ \ksl_j \ksl_l ] = 2k_j\cdot k_l = s_{jl} \,.
\label{spaspbeq}
\end{equation}
We will also use the notation
\begin{eqnarray}
\sand{a}.{K_{r\cdots s}}.{b} &=&
\bar{u}_-(k_a) \gamma_\mu u_-(k_b) \, K_{r\cdots s}^\mu
= \sum_{i=r}^{s}
\spa{a}.{i} \spb{i}.{b} \, ,
\label{aKbdef}\\
\sandmp{a}.{K_{r \cdots s} c }.{b} &=& \sum_{i=r}^{s}
\spa{a}.{i} \spb{i}.{c} \spa{c}.{b} \, , 
\label{aKcbdef}
\end{eqnarray}
where the sums run over cyclically ordered labels between $r$ and $s$,
as well as,
\begin{equation}
s_{r\cdots s} = K_{r\cdots s}^2 \, , 
\hskip 1 cm  s_{jlm} = (k_j+k_l+k_m)^2 \,.
\label{sjlmdef}
\end{equation}

Parity acts on a helicity amplitude by flipping the sign of all 
helicities.  This symmetry may be implemented by exchanging the 
left- and right-handed spinor products in the amplitude, 
$\spa{j}.{l} \leftrightarrow \spb{l}.{j}$.

For real momenta, $\lambda_i$ and $\tlambda_i$ are complex conjugates
of each other.  Therefore the spinor products are complex square roots
of the Lorentz products,
\def\vph{\vphantom{A}}
\begin{equation}
\spa{j}.{l}
= \sqrt{\vph{}s_{jl}} e^{i\phi_{jl}} \,,
\qquad
\spb{j}.{l} = \pm \sqrt{\vph{}s_{jl}} e^{-i\phi_{jl}} \,.
\label{spinorprodtwo}
\end{equation}
In this case, if all the $s_{jl}$ vanish, then so do all the spinor
products.  However, for complex momenta, \eqn{spinorprodtwo}
does not hold.  In fact, it is possible to choose all three left-handed 
spinors to be proportional, 
$\tlambda_1 = c_1 \tlambda_3$, $\tlambda_2 = c_2 \tlambda_3$, 
while the right-handed spinors are not proportional,
but obey the relation,
$c_1 \lambda_1 + c_2 \lambda_2 + \lambda_3 = 0$,
which follows from momentum conservation, $k_1+k_2+k_3 = 0$,
and \eqn{kfact}.  Then 
\begin{equation}
\spb1.2 = \spb2.3 = \spb3.1 = 0,
\label{threeptcomplex}
\end{equation}
while $\spa1.2$, $\spa2.3$ and $\spa3.1$ are all nonvanishing.

For this kinematical choice, the tree-level primitive amplitude for
two negative helicities and one positive helicity, $A_3^\tree$,
is nonsingular, even though all momentum invariants $s_{jl}$ vanish
according to \eqn{spaspbeq}.   (We assign helicities to particles
under the assumption that they are outgoing; if a particle is
incoming, its true helicity is opposite to the label.)
For three gluons, $A_3^\tree$ can be evaluated using the 
three-gluon-vertex in the color-ordered Feynman 
rules~\cite{OneLoopReview}, as
\begin{eqnarray}
A_3^\tree(1^-,2^-,3^+) &=& 
{i\over\sqrt{2}} \biggl[ 
\pol_1^-\cdot\pol_2^- \, \pol_3^+\cdot(k_1-k_2) 
+ \pol_2^-\cdot\pol_3^+ \, \pol_1^-\cdot(k_2-k_3)
\nonumber\\ 
&&\hskip0.5cm
+\ \pol_3^+\cdot\pol_1^- \, \pol_2^-\cdot(k_3-k_1) \biggr] \,.
\label{threept1}
\end{eqnarray}
In the spinor-helicity formalism~\cite{SpinorHelicity,TreeReview}, 
the polarization vectors $\pol_i^{\pm}$ are expressed as,
\begin{equation}
\pol_i^{\pm,\mu}\ =\ \pol^{\pm,\mu}(k_i,q_i)\ =\ 
 \pm { \langle q_i^\mp | \gamma^\mu | k_i^\mp \rangle
  \over \sqrt{2} \langle q_i^\mp | k_i^\pm \rangle } \,,
\label{poldefn}
\end{equation}
in terms of null reference momenta $q_i$, which may be chosen to 
simplify the computation.  In \eqn{threept1},
if we choose $q_2=q_1$ and $q_3=k_1$, then 
$\pol_1^-\cdot\pol_2^- = \pol_3^+\cdot\pol_1^- = 0$.
Upon using a Fierz rearrangement and momentum conservation,
the lone surviving term in \eqn{threept1} becomes,
\begin{eqnarray}
A_3^\tree(1^-,2^-,3^+)\ &=&\ 
i \sqrt{2}\ \pol_2^-\cdot\pol_3^+ \ \pol_1^-\cdot k_2
\nonumber\\
&=&%
%%%%% begin : a3mmp1
\ i\ { \spb{q_1}.{3} \spa1.2 \over \spb{q_1}.2 \spa1.3 }
    \ { \spb{q_1}.2 \spa2.1 \over \spb{q_1}.1 }
%%%%% end : a3mmp1
\nonumber\\
&=&%
%%%%% begin : a3mmp2
\ i\ { {\spa1.2}^2 \over \spa3.1 }
 \, { \spb{q_1}.{3} \over \spb{q_1}.1 } \, { \spa3.2 \over \spa3.2 }
%%%%% end : a3mmp2
\nonumber\\
&=&%
%%%%% begin : a3mmp3
 -i\ { {\spa1.2}^2 \over \spa3.1 }
 \, { \spb{q_1}.{1} \spa1.2 \over \spb{q_1}.1 \spa3.2 } \,,
%%%%% end : a3mmp3
\label{threept2}
\end{eqnarray}
or
\begin{equation}
A_3^\tree(1^-,2^-,3^+)
\ =\ 
%%%%% begin : a3mmp4
i { {\spa1.2}^4 \over \spa1.2 \spa2.3 \spa3.1 } \,.
%%%%% end : a3mmp4
\label{threeptfinal}
\end{equation}

\Eqn{threeptfinal} is the first member of the sequence of
Parke-Taylor, or maximally-helicity-violating (MHV), tree
amplitudes for $n$ gluons~\cite{ParkeTaylor,BGSix,MPX},
\begin{equation}
A_n^{{\rm tree\,MHV},\,jk}\ \equiv\ 
A_n^\tree(1^+,\ldots,j^-,\ldots,k^-,\ldots,n^+)
\ =\ i { {\spa{j}.{k}}^4 \over \spa1.2 \spa2.3 \cdots \spa{n}.1 } \,.
\label{MHVtree}
\end{equation}
In this expression, only gluons $j$ and $k$ have negative helicity;
the remaining $(n-2)$ gluons have positive helicity.
For $n \geq4$, these amplitudes are well-defined for real momenta.
However, for $n=3$, formula~(\ref{threeptfinal}) only makes sense
for complex momenta of the type~(\ref{threeptcomplex}).

There is a class of complex momenta conjugate to \eqn{threeptcomplex},
for which 
\begin{equation}
\spa1.2 = \spa2.3 = \spa3.1 = 0,
\label{threeptcomplexconj}
\end{equation}
while $\spb1.2$, $\spb2.3$ and $\spb3.1$ are all nonvanishing.
Such momenta make the parity-conjugate three-point amplitude,
\begin{equation}
A_3^\tree(1^+,2^+,3^-)
\ =\ 
%%%%% begin : a3ppm1
-i { {\spb1.2}^4 \over \spb1.2 \spb2.3 \spb3.1 } \,,
%%%%% end : a3ppm1
\label{threeptfinalconj}
\end{equation}
well-defined.  When the amplitude $A_3^\tree(1^-,2^-,3^+)$
appears in the `wrong' kinematics~(\ref{threeptcomplexconj}),
it should be set to zero, because more vanishing spinor products
appear in the numerator than in the denominator.

For any on-shell complex momenta, the all-positive amplitude
vanishes, 
\begin{equation}
A_3^\tree(1^+,2^+,3^+)\ =\ 0.
\label{threeplusvanish}
\end{equation}
This result follows easily by choosing all reference momenta $q_i$ 
to be the same,
which forces $\pol_i^+\cdot\pol_j^+$ to vanish for all pairs $i,j$.
The same result holds for $A_3^\tree(1^-,2^-,3^-)$, of course.
In a similar fashion, the three-point amplitude for 
a pair of massless quarks (which must carry opposite helicity)
plus one gluon is well-defined and nonzero using the 
kinematics~(\ref{threeptcomplex}) if the gluon helicity is negative,
and using the kinematics~(\ref{threeptcomplexconj}) if it is positive.
It vanishes for the `wrong' kinematics.  These rules will
become important later for evaluating 
generalized unitarity cuts and recursive diagrams
containing three-point vertices.

%%%%%%%%%%%%%%%%%%%%%%%%%%%%%%%%%%%%%%

\subsection{Complex Kinematics to Utilize Factorization Information}
\label{KinFactSubsection}

We have seen how suitable complex kinematics can simplify
the structure of the three-point amplitude.  Much more significantly,
however, complex kinematics allow the exploration of generic
factorization singularities of on-shell amplitudes, and the use of
factorization information to reconstruct the amplitude,
as was recognized at tree-level by Britto, Cachazo, Feng and 
Witten (BCFW)~\cite{BCFW}.
We will review the BCFW recursion relations in 
the next section.  Here we shall just discuss
the simple example of the Parke-Taylor amplitudes.
The idea is to embed a tree amplitude $A_n^\tree$ into
a one-complex-parameter family of on-shell amplitudes $A_n^\tree(z)$.
The rationale 
for introducing the complex parameter $z$ is that it allows
us to apply the full power of complex variable theory to reconstruct
amplitudes from their poles. 
The simplest way to introduce this parameter is by modifying, or `shifting' 
the momenta of just two of the $n$ partons, in a way that 
keeps them on shell.  Because of \eqn{kfact}, the two momenta
will automatically remain on shell if we shift the spinor variables,
and then define the shifted momenta to be the products of the new
left- and right-handed spinors.

We define the $\Shift{j}{l}$ shift to be~\cite{BCFW},
\begin{equation}
\tlambda_j \rightarrow \tlambda_j - z\tlambda_l \,,
\qquad\quad
\lambda_l \rightarrow \lambda_l + z\lambda_j \,,
\label{SpinorShift}
\end{equation}
where $z$ is a complex parameter.  The shift leaves
untouched $\lambda_j$, $\tlambda_l$, and the spinors
for all the other particles in the process.
Under the shift, the corresponding momenta shift as,
\begin{eqnarray}
&k_j^\mu &\rightarrow k_j^\mu(z) = k_j^\mu - 
      {z\over2}{\sand{j}.{\gamma^\mu}.{l}},\nonumber\\
&k_l^\mu &\rightarrow k_l^\mu(z) = k_l^\mu + 
      {z\over2}{\sand{j}.{\gamma^\mu}.{l}} \,,
\label{MomShift}
\end{eqnarray}
which preserves their masslessness, $k_j^2(z) = 0 = k_l^2(z)$,
as well as overall momentum conservation.

\begin{figure}[t]
\centerline{\epsfxsize 1.9 truein\epsfbox{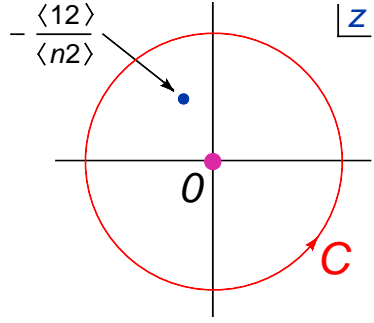}}
\caption{Analytic structure of $A_n^{{\rm tree\,MHV},\,jn}(z)/z$ under 
the $\Shift{n}{1}$ shift.}
\label{MHVCauchyFigure}
\end{figure}

Suppose we apply the $\Shift{n}{1}$ shift,
$\tlambda_n \rightarrow \tlambda_n - z\tlambda_1$, 
$\lambda_1 \rightarrow \lambda_1 + z\lambda_n$,
to the MHV amplitude~(\ref{MHVtree}), for the case $k=n$
(and $j\neq1$). Because the formula contains only right-handed $\lambda_i$
spinors,
the only induced $z$ dependence arises from terms containing
$\lambda_1$. The spinor product $\spa1.2$ is shifted to 
$\spa1.2 \to \spa1.2 + z \spa{n}.2$
The spinor product $\spa{n}.1$ is unaffected,
because $\spa{n}.1 \to \spa{n}.1 + z \spa{n}.{n}$,
but $\spa{n}.{n} = 0$ by antisymmetry.
Thus the MHV amplitude becomes
\begin{equation}
A_n^{{\rm tree\,MHV},\,jn}(z)\ 
\ =\ i { {\spa{j}.{n}}^4 \over 
(\spa1.2 + z \spa{n}.2) \spa2.3 \cdots \spa{n}.1 } \,.
\label{MHVtreez}
\end{equation}
If we divide this shifted amplitude by $z$, we get a function with
two poles at finite $z$, one at the origin and one at
$z=-\spa1.2/\spa{n}.2$, as shown in \fig{MHVCauchyFigure}.
The function behaves like $1/z^2$ as $z\to \infty$,
so the integral around the circle $C$ at infinity vanishes,
\begin{equation}
0 = \oint_C {dz \over 2\pi i} { A_n^{{\rm tree\,MHV},\,jn}(z) \over z } \,.
\label{CauchyVanish}
\end{equation}
Cauchy's theorem then guarantees that the two poles at finite $z$
have equal and opposite residue.

%%%%%% FIGURE %%%%%%%%%%%
%%%%%%%%%%%%%%%%%

The residue of the pole at the origin is simply the
MHV amplitude we want to compute,
$A_n^{{\rm tree\,MHV},\,jn}(0) = A_n^{{\rm tree\,MHV},\,jn} \equiv A_n$.
The residue of the second pole is determined by the factorization
properties of the amplitude, because it occurs at the point that
the intermediate momentum 
$K_{12}(z)\ \equiv\ k_1(z) + k_2$
goes on shell, 
$K_{12}^2(z) = (\spa1.2 + z \spa{n}.2) \spb2.1\ \to\ 0$.
That is, at the value $z = z_{12} \equiv -\spa1.2/\spa{n}.2$ we have,
\begin{eqnarray}
\spash\hat{1}.2\ &=&\ \spa1.2 + z_{12} \spa{n}.2 \ =\ 0,
\nonumber\\
\spbsh\hat{1}.2\ &=&\ \spb1.2 \ \neq\ 0,
\label{MHVBCFdiagram}\\
s_{\hat{1}2}\ &=&\ \spash{\hat{1}}.{2} \, \spbsh{2}.{\hat{1}}\ =\ 0.
\nonumber
\end{eqnarray}
A hat on a kinematic variable means that a shift of the
form~(\ref{SpinorShift}) has been applied to the variable, 
with $z$ then fixed to the value that puts an intermediate momentum
on shell.
The key point is that, because an intermediate state goes on shell,
the amplitude factorizes at this point into the product of two lower-point
amplitudes, multiplied by the diverging propagator.
In general there is a sum over the helicity $h$ of the intermediate
state, $h=\pm1$ for intermediate gluons. 

Equating the residue at $z = z_{12}$
to the negative of the one at the origin, 
we have
\begin{eqnarray}
A_n^\tree\ &=&\ \sum_{h = \pm}
A_{n-1}^\tree(\hat{K}_{12}^h,3^+,\ldots,j^-,\ldots\hat{n}^-)
\nonumber\\
&&\hskip0cm \times 
 \Biggl[ - \Res_{z = z_{12}} \; 
 \Bigl({ 1 \over z } {i\over \hat{K}_{12}^2(z)}\Bigr) \Biggr]
A_3^\tree(\hat{1}^+,2^+,-\hat{K}_{12}^{-h})
\nonumber\\
&=&\ 
A_{n-1}^\tree(\hat{K}_{12}^+,3^+,\ldots,j^-,\ldots\hat{n}^-)
{i\over s_{12}}
A_3^\tree(\hat{1}^+,2^+,-\hat{K}_{12}^{-}) \,.
\label{MHVBCFdiag2}
\end{eqnarray}
In the second step we evaluated the residue in brackets,
and used the vanishing of $A_3^\tree(1^+,2^+,3^+)$ in
\eqn{threeplusvanish} to reduce the helicity sum to a single term.

\Eqn{MHVBCFdiag2} is the prototype for the
tree-level BCFW recursion relation reviewed in 
\sect{TreeRecursionSection}.  In general, there will
be several terms on the right-hand side, corresponding to
different nontrivial factorization channels which can
be probed for suitable values of $z$.
The MHV amplitudes are unique in having no multi-particle 
poles, which is related to the vanishing of $n$-gluon amplitudes
with fewer negative helicities,
\begin{equation}
A_n^\tree(1^\pm,2^+,3^+,\ldots,n^+) = 0.
\label{npointvanish}
\end{equation}
For this reason, the recursion relation~(\ref{MHVBCFdiag2})
in the MHV case has but a single term.

We can further evaluate \eqn{MHVBCFdiag2} using the explicit 
form~(\ref{MHVtree}) of the MHV amplitudes, as a simple exercise
in manipulating the hatted variables that appear, and as an
on-shell recursive proof of the formula.  (As an historical note,
the first proof of \eqn{MHVtree} employed recursion
relations of the off-shell variety~\cite{BGRecursion}.)
Inserting the forms of the $(n-1)$-point and three-point amplitudes,
\eqn{MHVBCFdiag2} becomes,
\begin{equation}
A_n^\tree = 
%%%%% ignore begin : An1[n_,j_]
{ i \, {\spash{j}.{\hat{n}}}^4 
\over \spash{\Kh_{12}}.3 \spa3.4 \cdots \spash{(n-1)}.{\hat{n}} 
\spash{\hat{n}}.{\Kh_{12}} \vphantom{\hat{K}} }
{ i \over s_{12} }
{ - i \, {\spbsh{\hat{1}}.2}^3 
\over \spbsh2.{(-\Kh_{12})} \spbsh{(-\Kh_{12})}.{\hat{1}} 
  \vphantom{\hat{K}} } \,.
%%%%% ignore end : An1[n_,j_]
\label{MHVBCFdiag3}
\end{equation}
We need to continue the spinors for $(-\Kh_{12})$
into those for $\Kh_{12}$.  Because a pair of such spinors appears,
we pick up a minus sign. (Determining the sign for the case of an 
intermediate quark line is more subtle.)

Because the $\Shift{n}{1}$ shift leaves $\tlambda_1$ and $\lambda_n$ alone,
we can let $\hat{n}\rangle\to n\rangle$ and $\hat{1}] \to 1]$
in \eqn{MHVBCFdiag3}.  
Because the shifted momentum is proportional to $\tlambda_1\lambda_n$,
a hatted momentum appearing in a right-handed spinor product 
with $n$, or in a left-handed spinor product with $1$, can have its 
hat removed as well.  For this reason, it is very convenient to use 
factors of $\spash{n}.{\Kh_{12}}$ and $\spbsh{\Kh_{12}}.1$
to clean up other spinor products containing $\Kh_{12}$,
inserting them into the numerator and denominator of expressions as
needed.  In the present case, the necessary factors are already present.
\Eqn{MHVBCFdiag3} becomes,
\begin{eqnarray}
A_n^\tree &=& 
%%%%% ignore begin : An2
i \, { {\spa{j}.{n}}^4 \, {\spb1.2}^3 
\over \spa3.4 \cdots \spa{(n-1)}.{n}
\, \langle n^- | \Kh_{12} | 2^- \rangle
\, \langle 3^- | \Kh_{12} | 1^- \rangle }
{ 1 \over \spa1.2 \spb1.2 }
%%%%% ignore end : An2
\nonumber\\
&=& 
%%%%% ignore begin : An3
i \, { {\spa{j}.{n}}^4 \, {\spb1.2}^3 
\over \spa3.4 \cdots \spa{(n-1)}.{n}
\, \spa{n}.1 \spb1.2 \, \spa3.2 \spb2.1 }
{ 1 \over \spa1.2 \spb1.2 }
%%%%% ignore end : An3
\nonumber\\
&=& 
%%%%% ignore begin : An4
i \, { {\spa{j}.{n}}^4 
\over \spa1.2 \spa2.3 \spa3.4 \cdots \spa{(n-1)}.{n} \spa{n}.1 } \,.
%%%%% ignore end : An4
\label{MHVBCFdiag4}
\end{eqnarray}
The final form indeed matches the expression~(\ref{MHVtree}),
thus confirming it recursively.

%%%%%%%%%%%%%%%%%%%%%%%%%%%%%%%%%%%%%%

\subsection{Complex Kinematics for Generalized Unitarity}
\label{KinGenUSubsection}

\Fig{complexkinFigure}(b) shows a second class of configurations for which
complex kinematics are useful, namely generalized unitarity conditions,
which will be discussed in more detail in \sect{UnitaritySection}.  At one
loop, conventional unitarity constraints on amplitudes are analyzed by
putting two intermediate states on shell.  For example, if legs 1, 2, 3
and 4 in \fig{complexkinFigure}(b) are outgoing, and legs 5, 6, 7 and 8 are
incoming, then the conventional cut in the 1234 channel is computed by
imposing $\ell_1^2 = \ell_5^2 = 0$.  This constraint can be realized with all
momenta real in Minkowski space.  It can then be interpreted as a $4 \to
2$ particle scattering process, followed by a $2 \to 4$ particle
scattering.

Generalized unitarity corresponds
to requiring {\it more} than two internal particles to be on shell.
Often these constraints cannot be realized with real Minkowski momenta.
Suppose we try to add the condition $\ell_3^2 = 0$ to the standard
cut constraints $\ell_1^2 = \ell_5^2 = 0$ in \fig{complexkinFigure}(b).
The problem is that $1\to3$ processes are forbidden for real,
non-collinear massless momenta, although $2\to2$ processes are 
allowed.  After setting $\ell_3^2 = 0$, \fig{complexkinFigure}(b)
contains a four-point subamplitude in 
which leg $\ell_1$ is incoming, and legs 1 and 2 are outgoing
with $s_{12}$ nonzero.  We can arrange for this to be a real Minkowski
process by taking $\ell_3$ to be incoming.  But then the subamplitude
below it in the figure has only leg $\ell_5$ incoming,
and legs $\ell_3$, 3 and 4 outgoing.  It cannot correspond to a
real on-shell process, as long as $s_{34}$ is nonzero.

Similarly, we cannot impose $\ell_7^2=0$ on top of $\ell_1^2 = \ell_5^2 = 0$,
in the process $5\,+\,6\,+\,7\,+\,8 \to 1\,+\,2\,+\,3\,+\,4$,
and still have real momenta.  Typically then, one needs to allow
for complex momenta in order to implement generalized cut conditions.
As an even simpler example, if we consider a quadruple cut with fewer 
than the eight external momenta shown in \fig{complexkinFigure}(b), 
then at least one tree amplitude will have only three external legs,
and this fact alone dictates complex momenta.

On the other hand, generalized cut conditions {\it can} sometimes 
be satisfied with only real momenta.
In \fig{complexkinFigure}(b), suppose now that the scattering process
has legs 4, 5, 6 and 7 incoming, and legs 8, 1, 2 and 3 outgoing.
Then it is possible to solve all four internal constraints, 
$\ell_1^2 = \ell_3^2 = \ell_5^2 = \ell_7^2 = 0$, with real Minkowski momenta,
corresponding to scattering that proceeds from the lower left to
the upper right of the figure; that is, $5\,+\,6 \to \ell_7\,+\,\ell_5$, 
followed by $\ell_7\,+\,7 \to 8\,+\,\ell_1$ and 
$4\,+\,\ell_5 \to \ell_3\,+\,3$,
followed by $\ell_3\,+\,\ell_1 \to 1\,+\,2$.

%%%%%%%%%%%%%%%%%%%%%%%%%%%%%%%%%%%%%%%

\section{On-Shell Recursion for Tree Amplitudes}
\label{TreeRecursionSection}

\subsection{General Framework}

In this section we describe the construction of tree amplitudes via
on-shell recursion.  As alluded to in the previous section, the BCFW
recursion relation~\cite{BCFRecursion,BCFW} is based on introducing a
complex-parameter-dependent shift of two of the external massless
spinors, as given in \eqn{SpinorShift}.  The construction of tree
amplitudes via on-shell recursion essentially amounts to generalizing
and reversing the steps in \sect{KinFactSubsection}.  Instead of
starting with a known amplitude, and verifying its analytic properties
under the parameter-dependent shift, we use such properties to
systematically construct unknown amplitudes.

Following the same procedure as for the MHV case in \eqn{MHVtreez}, for
generic amplitudes we define an analytically continued amplitude,
\begin{equation}
A(z)\ =\ A(k_1,\ldots,k_j(z),k_{j+1},\ldots,k_l(z),\ldots,k_n),
\end{equation}
which remains on-shell, but depends on the complex parameter $z$.
If $A$ is a tree amplitude, then $A(z)$ is a rational function of $z$.
The physical amplitude is given by $A(0)$. 

Following the MHV case~(\ref{CauchyVanish}) consider the
contour integral,
\begin{equation}
{1\over 2\pi i} \oint_C {dz\over z}\,A(z) \,,
\end{equation}
where the contour is taken around the circle at infinity.
If $A(z)\rightarrow 0$ as $z\rightarrow\infty$, the contour integral
vanishes and we obtain
a relationship between the  physical amplitude, at $z=0$, and 
a sum over residues for the poles of $A(z)$, located at $z_{\alpha}$,
\begin{equation}
A(0) = -\sum_{{\rm poles}\ \alpha} \Res_{z=z_\alpha}  {A(z)\over z}\,.
\label{TreeResidueSum}
\end{equation}
%

%%%%%% FIGURE %%%%%%%%%%%
\begin{figure}[t]
\centerline{\epsfxsize 2.2truein\epsfbox{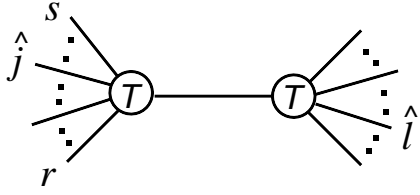}}
\caption{Diagrammatic representation of one term in the tree-level recursion
relation.  The label `$T$' refers to tree vertices, which are on-shell
lower-point amplitudes. The momenta $\hat{\jmath}$ and $\hat{l}$ undergo the
$\Shift{j}{l}$ shift in \eqn{SpinorShift}. The two shifted legs must be 
attached to separate tree vertices.}
\label{TreeGenericFigure}
\end{figure}
%%%%%%%%%%%%%%%%%

To determine the residues at each pole, we use the general
factorization properties that any amplitude must satisfy as an
intermediate momentum $K^\mu$ goes on shell, $K^2 \to 0$.
In general, the residue is given by a product of lower-point
on-shell amplitudes.
Only a subset of the possible factorization limits for an
amplitude are explored by the $z$-dependent shift.
Poles in the $z$ plane can develop in any channel that has
leg $j$ on one side of the pole and leg $l$ on the other side,
because the intermediate momentum is $z$-dependent.  Thus we can solve
the on-shell condition,
\begin{eqnarray}
\hskip -.6 cm 
0\ &=&\ [K_{r\cdots s}(z)]^2 
\ =\ \Bigl(k_r + k_{r+1} + \cdots + k_j(z) 
       + \cdots + k_s \Bigr)^2
\nonumber\\
\ &=&\ K_{r\cdots s}^2 - z \sand{j}.{K_{r\cdots s}}.l\,.
\label{solvevanish}
\end{eqnarray}
The solution is 
\begin{equation}
z_{rs} = {K_{r\cdots s}^2 \over \sand{j}.{K_{r\cdots s}}.l } \,.
\end{equation}

A contribution to the recursion relation from this residue
is illustrated diagrammatically in \fig{TreeGenericFigure}.
To get the precise form of the contribution, using \eqn{TreeResidueSum},
we need to evaluate the residue (as we did in \eqn{MHVBCFdiag2} for a 
special case),
\begin{equation}
- \Res_{z = z_{rs}} \; 
 \Bigl({ 1 \over z } {i\over K_{r\cdots s}^2(z)}\Bigr)
\ =\ {i \over K_{r\cdots s}^2}  \,.
\label{GenResidue}
\end{equation}
The final form of the tree-level recursion relation 
is~\cite{BCFRecursion,BCFW}
\begin{equation}
A(0) = \sum_{r,s,h} 
 A^h_L(z = z_{rs}) { i \over K_{r\cdots s}^2 } A^{-h}_R(z = z_{rs}) \,.
\label{BCFWRepresentation}
\end{equation}
Generically we have a double sum, labeled by $r,s$, over recursive
diagrams, with legs $j$ and $l$ always appearing on opposite 
sides of the pole.  
There is also a sum over the helicity $h$ of the intermediate state.
The squared momentum associated with the pole, $K_{r\cdots s}^2$, 
arising from \eqn{GenResidue}, is evaluated in the unshifted kinematics.
The on-shell tree amplitudes $A_L$ and $A_R$ are evaluated in kinematics
that have been shifted by \eqn{SpinorShift}, with $z=z_{rs}$.
The shifted momenta for such kinematics are indicated by hats.

\Eqn{BCFWRepresentation} may alternatively be derived
by expressing $A(z)$ as a sum over poles multiplied by their residues
(under our assumption that $A(z) \rightarrow 0$ 
as $z \rightarrow \infty$).
This representation, valid for all $z$, is 
\begin{equation}
A(z) = \sum_{r,s,h} 
 A^h_L(z = z_{rs}) { i \over 
K_{r\cdots s}^2 - z \sand{j}.{K_{r\cdots s}}.l } 
A^{-h}_R(z = z_{rs})  \,.
\end{equation}
By setting $z=0$ in this formula, we recover the recursion
relation~(\ref{BCFWRepresentation}).

To derive the recursion relation, we assumed that the amplitude 
$A(z)$ vanishes as $z\to\infty$.  If all the external particles
are gluons, then the validity of this assumption
depends only on the helicity of the two shifted legs. 
There are four different cases, which we can label by
$\Shift{h_j}{h_l} = \Shift{\pm}{\pm}$.
A shift of type $\Shift{+}{-}$ does not generally make $A(z)$
vanish at infinity.
As an example, the MHV amplitude~(\ref{MHVtree}) behaves as 
either $z^2$ or $z^3$ as $z\to\infty$, because of the factor of 
${\spa{j}.{k}}^4$ in the numerator.  The case $\Shift{-}{+}$ is
the simplest to analyze because 
each individual Feynman diagram vanishes separately~\cite{BCFW}.
Shifts of type $\Shift{-}{-}$ and $\Shift{+}{+}$ also make 
the amplitude vanish as $z\rightarrow\infty$,
but here cancellations between Feynman diagrams are required.
The vanishing behavior has been proven using generalizations
of the shift~(\ref{SpinorShift}) that affect three or more
momenta~\cite{BGKS,SW}.  

The on-shell recursion relation~(\ref{BCFWRepresentation}) 
contains spinor products involving hatted momenta.  
For the purposes of numerical evaluation, we can leave the amplitudes
in this form, because the complex hatted momenta are built
from well-defined spinors, whose inner products can be 
computed from \eqns{spinorproddefa}{spinorproddefb}.
However, for analytic purposes
it is useful to eliminate the hatted momenta in favor of 
external momenta.  We can make use of the following relations,
\begin{eqnarray}
&& \spash{\hat \jmath}.{a} = \spa{j}.a\,, \hskip 1 cm
   \spbsh{\hat \jmath}.{a} = \spb{j}.a - z_{rs} \spb{l}.a \,, \nn \\
&& \spbsh{\hat l}.{a} = \spb{l}.a \,, \hskip 1.3 cm
   \spash{\hat l}.{a} = \spa{l}.a + z_{rs} \spa{j}.a \,,  \nn \\
&&   \spbsh{\Kh_{r\cdots s}}.{a} = 
         {\sand{j}.{\Kh_{r\cdots s}}.{a} \over \vK \spash{j}.{\Kh_{r\cdots s}}}
   = {\sand{j}.{K_{r\cdots s}}.{a} \over \vp \spash{j}.{\Kh_{r\cdots s}}}
                                                              \,, \nn \\
&&  \spash{a}.{\Kh_{r\cdots s}} = 
       {\sand{a}.{\Kh_{r\cdots s}}.{l} \over \vK \spbsh{\Kh_{r\cdots s}}.{l}}
  =  {\sand{a}.{K_{r\cdots s}}.{l} \over \vK \spbsh{\Kh_{r\cdots s}}.{l}}
                                                              \,, \nn\\
&& \spbsh{\Kh_{r\cdots s}}.{\hat \jmath} = 
               { - K^2_{r \cdots s} + 2 k_j \cdot K_{r\cdots s}
                       \over \vK \spash{j}.{\Kh_{r\cdots s}}}\,, \nn\\
&& \spash{\Kh_{r\cdots s}}.{\hat l} = 
              - {K^2_{r \cdots s} + 2 k_l \cdot K_{r\cdots s}
                       \over \vK \spbsh{\Kh_{r\cdots s}}.{l}}\,.
\label{Simplifications}
\end{eqnarray}
To simplify the expressions we used, for example, 
\begin{equation}
\sand{j}.{\Kh_{r\cdots s}}.{a} = \sand{j}.{K_{r\cdots s}}.{a} - 
      {z_{rs} \over 2} \sand{j}.{\gamma_\mu}.{a} \sand{j}.{\gamma^\mu}.l \,,
\end{equation}
where the last term vanishes by a Fierz identity.  
The product $A_L^h \times A_R^{-h}$ is homogeneous in the spinors
carrying the intermediate momentum $\Kh_{r\cdots s}$.
Because of this, any remaining factors of $\spash{j}.{\Kh_{r\cdots s}}$ 
and $\spbsh{\Kh_{r\cdots s}}.l$ can be paired up to give factors of
\begin{eqnarray}
\spash{j}.{\Kh_{r\cdots s}}\spbsh{\Kh_{r\cdots s}}.l 
= \sand{j}.{K_{r\cdots s}}.l \,.
\end{eqnarray}

Recursive diagrams containing three-point amplitudes often vanish
because the `wrong' kinematics are present.  
In general, if a $\Shift{j}{l}$ shift is used, and the recursive
diagram contains a three-vertex with two positive helicities, 
one of which is $j$, then the diagram vanishes.
The reason is that the spinor $\lambda_j$ is unaffected by the shift,
so its product with the spinor for the other external leg $a$ in the 
three-point amplitude, $\spa{j}.{a}$, remains nonvanishing.
Therefore $\spb{j}.{a}$, and all of the left-handed spinor products,
must vanish, and so the three-vertex with two positive helicities
vanishes, as discussed in \sect{KinematicsSection}.
Similarly, three-vertices with two negative helicities
can also be dropped, when one of the three legs is $l$.

%%%%%%%%%%%%%%%%%%%%%%%%%%%%%%%%%%%%%%%%%

\subsection{Tree-level examples}

%%%%%% FIGURE %%%%%%%%%%%
\begin{figure}[t]
\centerline{\epsfxsize 3.7 truein\epsfbox{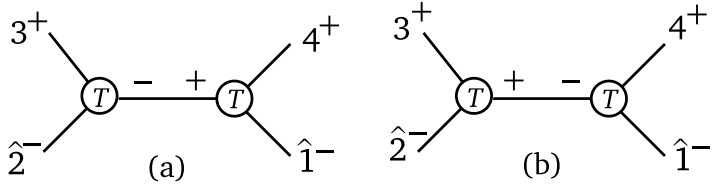}}
\caption{The two on-shell recursive diagrams obtained from the
$\Shift{1}{2}$ shift. The label `$T$' in the blobs indicates that the
vertices are tree amplitudes.  Diagram (a) vanishes, as explained in the
text.}
\label{FourPtTreeFigure}
\end{figure}
%%%%%%%%%%%%%%%%%

As a first example, consider the amplitude
$A_4^\tree(1^-,2^-,3^+,4^+)$.  This amplitude can be
constructed recursively from the three-point amplitudes given in
\eqns{threeptfinal}{threeptfinalconj}.  
As discussed above, for the $\Shift12$ shift (a $\Shift{-}{-}$ shift)
the amplitude vanishes for large $z$.
Using this shift, there are two potential terms in the recursion relation, 
corresponding to diagrams (a) and (b) in \fig{FourPtTreeFigure}.
\begin{equation}
A_4^\tree(1^-,2^-,3^+,4^+) = D_4^{\rm (a)} + D_4^{\rm (b)} \,.
\end{equation}
The first of these diagrams, 
\begin{eqnarray}
D_4^{\rm (a)} &=& A_3^\tree(\hat 2^-, 3^+, -\Kh_{23}^-) \,
{i \over s_{23}} \, A_3^\tree(4^+, \hat 1^-, \Kh_{23}^+) \,,
\end{eqnarray}
vanishes because of the `wrong' kinematics discussed above.

Now evaluate diagram (b), using \eqns{threeptfinal}{threeptfinalconj},
with an overall minus sign for the continuation $-\Kh_{23}\to\Kh_{23}$,
\begin{eqnarray}
D_4^{\rm (b)} &=& A_3^\tree(\hat 2^-, 3^+, -\Kh_{23}^+) \,
{i \over s_{23}} \, A_3^\tree(4^+, \hat 1^-, \Kh_{23}^-) \nn \\
&=&
%%%%% begin : D4
- i {\spbsh{3}.{\Kh_{23}}^3 \over \vK \spbsh{\hat 2}.{3} 
                   \spbsh{\Kh_{23}}.{\hat 2}} \, {1 \over s_{23} \vK} \,
  { \spash{\hat1}.{\Kh_{23}}^3 
\over  \vK \spash4.{\hat 1} \spash{\Kh_{23}}.{4}  } \,.
%%%%% end : D4
\end{eqnarray}
This form is already satisfactory for the purpose of 
evaluating the amplitude numerically.
It is, however, a useful exercise to eliminate hatted momenta
in favor of unhatted external momenta. Applying the 
substitutions~(\ref{Simplifications}) and simplifying the expression 
for diagram (b), we find
\begin{equation}
A_4^\tree(1^-,2^-,3^+,4^+)
\ =\ i \, { {\spa{1}.{2}}^4 \over \spa1.2 \spa2.3 \spa3.4 \spa4.1 } \,,
\label{MHVtree4}
\end{equation}
in agreement with the MHV formula~(\ref{MHVtree}).

Interestingly, using the on-shell recursion relations, the four-point
amplitude, and indeed all tree amplitudes, can be constructed from the
on-shell three-vertices.  The four-point vertex in the Yang-Mills
Feynman rules is unnecessary.  On the other hand, the four-point
vertex is related by gauge invariance to the three-point vertex.
Gauge invariance is necessary to decouple unphysical states.
The recursion relations rely on the fact that such states are 
not present in factorization limits.  In this way, they implicitly
incorporate the correct four-point vertex.

%%%%%% FIGURE %%%%%%%%%%%
\begin{figure}[t]
\centerline{\epsfxsize 3.7 truein\epsfbox{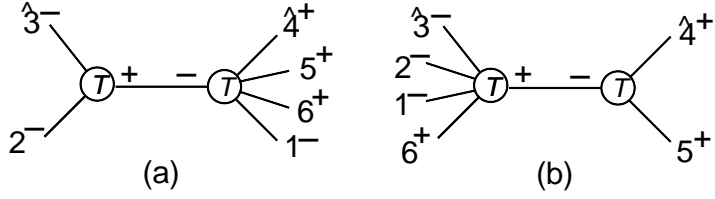}}
\caption{The nonvanishing recursive diagrams for
$A_6^\tree(1^-, 2^-, 3^-, 4^+, 5^+, 6^+)$, using a $\Shift34$ shift.
}
\label{NMHV6PtRecursFigure}
\end{figure}
%%%%%%%%%%%%%%%%%

Consider now the less trivial example of a six-point next-to-MHV
(NMHV) amplitude, $A_6^\tree(1^-, 2^-, 3^-, 4^+, 5^+, 6^+)$. Using a
$\Shift34$ shift (a $\Shift{-}{+}$ shift) yields the two recursive
diagrams in \fig{NMHV6PtRecursFigure},
\begin{equation}
A_6^\tree(1^-, 2^-, 3^-, 4^+, 5^+, 6^+) = D_6^{\rm (a)} + D_6^{\rm (b)} \,.
\label{A6DaDb}
\end{equation}
The first of these diagrams gives, 
\begin{eqnarray}
D_6^{\rm (a)} &=& A_3^\tree(2^-,\hat 3^-, -\Kh_{23}^+) {i \over s_{23}} 
          A_5^\tree(\hat 4^+, 5^+, 6^+, 1^-, \Kh_{23}^-) \\
&=&
%%%%% begin : D6a1
 i \, {\spash2.{\hat 3}^3 
         \over \spash{\hat 3}.{\Kh_{23}} \spash{\Kh_{23}}.2\vK} 
       \, {1 \over s_{23}\vK} \,
       {\spash1.{\Kh_{23}}^3 
       \over \spash{\hat 4}.5 \spa5.6 \spa6.1 \spash{\Kh_{23}}.{\hat 4}
             \vK} \,.
%%%%% end : D6a1
\end{eqnarray}
Applying \eqn{Simplifications}, we may rewrite this expression in terms
of unhatted momenta, using the definitions in \eqns{aKbdef}{sjlmdef},
\begin{eqnarray}
D_6^{\rm (a)} &=& 
%%%%% begin : D6a2
i \, 
  {\spa2.3^3 \over \sand{3}.{\Kinsl_{23}}.4 \sand2.{\Kinsl_{23}}.4 } 
       \, {1 \over \spa2.3 \spb3.2} \nn\\
&& \hskip 1 cm \times
      {\sand1.{\Kinsl_{23}}.4 ^3 
       \over (\spa4.5 + \spb2.3 \spa3.5/\spb2.4)
                     \spa5.6 \spa6.1 s_{234} } \nn \\
%%%%% end : D6a2
%
&=&
%%%%% begin : D6a3
 i \, {\spa2.3^2 \over \spa3.2\spb2.4 \spa2.3 \spb3.4 } 
       \, {1 \over \spb3.2} \nn\\
&& \hskip 1 cm \times
      {\spb2.4 \sand1.{\Kinsl_{23}}.4 ^3 
       \over \sand5.{\Kinsl_{34}}.2 \spa5.6 \spa6.1 s_{234} } \nn \\
%
%%%%% end : D6a3
&=&
%%%%% begin : D6a4
  i \, {\sand1.{\Kinsl_{23}}.4^3 \over 
            \spb2.3 \spb3.4 \spa5.6 \spa6.1
                           \sand5.{\Kinsl_{34}}.2  s_{234} } \,.
%%%%% end : D6a4
\end{eqnarray}
Similarly, diagram (b) in \fig{NMHV6PtRecursFigure} is given by
\begin{eqnarray}
D_6^{\rm (b)} &=& A_5^\tree(6^+, 1^-, 2^-, \hat{3}^-, \Kh_{45}^+) 
{i \over s_{45}}  A_3^\tree(\hat4^+, 5^+, -\Kh_{45}^-) \nn \\
&=&
%%%%% begin : D6b1
 i \, {\sand3.{\Kinsl_{12}}.6^3 \over 
\spb6.1 \spb1.2 \spa3.4 \spa4.5  \sand5.{\Kinsl_{61}}.2 s_{345}} \, .
%%%%% end : D6b1
\end{eqnarray}

It is interesting that this kind of representation of the amplitude 
is intimately
connected~\cite{RSVTree} to the forms in which tree amplitudes appear
in the infrared singularities of certain one-loop amplitudes~\cite{Neq47pt}.  
This feature is related to the appearance of
denominator factors such as $\sand5.{K_{34}}.2$ 
at one loop, where they arise in the reduction of various loop
integrals to a basic set of integrals.  They can be thought of as
spinor `square roots' of certain Gram determinants.

The expression $\sand5.{K_{34}}.2$ vanishes 
on a subspace of phase space.
For example, when $K_{34}^\mu$ is any linear combination of 
$k_2^\mu$ and $k_5^\mu$, it vanishes using the massless Dirac equation,
$\ksl u_\pm(k) = 0$.  Note that $\sand5.{K_{34}}.2 = -\sand5.{K_{61}}.2$ 
by momentum conservation, and so the latter form also vanishes on the
same subspace.  This subspace does not correspond to physical factorizations
of the amplitude, so
$D_6^{\rm (a)}$ and $D_6^{\rm (b)}$
each have spurious singularities on it.
However, their sum, the full amplitude $A_6^\tree$, is nonsingular.
In principle, a numerical program should check for small values
of such spurious denominator factors, in order to avoid round-off errors.
(Near a spurious singularity for the above representation~(\ref{A6DaDb}),
one could for example make use of a different shift~(\ref{SpinorShift}), 
whose spurious singularities are located elsewhere.)
On the other hand, the singularity is fairly mild, because only 
one power of $\sand5.{K_{34}}.2$ appears in the denominator.

Curiously, the introduction of denominators such as $\sand5.{K_{34}}.2$,
gives a much more compact representation of amplitudes (for $n\geq7$)
than forms
without such denominators.  The compactness is basically due to the
more manifest factorization properties of amplitudes constructed
via on-shell recursion relations.  For example, the
representation~(\ref{A6DaDb}) makes manifest the correct
behavior $A_6^\tree \sim1/\sqrt{\vph{}s_{i(i+1)}}$
as any pair of adjacent momenta become collinear, $k_i \parallel k_{i+1}$,
because the spinor products are square roots of momentum invariants,
as described in \eqn{spinorprodtwo}.

%%%%%%%%%%%%
\subsection{Generalizations}

Many applications of these techniques have already been carried out
at tree level.  In the case of $n$-gluon amplitudes, a closed-form 
formula for the `split helicity' configuration, 
$A_n^\tree(1^-, 2^-, \ldots, i^-, (i+1)^+,\ldots,n^+)$, 
has been constructed recursively, using essentially the same
shift described above for 
$A_6^\tree(1^-, 2^-, 3^-, 4^+, 5^+, 6^+)$~\cite{SplitHelicityTree}.  
The recursion relations have been
implemented numerically, and the computer time required for their
evaluation is competitive with other methods~\cite{DTW}.

In many circumstances it can be useful to generalize the shift
to act on more than two spinor variables.  For example, 
using the shift,
\begin{eqnarray}
\tlambda_{1} \rightarrow \tlambda_{1} + z \spa2.3 \teta \,, \hskip 1 cm 
\tlambda_{2} \rightarrow \tlambda_{2} + z \spa3.1 \teta \,, \hskip 1 cm 
\tlambda_{3} \rightarrow \tlambda_{3} + z \spa1.2 \teta \,,
\label{RisagerShift}
\end{eqnarray}
for the case where gluons 1, 2 and 3 are of negative helicity and
the rest are of positive helicity, and $\teta$ is an arbitrary
left-handed spinor, the recursive diagrams that are generated~\cite{Risager} 
are in one-to-one correspondence with the MHV construction of
Cachazo, Svr\v{c}ek and Witten~\cite{CSW,CSLectures}.  The MHV construction
was developed prior to the BCFW recursion relations.  
It provides a diagrammatic representation of tree amplitudes 
which makes manifest their remarkable
twistor-space properties~\cite{WittenTopologicalString,CSW}.
It is quite interesting that these two different approaches
can be directly connected.

Other types of shifts of multiple legs are useful because they can lead to
improved behavior of the shifted amplitudes as $z \rightarrow \infty$.
Such a shift was used to construct a recursion relation for the 
one-loop amplitudes $A_n^\oneloop(1^+,2^+,\ldots,n^+)$ containing
$n$ identical-helicity gluons~\cite{OnShellOneLoop}.
As mentioned above, multiple shifts can also be used to construct
proofs of the proper large $z$ behavior for tree amplitudes $A(z)$,
under a standard shift of the form~(\ref{SpinorShift})~\cite{BGKS,SW}.
  
The BCFW recursive analysis of $n$-gluon amplitudes was quickly extended
to amplitudes with massless external quarks as well as gluons~\cite{LWdeFZ}.
Recursion relations have also been established 
for tree-level amplitudes containing massive external
particles, such as electroweak vector bosons, Higgs bosons,
heavy quarks and squarks~\cite{BGKS,MassiveRecursion}.
Processes in abelian theories such as QED can also be handled
recursively~\cite{OSQED}.
The general recursive construction of all tree-level amplitudes in
QCD with massive quarks has been described recently by Schwinn and 
Weinzierl~\cite{SW}.  In particular, they enumerate all possible standard 
shifts that lead to a vanishing amplitude $A(z)$ as $z\to\infty$.

%%%%%%%%%%%%%%%%%%%%%%%%%%%%%%%%%%%%%%%%%%%%%%%%%%%%%%%%%%

\section{The Unitarity-Based Method}
\label{UnitaritySection}

Our goal is to compute a variety of one-loop amplitudes efficiently.  
For pedagogical reasons, we focus on higher-multiplicity multi-gluon
amplitudes.  
Conventional Feynman-diagram methods in gauge theories involve unphysical
states inside diagrams.  This renders computations vastly more complicated
than final results, because most of the computational effort is devoted to
manipulating unphysical information which ultimately cancels out.

On-shell methods, as discussed in the Introduction,
restrict states used in a calculation to physical
states.  For external gluons, the spinor-helicity basis imposes
this restriction efficiently.  We must also impose the
restriction on internal states.  We then rely
 on factorization, unitarity, and the existence of a representation in terms
of Feynman integrals in order to compute amplitudes.  
We will also make use of several simplifications for
the purely-massless amplitudes we are considering in this review.

%%%%%%%%%%%%%%%%%%%%%%%%%%%%%%%%

\subsection{Structure of the Amplitude}
\label{UniStructureSubsection}

The first simplification comes from the use of color ordering, as mentioned
in~\sect{KinematicsSection}.  This reduces our problem to that of computing
a color-ordered one-loop amplitude, in which external legs are ordered
cyclicly.  We can correspondingly introduce color-ordered Feynman diagrams,
in which each vertex inherits an ordering from the overall ordering
of the diagram.
Although we do not want to use Feynman diagrams, not even
the smaller set of color-ordered ones, to perform explicit 
computations, we can nonetheless imagine performing a gedanken
calculation.  By considering which Feynman diagrams would contribute
to the amplitudes, and properties of the one-loop integrals they would 
generate, we shall arrive at a compact set of loop-integral functions,
in terms of which we will ultimately express the amplitudes.  

\begin{figure}[t]
\centerline{\epsfxsize 5.0truein\epsfbox{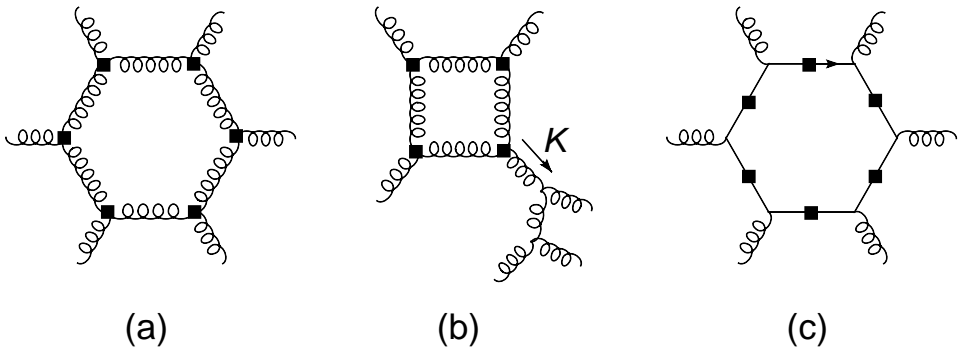}}
\caption{(a) A ring diagram, for a gluon propagating in the loop.
It contains the maximal possible powers of loop momenta in the numerator.
Each black square, at a trivalent vertex, contributes one factor of 
loop momentum.  (b) A non-ring diagram, in which some of the external
legs attach to a tree which attaches to the loop.  The momentum $K$
is one of the momenta for the corresponding box integral.
(c) A ring diagram for a quark in the loop;
in this case the loop-momentum factors are associated with the
fermion propagators.}
\label{RingDiagramsFigure}
\end{figure}

In a gauge theory, we have both trivalent and tetravalent vertices.  
Let us first consider the diagrams built only out of gluons.  Any diagram with
four-point vertices can be obtained (possibly in more than one way) from
a `parent' diagram with all four-point vertices replaced by pairs of
three-point vertices connected by a propagator.  
Diagrams with the maximal number of propagators inside the loop are 
thus `ring' diagrams,
built entirely out of three-point vertices; moreover,
all external legs attach directly to vertices inside the loop,
as shown in \fig{RingDiagramsFigure}(a).
Accordingly, when computing an $n$-point amplitude, we must
consider loop integrals with up to $n$ external legs.  We must also
consider diagrams with fewer external legs attached to directly to
the loop, but rather attached to a tree which in turn attaches to the loop.
These will give rise to loop integrals with fewer than $n$ legs, but
where some of the legs have momenta $K$ 
which are sums of the original external
momenta, as depicted in \fig{RingDiagramsFigure}(b).
The original momenta may be massless (for gluons or massless
quarks) or massive (for colorless heavy particles).
The sums of momenta are in either case
no longer massless; $K^2$ may be either positive
or negative.  That is, we must also consider loop integrals with external
masses; but we may take all {\it internal\/} masses to be zero.

%%%%%% FIGURE %%%%%%%%%%%
%%%%%%%%%%%%%%%%%

Each trivalent vertex contains terms proportional
to the momenta flowing through it.  If we consider the vertices within
the loop, some of the terms contain a factor of the external momenta (or
sums of external momenta).  Others contain a factor of the loop
 momentum $\ell$.  
The latter give rise to tensor integrals, in which tensors
in the loop-momentum appear in the numerator.  The maximal power of
loop momentum arises when every vertex in the loop gives us one power,
so that we get an $n^{\rm th}$-rank $n$-point tensor integral.  The
(tensor) indices on the loop momenta are contracted into external momenta
or polarization vectors.  
We will refer to integrals containing no powers of the loop
momentum in the numerator as scalar integrals.

The above analysis holds equally well for gluons circulating in the loop
as it does for scalar particles circulating there.  (In fact, the 
contribution of a scalar particle in the adjoint representation of
the gauge group accounts completely for the leading tensor-integral 
part of the gluon contribution.)  For contributions of quarks in the loop,
the counting starts out a bit differently.  There are no
powers of momenta at the vertices, of course, but the fermion
propagator, $1/\ellsl = \ellsl/\ell^2$, supplies a factor $\ellsl$
proportional to the loop momentum.  As the number of
vertices in the loop is the same as the number of propagators, fermionic
contributions again give us integrals which range up to
$n^{\rm th}$-rank $n$-point tensor integrals,
as illustrated in \fig{RingDiagramsFigure}(c).

We now wish to organize all of the loop integrals occurring in an 
amplitude, by reducing them down to some basic set.
Tensor integrals can be reduced to scalar integrals, containing 
the same or fewer propagators, using traditional 
Brown--Feynman~\cite{BrownFeynman}
or Passarino--Veltman~\cite{PassarinoVeltman}
techniques, or using more recent spinorial 
techniques~\cite{Vermaseren,Pittau,Weinzierl,OPP,AguilaPittau}.  
All these techniques
effectively use Lorentz invariance to re-express integrals over powers of
the loop momentum in terms of the metric and external momenta.

We will take all external vectors --- both momenta and polarization
vectors --- to be strictly four dimensional.  (For the polarization
vectors, this is equivalent to adoption of the four-dimensional helicity
scheme (FDH)~\cite{FDH,FDH2}.)
For five- or higher-point integrals,
any numerators can be expressed entirely in terms
of differences of propagators and external Lorentz invariants.
Each numerator factor has the form $\ell\cdot E$, where
$\ell$ is the loop momentum and $E$ is one of the external vectors.
(A factor of $\ell^2$ would cancel a propagator, immediately reducing
the integral to a lower-point one.)  With five or more external
momenta $k_i$, satisfying momentum conservation, $\sum_i k_i = 0$, 
we can use four of them (massive or massless)
as a basis to express any four-dimensional vector $E$.
Then $\ell\cdot E$ becomes a linear combination of dot products
$\ell\cdot k_j$, which in turn are differences of propagator
denominators and invariants built from external momenta.  
After canceling propagators, we choose
a new set of basis vectors for the daughter integrals, and repeat
the process.  Each step reduces the degree of the numerator by one,
and may in addition reduce the number of external legs of the integrals by one.

For four- or lower-point integrals, we do not have enough
independent external momenta to form a basis.  We can instead use
Lorentz invariance as above to re-express the integrals in terms of
numerators involving the external
momenta~\cite{BrownFeynman,PassarinoVeltman}.  
This reduction procedure again generates integrals with fewer 
numerator powers of the loop momentum,
and possibly fewer external legs as well.
At the end of this reduction, we are left only with 
scalar integrals having trivial numerators, 
with up to $n$ external momenta.  

Another option is to use not only external momenta in the basis,
but also complex momenta built out of the associated spinors. 
The Lorentz products of the loop momentum with these complex momenta
cannot cancel propagators; but a judicious choice will result
in their integrals vanishing.  Instead of expanding the external
momenta in this basis, we could also choose to expand the four-dimensional
components of the loop momentum.  As we shall see in an example,
this is particularly useful for evaluating one- and two-mass
triangle and bubble integral contributions.  In the case
of a one-mass triangle with massless legs $k_1$ and $k_2$, 
one can use the basis given by 
del~Aguila and Pittau~\cite{AguilaPittau}
(see also refs.~\cite{Pittau,Weinzierl}),
\begin{equation}
v_1^\mu = k_1^\mu,\qquad
v_2^\mu = k_2^\mu,\qquad
v_3^\mu = \sand1.{\gamma^\mu}.2,\qquad{\rm\ and\ }
v_4^\mu = \sand2.{\gamma^\mu}.1\,.
\label{OneMassTriangleBasis}
\end{equation}
For a two-mass triangle with massive legs $K_1$ and $K_2$, and
massless leg $k_3$,
\iffalse
and a loop momentum carried by the line connecting one of the massive
legs to the massless leg, 
\fi
a different basis is more appropriate,
\begin{equation}
v_1^\mu = K_1^\mu,\quad
v_2^\mu = K_2^\mu,\quad
v_3^\mu = \sandpm3.{\Kinsl_1\gamma^\mu}.3,\quad{\rm\ and\ }
v_4^\mu = \sandmp3.{\Kinsl_1\gamma^\mu}.3\,.
\label{TwoMassTriangleBasis}
\end{equation}

We can further 
reduce~\cite{Melrose,Vermaseren,BDKPentagon,OtherPentagons,DDReduction}
all six- or higher-point scalar
integrals to linear combinations of five- or lower-point integrals.  This
reduction is true to all orders in the dimensional regulator $\e=(4-D)/2$.
After the reduction, the five-point integrals that remain are all finite.
They can be thought of as scalar pentagon integrals evaluated in six
dimensions; such integrals are free of both ultraviolet and infrared 
divergences.  They also appear multiplied by at least one explicit 
power of $\e$.  If we were interested in computing amplitudes to 
$\Ord(\e)$ or beyond, we would need to evaluate such pentagon integrals.
If we are only interested in computing amplitudes to $\Ord(\e^0)$, 
however, we may drop them.

%%%%%% FIGURE %%%%%%%%%%%
\begin{figure}[t]
\centerline{\epsfxsize 3.3truein\epsfbox{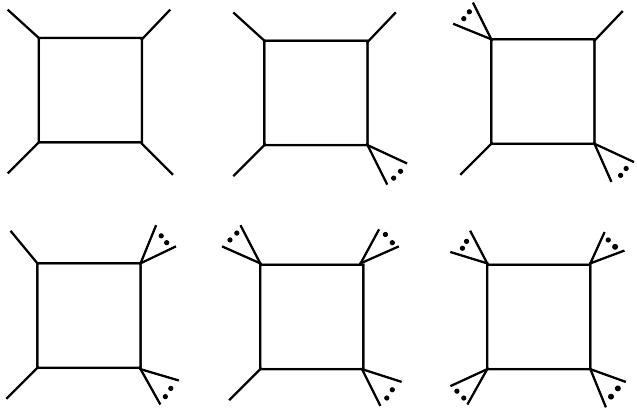}}
\caption{The possible box integrals that can appear in one-loop
amplitudes. }
\label{BoxesFigure}
\end{figure}
%%%%%%%%%%%%%%%%%

%%%%%% FIGURE %%%%%%%%%%%
\begin{figure}[t]
\centerline{\epsfxsize 3.9truein\epsfbox{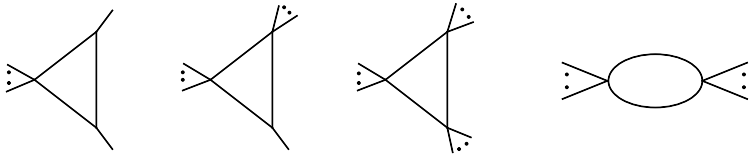}}
\caption{The possible triangle and bubble integrals that can appear
at one loop.   }
\label{TriBubsFigure}
\end{figure}
%%%%%%%%%%%%%%%%%

Consequently, all amplitudes can ultimately be expressed in terms of
ten different types of integrals: boxes with up to four external massive
legs, triangles with up to three external massive legs, and bubbles.  
These integrals are shown in
\figs{BoxesFigure}{TriBubsFigure}.  All internal propagators are
massless, but each external momentum can be either massless or
massive.  Massive external momenta correspond to sums of the original massless or
massive momenta of the amplitude.  These integrals are all computed in
dimensional regularization, which regulates both the ultraviolet
divergences (which show up as single poles in bubble integrals) and
the infrared divergences (which show up as double or single poles in
the box and triangle functions).

Any color-ordered amplitude we wish to compute can therefore be expressed
in terms of a standard basis of integrals,
\begin{equation}
A_n^\oneloop = \sum_{j\in B} c_j \I_j\,.
\label{BasisEquation}
\end{equation}
For a given process, the integrals $\I_j$ in the basis $B$ are given
{\it a priori\/} by the set of all functions defined
in \figs{BoxesFigure}{TriBubsFigure},
for all possible cyclicly-ordered combinations of momenta. 
These integrals have been tabulated, for example, in the first 
appendix in ref.~\cite{Neq1Oneloop}. 
The coefficients $c_j$ are rational functions of the
external momenta and polarization vectors.  Using the spinor-helicity
basis, we express the $c_j$ in terms of spinor products constructed
from spinors corresponding to external momenta.  The coefficients do in general 
depend on $\e$ as well.  Rational terms in amplitudes arise from 
$\e$ terms striking ultraviolet poles in the integrals.  (With the
basis shown in \figs{BoxesFigure}{TriBubsFigure}, the only ultraviolet
pole is in the bubble integral.)

%%%%%%%%%%%%%%%%%%%%%%%%%%%%%%%%

\subsection{Unitarity}
\label{UniUnitaritySubsection}

The conservation of probability is a fundamental requirement of
any consistent field theory.  It implies the unitarity of the scattering
matrix $S$.  If we examine the non-forward part of the scattering matrix,
$T=-i (S-1)$, unitarity implies that
\begin{equation}
-i (T-T^\dagger) = T^\dagger T\,.
\label{BasicUnitarity}
\end{equation}
The implicit sum on the right-hand side goes over all possible 
physical states, intermediate between the processes defined
by $T^\dagger$ and $T$.
At one-loop order, only two-particle intermediate states are possible.
There is a phase-space integral over the intermediate on-shell 
momenta, which amounts to an integral over the solid angle of
one of the two particles, in the center-of-momentum frame.
There is also a discrete sum over allowed particle types. 
 
The left-hand side of \eqn{BasicUnitarity}
corresponds to a discontinuity in the scattering amplitude, that is
a branch cut in complex momenta.  This discontinuity gives the
absorptive part of an amplitude.

The right-hand side may be obtained from a loop amplitude by {\it cutting\/}
it.  In a single Feynman diagram,
the discontinuity in a given invariant or channel can be computed by
replacing the two propagators separating a set of legs carrying that
invariant from the rest of the diagram by a delta function,
\begin{equation}
{i \over p^2+i\varepsilon} \longrightarrow 2\pi\,\delta^{(+)}(p^2)\,.
\label{deltaplus}
\end{equation}
At the diagrammatic level, this replacement goes under the name of
the Cutkosky rules~\cite{Cutkosky}.  The pair of delta functions reduces
the loop integral to a phase-space integral.  We can of course also cut
sums of diagrams, taking care to throw away any contribution in which one or
both of the required propagators is missing.  (The missing
propagator prevents such terms from contributing to the discontinuity
in the target invariant.)  

The application of unitarity as an on-shell method of calculation turns
the cutting step around.  Instead of cutting one-loop amplitudes, we will
sew tree amplitudes together to form one-loop amplitudes.  In other words,
we will be reconstructing the dispersive parts of amplitudes from
the absorptive ones.  This reconstruction could in principle be 
accomplished by performing dispersion integrals; but we do {\it not\/} 
want to perform such integrals explicitly.  Rather, we rely on the 
existence of an underlying representation in terms of Feynman integrals 
to do the job.  For this purpose, we do not need an explicit basis of 
the kind discussed in the previous subsection, but having one
makes the reconstruction easier and more powerful.
In brief, we evaluate the cuts in each channel, and represent them
as linear combinations of cuts of integrals $\I_j$ in that channel,
\begin{equation}
\Im A_n^\oneloop = \sum_{j\in B} c_j \Im \I_j \,,
\label{cutchannel}
\end{equation}
in order to read off the coefficients $c_j$.

Consider the cut in a channel with momentum $K$ crossing the cut,
where $K$ is composed of $j$ external legs and $s=K^2$.
The cut contains delta functions imposing the on-shell 
condition~(\ref{deltaplus}).  By removing them, we promote the cut
to a full loop integral,
\begin{eqnarray}
&& \hskip -.9 cm 
(2\pi)^2 \! \sum_{\rm helicity}\int \frac{d^D\ell}{(2\pi)^D}\;
\delta^{(+)}(\ell^2) A_{n-j+2}^\tree(K-\ell,\ldots,\ell)
 \delta^{(+)}( (\ell-K)^2 ) A_{j+2}^\tree(-\ell,\ldots,\ell-K) \nn\\
&& \longrightarrow 
\sum_{\rm helicity}\int \frac{d^D\ell}{(2\pi)^D}\;
{i\over\ell^2} A_{n-j+2}^\tree(K-\ell,\ldots,\ell) {i \over (\ell-K)^2} 
A_{j+2}^\tree(-\ell,\ldots,\ell-K)
\,.
\label{basiccut}
\end{eqnarray}
Here the tree amplitudes are on shell, and the sum is taken over the
possible helicities of the internal legs crossing the cut.  
In the applications we will
describe in this review, the cut momenta entering the tree amplitudes
in \eqn{basiccut} are taken to be four-dimensional.  
Accordingly, we can use the spinor-helicity method to the fullest 
in evaluating them.

The propagators in \eqn{basiccut} are the same whether the particle
crossing the cut is a scalar, a fermion, or a gauge boson.  The
corresponding helicity projector is already present in the on-shell
tree amplitudes.   The loop integral, in contrast, is evaluated in
$D$ dimensions.  The integral reductions discussed in the
previous subsection are algebraic manipulations of the integrand.  
To perform them on expressions given in terms of spinor products,
we can complete the spinor products to form scalar-propagator 
denominators.  For example, if external momentum $k_1$ appears
right after the loop momentum $\ell$, then we may write,
\begin{equation}
{ 1 \over \spa{\ell}.{k_1}}
 = - { \spb{\ell}.{k_1} \over 2\ell\cdot k_1 }
 = { \spb{\ell}.{k_1} \over (\ell-k_1)^2 } \,.
\end{equation}
If all denominators are converted to scalar propagators, then the
numerators can be rewritten as functions of the loop momentum (and not
merely of spinors carrying the on-shell cut loop momentum).
Performing the integral reductions on this expression will decompose
the integrand into a sum of terms, each corresponding to an integral
in the basis~\eqn{BasisEquation}.  We can then read off the
coefficient of any integral from the term with propagators
corresponding to the integral.  In practice, applying the full
reduction machinery is not necessary; appropriate partial-fractioning
of the integrand with respect to the loop momentum accomplishes the
same goal.  As outlined below, an alternative approach for computing 
the coefficients, developed by Britto {\it et al.}~\cite{BBCF,BFM}, 
transforms the cut integral into a contour integral in spinor
variables, which may be evaluated by residue extraction.

In a general amplitude, we cannot detect all terms by looking in a single
channel.  For this reason, we have to look at all channels.  Some integrals
contributing to the amplitude will appear only in a single channel; we
can simply read off the coefficient in that channel.  Other integrals
will show up in more than one channel; we can read off their coefficients
in any of the channels.  We must take only a single copy of the integral
in the latter case.

Within the basis used in \eqn{BasisEquation}, cutting the amplitude
in a given channel isolates those integrals that have a discontinuity
in it.  The product of tree amplitudes $A_{n-j+2}^\tree$, $A_{j+2}^\tree$ 
can be equated to a linear combination of cuts of a subset of integrals 
in the basis, as in \eqn{cutchannel}.
Because the linear combination typically contains more than one integral, 
we have to identify the different terms on the left-hand side.
One way to do this is by performing a partial-fraction decomposition,
and then isolating different terms according to the propagators they 
contain.  We will refine this isolation in the next subsection.

One might worry that were we to use a dispersion integral to perform
this reconstruction, the results we found would suffer from an 
additive ambiguity because of ultraviolet divergences in the integral.
That is, one can add a branch-cut-free or rational
function to the amplitude.  Such rational terms are indeed present in
QCD amplitudes.   They cannot be computed using the four-dimensional
unitarity approach described above.  In the four-dimensional case,
the coefficients in the basis~(\ref{BasisEquation}) are understood 
to have $\e$ set to zero, and the basis only serves as one for the 
cut-containing parts of the amplitude.

As an aside, we note that in supersymmetric theories at
one loop, rational terms are tightly linked to cut-containing
terms.  They can be deduced entirely from the integral functions 
determined within the four-dimensional approach.  There are no missing terms.
(In theories with $\NeqFour$ supersymmetry, there are strong indications 
that the same result is even true at higher loops.)  This result reflects the
improved ultraviolet behavior of supersymmetric theories.

There are two distinct on-shell methods for computing these rational terms, 
implicitly fixing this remaining ambiguity.  (For other analytic
approaches, see refs.~\cite{XYZ,BinothRat}.) One method, to be discussed 
in~\sect{LoopRecursionSection}, is to extend on-shell methods by making
use of another universal property of amplitudes, that of factorization.

The second method stretches slightly our notion of a physical state
from four to $D$ dimensions~\cite{DDimU,ABFKM}.  
If we take the sum over intermediate
states in $D$ dimensions instead of four dimensions, we recover
the entire amplitude with no rational ambiguity.  These intermediate
states are massless and transverse but in a $D$-dimensional rather
than a four-dimensional sense.  The latter method adds a third role
to dimensional regularization, beyond regulating both ultraviolet
and infrared divergences.

There are two different ways of understanding how $D$-dimensional unitarity
can capture the complete amplitude.  From the point of view of
dispersion integrals, dimensional regularization makes the absorptive
part of the amplitude ultraviolet convergent, and hence eliminates
any additive ambiguity.  From a more concrete point of view,
every term in an amplitude must contain a power of $(-s)^{-\e}$, 
where $s$ is an invariant,
in order to compensate the dimension of the coupling and keep the
amplitude of fixed dimension as we change $\e$.  (In an amplitude
with only massless particles propagating internally, there are
no masses to supply the required dimension.)  The invariants $s$
that appear can be different in different terms of the amplitude.
If we now expand the amplitude beyond $\Ord(\e^0)$, the epsilonic
power $(-s)^{-\e}$ will give rise to a term,
\begin{equation}
1-\e \ln(-s)+\cdots\,,
\end{equation}
which contains a discontinuity for $s>0$.  Hence it will be detected
in the unitarity cut in the $s$ channel.  
The rational terms can then be obtained by
truncating the result to $\Ord(\e^0)$.  The underlying Feynman-integral
representation forces them to come along with the $\Ord(\e)$ cut terms.

Because the $D$-dimensional unitarity method effectively computes the
amplitude to all orders in $\e$, it goes beyond what is truly needed
for collider applications.  It does not allow full use of the
spinor-helicity basis at the early stages of a calculation, though of
course we can still make use of the on-shell conditions in $D$
dimensions.  Accordingly, it tends to require more computational
effort than the combination of the four-dimensional unitarity method
with the additional techniques for rational parts discussed
in~\sect{LoopRecursionSection}.  Nonetheless it is conceptually
useful, and can be useful in practice for lower-point amplitudes
needed as starting points for the recursive techniques we review
later.  It is necessary to use this version of unitarity, of course,
when one needs a quantity, such as a splitting amplitude, to higher
order in $\e$.  Recent work~\cite{ABFKM,BFMassive} has shown how to
apply spinor methods in this context as well, and should widen the
applicability of $D$-dimensional methods.

%%%%%%%%%%%%%%%%%%%%%%%%%%%%%%%%

\subsection{Generalized Unitarity}
\label{UniGenUnitaritySubsection}

Cutting propagators in an amplitude selects only those contributions
that have the propagators present in the first place.  We can think
of this in terms of the original Feynman diagrams for our target amplitude,
where it reflects the observation that a diagram with a relevant
propagator missing cannot contribute to the discontinuity in the
desired channel.  We can also think of this in terms of the integrals
in the basis.  Each original Feynman diagram will produce contributions
to the coefficients of several basis integrals.  If the original diagram
is missing a required propagator, none of the corresponding basis integrals
will have it, and no contribution to the discontinuity will arise.  Even
when the original diagram does have a required propagator, it will be
present only in a subset of the descendant basis integrals.  Cutting
pairs of propagators thus winnows the set of basis integrals down to those in
which the pair is present.  When we sew together two tree amplitudes
to form the cut in a given channel, only those integrals that possess
both cut propagators can show up in the result, and accordingly the given
cut contributes only to the coefficients of those integrals.

As discussed in the previous subsection, we must in general perform
analytic simplifications on the cut expression in order to isolate the
contributions of different integrals.  This procedure will sort terms into
contributions corresponding to different integrals according to the
propagator denominators present, and will remove contributions to
lower-point integrals.  If we require the presence
of additional propagators beyond the pair isolating a given channel,
only a subset of terms will remain.  That is, fewer integrals can
contribute.  Requiring the presence of additional propagators, or
equivalently cutting them, goes under the name of `generalized
unitarity'.  It corresponds, in the old-fashioned language of
dispersion relations, to extracting the leading discontinuity of an
amplitude~\cite{ELOP}.  In addition to isolating a smaller number of
candidate integrals to consider, the additional on-shell condition
splits one of the two tree amplitudes into smaller trees.  We are thus
sewing smaller and simpler expressions together, and will in general
need to perform less algebra to extract coefficients.

The ultimate refinement of this procedure comes when we require enough
propagators to isolate a {\it single\/} integral.  This is possible
for the box integrals.  Maximal generalized unitarity --- cutting four
propagators --- isolates the coefficient of a single box integral
as a product of four tree amplitudes, as illustrated 
in \fig{complexkinFigure}(b).
In addition, when we take the cut momenta to be strictly
four-dimensional, the four delta functions freeze the loop momentum
entirely~\cite{BCFGeneralized},
\begin{eqnarray}
&& \hskip -1.1 cm \int \mskip -2mu \frac{d^D\ell}{(2\pi)^D} 
\biggl[\frac{f(\ell)}{[\ell^2+i\varepsilon]
         [(\ell-K_1)^2+i\varepsilon]
         [(\ell-K_{12})^2+i\varepsilon]
         [(\ell-K_{123})^2+i\varepsilon]}
\biggr]\bigg|_{\ell^D\rightarrow\ell^4} \nonumber\\
&\longrightarrow&\int \mskip -2mu\frac{d^D\ell}{(2\pi)^D} 
\bigl[f(\ell)
\delta^{(+)}(\ell^2)
\delta^{(+)}((\ell-K_1)^2)
\delta^{(+)}((\ell-K_{12})^2)
\delta^{(+)}((\ell-K_{123})^2)
\bigr]\Big|_{\ell^D\rightarrow\ell^4}
\nonumber\\
&=&{1\over2}\sum_{\rm solutions} f(\ell) \,.
\end{eqnarray}
\def\ellm{\ell^{(m)}}
The sum is over the discrete number of solutions to the
simultaneous on-shell equations imposed by the four-dimensional delta 
functions.
The coefficient of the desired integral is then simply the product of the four
tree amplitudes, summed over these solutions for the cut loop momenta $\ellm$,
\begin{eqnarray}
 c_j &=& {1\over2} \sum_{m=1,2}
 A_{n_1}^\tree(\ldots,-\ellm,\ellm-K_1,\ldots)
 A_{n_2}^\tree(\ldots,-\ellm+K_1,\ellm-K_{12},\ldots)
\nonumber\\
&&\hskip0.1cm \times
 A_{n_3}^\tree(\ldots,-\ellm+K_{12},\ellm-K_{123},\ldots)
 A_{n_4}^\tree(\ldots,-\ellm+K_{123},\ellm,\ldots)\,,
\nonumber\\
&&{~} \label{boxcoeff}
\end{eqnarray}
where the implicit momenta are all external, and $\sum_i n_i = n+8$.
The sum is normalized by the total number of solutions, which turns out 
to be two, including solutions for which $f$ may happen to vanish.
The complete freezing of the loop
momentum means that no further algebra is required, and the coefficient
could even be evaluated purely numerically.

For many applications of \eqn{boxcoeff}, including all for less than
eight external (massless) legs, there would seem to be 
a catch: some of the tree amplitudes, corresponding to massless
external legs, will be all-massless three-point amplitudes.
Naively, these amplitudes would vanish.  The quadruple cuts would 
vanish along with them, swallowing the coefficients with them.  
As we have seen in~\sect{KinematicsSection},
however, using complex momenta we can obtain non-zero values for these
amplitudes.  We can then treat them in the same way as higher-point 
amplitudes in the quadruple cuts, and recover the box coefficients 
simply from sums over appropriate products of tree amplitudes.  
Indeed, as we shall see, because of the special properties of three-point 
amplitudes, the computation of coefficients of box integrals with 
massless external legs is even simpler than that of four-mass boxes.

In an approach based on maximal use of generalized unitarity, we will
compute a larger number of cuts; but each cut will be simpler, and
will give the coefficient of a single integral directly.  One starts
with the box integrals, computing a separate quadruple cut for each
different one.  One turns next to the three-mass triangles, which
can be isolated through triple cuts as was done in ref.~\cite{Zqqgg}.
The integrands emerging from
triple cuts in general will also contain contributions to those
box integrals sharing the same cuts.  These contributions, and
box-like terms which vanish upon loop integration, must be removed
in order to extract the coefficient of the three-mass triangle.  This
can be done, for example, using the decomposition proposed
by Ossola {\it et al.}~\cite{OPP}, or by other methods.  Further developments
here are possible and desirable (see {\it e.g.} ref.~\cite{FordeTriBub}).
The spinor-residue approach of
refs.~\cite{BBCF,BFM} outlined below can also be used for this evaluation.
The remaining terms come from
one- and two-mass triangles as well as bubble integrals.  
The analytic forms of the one- and two-mass triangle integrals
imply that the remaining terms can all be written as sums of bubble 
integrals, with coefficients that depend on external invariants and on $\e$.
Accordingly, they should be treated together.  This part of the 
computation will make use of standard cuts, evaluated as described 
in the previous \sect{UniUnitaritySubsection}.

%%%%%%%%%%%%%%%%%%%%%%%%%%%%%%%%

\subsection{A Box Example}
\label{UniBoxExampleSubsection}

%%%%%% FIGURE %%%%%%%%%%%
\begin{figure}[t]
\centerline{\epsfxsize 1.4 truein\epsfbox{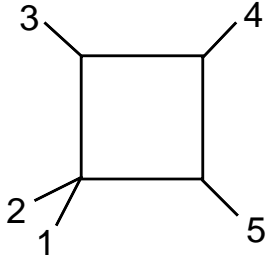}}
\caption{A box integral that can occur for five-point amplitudes with
legs following the 12345 ordering. The other four possible ones are given 
by cyclic permutations of this one.}
\label{Box5PtFigure}
\end{figure}
%%%%%%%%%%%%%%%%%

As an example, let us compute the coefficient of a box integral in
the pure gauge-theory amplitude $A_5^\oneloop(1^-,2^-,3^+,4^+,5^+)$.  
Because this amplitude contains five external massless legs,
the relevant box integrals, which are found by collapsing a single 
propagator on the pentagon ring diagram, have precisely one external
massive leg, and three massless ones.  One of these one-mass box integrals,
in which the massive leg is $K_{12} = k_1+k_2$, is shown in
\fig{Box5PtFigure}.  It is defined by
\begin{equation}
\I(K_{12}) =
\mu^{2\e} \int {d^{4-2\e}\ell \over (2\pi)^{4-2\e}}
{1 \over \ell^2 (\ell-K_{12})^2 (\ell-K_{123})^2 (\ell+k_5)^2} \,.
\label{IK12def}
\end{equation}
The other four boxes have the massive leg in turn being 
$K_{23}$, $K_{34}$, $K_{45}$ and $K_{51}$. 
The amplitude $A_5^\oneloop(1^-,2^-,3^+,4^+,5^+)$ is
antisymmetric under the reflection $(12345)\leftrightarrow (21543)$.
This symmetry relates the coefficient of $\I(K_{51})$ to that of
$\I(K_{23})$; and that of $\I(K_{45})$ to that of $\I(K_{34})$.

%%%%%% FIGURE %%%%%%%%%%%
\begin{figure}[t]
\centerline{\epsfxsize 4.7 truein\epsfbox{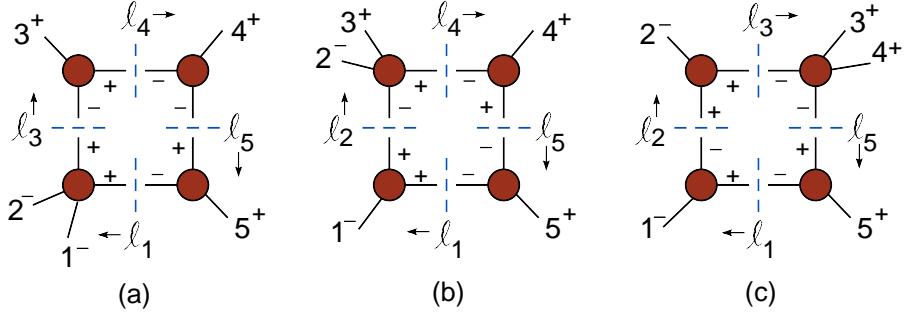}}
\caption{The quadruple cuts of of integrals $\I(K_{12})$, $\I(K_{23})$,
and  $\I(K_{34})$.  Only one 
helicity configuration contributes to each of the cuts.}
\label{QuadCutFigure}
\end{figure}
%%%%%%%%%%%%%%%%%

Let us perform the computation for
$\I(K_{12})$, using maximal generalized unitarity.  
When we cut all four propagators, 
as shown in \fig{QuadCutFigure}(a), and
restrict the cut momenta to four dimensions, we will be left with
three three-point amplitudes and one four-point amplitude.  If $\ell_4$
is the cut loop momentum in between $k_3$ and $k_4$, it is a momentum
entering into both adjacent three-vertices.  In order to obtain a
nonvanishing result, we must take it to be complex.  The on-shell
conditions on the first vertex require that
\begin{equation}
{\rm either\ }\quad\lambda_{\ell_4} \propto \lambda_{3}\qquad{\rm\ or\ }
\qquad \tlambda_{\ell_4} \propto \tlambda_{3}\,,
\end{equation}
and the on-shell conditions on the second vertex require that
\begin{equation}
{\rm either\ }\quad\lambda_{\ell_4} \propto \lambda_{4}\qquad{\rm\ or\ }
\qquad \tlambda_{\ell_4} \propto \tlambda_{4}\,.
\end{equation}
However, for generic external momenta, $s_{34} \neq 0$, and therefore 
$\lambda_{3}\not\propto \lambda_{4}$ and $\tlambda_{3}\not\propto\tlambda_{4}$.
Thus only two of the combined solutions are allowed,
\begin{equation}
(\lambda_{\ell_4} \propto \lambda_{3} {\rm\ and\ }
\tlambda_{\ell_4} \propto \tlambda_{4})
\qquad{\rm\ or\ }
\qquad (\tlambda_{\ell_4} \propto \tlambda_{3} {\rm\ and\ }
\lambda_{\ell_4} \propto\lambda_{4})\,.
\label{SolutionsI}
\end{equation}
This in turn implies that neighboring three-vertices must be
of opposite `type' --- if one is of the form $A_3^\tree(+{}+{}-)$ or a cyclic
permutation thereof, then its
neighbor must be of the form $A_3^\tree(-{}-{}+)$ 
or a cyclic permutation thereof.

The possible helicity configurations are further restricted by the
identity of the helicities attached to the massive leg.  The
vanishing of four-point tree amplitudes with zero or one positive helicity
(the parity conjugate of \eqn{npointvanish}) implies that
the helicities of both cut internal legs emerging from the four-point
vertex, $(-\ell_1)$ and $\ell_3$, must be positive.  Such a configuration
is only allowed for gluonic internal states; it vanishes for quarks
or scalars circulating in the loop.  (Massless fermions have
their helicity conserved along the fermion line, which translates
into a helicity flip between quark and anti-quark legs, 
in our all-outgoing helicity convention.  Scalars do not carry helicity, 
but  they can be considered to be complex, and then particle {\it vs.}
anti-particle plays the same role as helicity.)
The positive helicity of $(-\ell_1)$ and $\ell_3$, together with
the neighboring-three-vertex constraint,
fixes the helicities of the remaining three-point vertices
to be $A_3^\tree(-\ell_3^-, 3^+, \ell_4^+)$,
$A_3^\tree(-\ell_4^-,4^+,\ell_5^-)$, and 
$A_3^\tree(-\ell_5^+,5^+,\ell_1^-)$.
Accordingly, we have only a single solution, the first in
\eqn{SolutionsI}, to take into account.  The other solution
leads to vanishing three-point tree amplitudes, as discussed in
\sect{KinematicsSection}. 

To evaluate the coefficient, we must solve for $\ell_4$.  We only
need the solutions for $\lambda_{\ell_4}$ and $\tlambda_{\ell_4}$
up to an overall constant factor, because
the latter will cancel in the combination that appears in
the desired coefficient, $\lambda_{\ell_4}\tlambda_{\ell_4}$.
Three of the four on-shell equations,
\begin{equation}
\ell_4^2 = 0,\qquad \ell_3^2 = (\ell_4+k_3)^2 = 0,\qquad{\rm\ and\ }
\qquad \ell_5^2 = (\ell_4-k_4)^2 = 0\,,
\end{equation}
can be satisfied automatically by taking $\ell_4^\mu$ to have the form,
\begin{equation}
\ell_4^\mu = {\textstyle\frac{1}2} \xi_4\sand{3}.{\gamma^\mu}.4 \,.
\end{equation}
The constant $\xi_4$ is fixed by the last of the four on-shell equations,
\begin{equation}
\ell_1^2 = (\ell_4-K_{45})^2 = - \xi_4\sand3.5.4 + s_{45} = 0\,,
\end{equation}
to have the value $\xi_4 = \spa4.5/\spa3.5$.

The coefficient of the $K_{12}$ box is then,
\begin{eqnarray}
\hskip-1cm
c_{12} &=&
%%%%% ignore begin : quadcut1
\frac12 A_4^\tree(-\ell_1^+,1^-,2^-,\ell_3^+)
A_3^\tree(-\ell_3^-, 3^+, \ell_4^+)
A_3^\tree(-\ell_4^-,4^+,\ell_5^-)
A_3^\tree(-\ell_5^+,5^+,\ell_1^-)
%%%%% ignore end : quadcut1
\nonumber\\
&=& 
%%%%% begin : quadcut2
\frac12
\frac{\spa1.2^3}{\spa2.{\ell_3}\spa{\ell_3}.{(-\ell_1)}\spa{(-\ell_1)}.1}
\frac{\spb3.{\ell_4}^3}{\spb{\ell_4}.{(-\ell_3)}\spb{(-\ell_3)}.3}
\frac{\spa{\ell_5}.{(-\ell_4)}^3}{\spa4.{\ell_5}\spa{(-\ell_4)}.4}
\frac{\spb{(-\ell_5)}.5^3}{\spb5.{\ell_1}\spb{\ell_1}.{(-\ell_5)}}
%%%%% end : quadcut2
\nonumber\\
&=&
%%%%% begin : quadcut3
 -\frac12\frac{\spa1.2^3 \sandpm3.{\Ellin_{4}\Ellin_{5}}.5^3}
{\sand2.{\Ellin_{3}}.3\sand4.{\Ellin_{4}\Ellin_{3}\Ellin_{1}}.5
\sandmp1.{\Ellin_{1}\Ellin_{5}}.4}
%%%%% end : quadcut3
\,.
\end{eqnarray}
Using momentum conservation and the properties of the $\ell_i$, we can
simplify this expression to,
\begin{eqnarray}
c_{12} &=&
%%%%% begin : quadcut4
 \frac12\frac{\spa1.2^3\sand4.{\Ellin_{4}}.3^2\spb4.5^3}{\sand2.{\Ellin_{4}}.3
\spa3.4\spb4.5\spa1.5\sand4.{\Ellin_{4}}.5}
%%%%% end : quadcut4
\nonumber\\
&=&
%%%%% begin : quadcut5
 -\frac12\frac{\spa1.2^3 s_{34} s_{45}}{\spa2.3\spa3.4\spa4.5\spa5.1}
%%%%% end : quadcut5
\nonumber\\
&=&
%%%%% ignore begin : quadcut6
\frac{i}2 s_{34} s_{45} \, A_5^\tree(1^-,2^-,3^+,4^+,5^+)\,.
%%%%% ignore end : quadcut6
\label{K12coefficient}
\end{eqnarray}
The box integral multiplying this coefficient, defined in \eqn{IK12def},
has the Laurent expansion in $\e$,
\begin{eqnarray}
\I(K_{12}) 
&=& { - 2 i \, \cg \over s_{34} s_{45} } \biggl\{
-{1\over\e^2} \biggl[ 
    \biggl({\mu^2\over -s_{34}}\biggr)^{\e}
     +\biggl({\mu^2\over -s_{45}}\biggr)^{\e}
     -\biggl({\mu^2\over -s_{12}}\biggr)^{\e}
    \biggr]
\nonumber\\
&&\hskip1.2cm
  + \Li_2\biggl(1-{s_{12}\over s_{34}}\biggr)
  + \Li_2\biggl(1-{s_{12}\over s_{45}}\biggr)
  + {1\over2} \ln^2\biggl({-s_{34}\over -s_{45}}\biggr)
  + {\pi^2\over6}
\biggr\} 
\nonumber\\
&&\hskip0.0cm
+\ \Ord(\e) \,,
\end{eqnarray}
where the constant $\cg$ is defined by
\begin{equation}
\cg = {1\over(4\pi)^{2-\eps}}
  {\Gamma(1+\eps)\Gamma^2(1-\eps)\over\Gamma(1-2\eps)} \,.
\label{cgdefn}
\end{equation}

The coefficients of the other boxes also have only gluonic
contributions.  In the $K_{23}$ box, shown in~\fig{QuadCutFigure}(b),
the four-point vertex has opposite helicities for the internal legs;
but the requirement of opposite `type' for adjacent three-point
vertices requires the diagonally-opposite three-point vertex to have
identical helicities for its internal momenta.  This allows only
gluonic contributions.  In the $K_{34}$ box, both internal legs
attached to the four-point vertex containing $k_3$ and $k_4$ have
identical helicity, again allowing only gluonic contributions as shown
in \fig{QuadCutFigure}(c).  An explicit computation shows that the
coefficients of these two boxes are in fact similar to the coefficient
of the $K_{12}$ box~(\ref{K12coefficient}); they are equal to the tree
amplitude multiplied by the same constant and corresponding invariants 
($i s_{45} s_{51}/2$ for $\I(K_{23})$ and 
 $i s_{51} s_{12}/2$ for $\I(K_{34})$).
The gluon-loop contribution to the amplitude is then given by,
\begin{eqnarray}
\hskip-0.5cm
A_5^\oneloop(1^-,2^-,3^+,4^+,5^+)
&=& A_5^\tree(1^-,2^-,3^+,4^+,5^+) \, \cg \, \biggl\{
\nonumber\\
&&\hskip0.0cm
-{1\over\e^2} \biggl[ 
    \biggl({\mu^2\over -s_{34}}\biggr)^{\e}
     +\biggl({\mu^2\over -s_{45}}\biggr)^{\e}
     -\biggl({\mu^2\over -s_{12}}\biggr)^{\e}
    \biggr]
\nonumber\\
&&\hskip0.0cm
  + \Li_2\biggl(1-{s_{12}\over s_{34}}\biggr)
  + \Li_2\biggl(1-{s_{12}\over s_{45}}\biggr)
  + {1\over2} \ln^2\biggl({-s_{34}\over -s_{45}}\biggr)
  + {\pi^2\over6}
\nonumber\\
&&\hskip0.0cm
+\ \hbox{cyclic permutations} \biggr\}
\ \ +\ \hbox{triangles + bubbles}.
\end{eqnarray}
Next we discuss how to evaluate triangle and bubble contributions.

%%%%%%%%%%%%%%%%%%%%%%%%%%%%

\subsection{A Triangle Example}
\label{UniTriExampleSubsection}

In a five-point amplitude with all external legs massless, 
three-mass triangles cannot arise.  The remaining terms in the 
five-gluon amplitude are derived from one- or two-mass
triangles in addition to bubble integrals.  
These integrals can be expressed as,
\begin{eqnarray}
\I_2(s) &=& { i \, \cg \over \e(1-2\e)} 
 \biggl({\mu^2\over-s}\biggr)^\e \,,
\label{I2}\\
\I_3^{\rm 1m}(s) &=& - { i \, \cg \over \e^2} 
 \biggl({\mu^2\over-s}\biggr)^\e \,,
\label{I31m}\\
\I_3^{\rm 2m}(s_1,s_2) &=&  
- { \I_3^{\rm 1m}(s_1) - \I_3^{\rm 1m}(s_2) \over s_1-s_2}  \,.
\label{I32m}
\end{eqnarray}
That is, they are all linear combinations of bubble integrals.
Laurent expanding these expressions in $\e$, we see that
all remaining terms at order $\e^0$
will consist either of logarithms or logarithms squared.  
As an example, we will compute the coefficient of one of the 
logarithms in the internal-scalar contributions to the amplitude 
$A_5^\oneloop(1^-,2^-,3^+,4^+,5^+)$.  These terms will be required
in our recursive computation of the 
rational terms in \sect{LoopRecursionSection}.

To compute these terms, we can proceed in a variety of ways.
In order to detect bubble integrals, we must in any event include ordinary
cuts, which enforce the presence of just two propagators.  
We would proceed, for example, by forming the ordinary cut in
each of the three distinct channels --- $s_{12}$, $s_{23}$, and $s_{34}$.
Various box integrals also have cuts in these channels, so we would
have to subtract their contributions to the integrand.  What is
left will give us the remaining logarithms.  In the internal-scalar example
we are considering, this approach is particularly simple, because
as discussed above
the scalar and fermionic contributions have no box integrals, and
hence there is nothing to subtract.  (In principle, there could be
contributions that integrate to zero, requiring a nontrivial subtraction,
but that does not happen here.)

Helicity
conservation also means that in
both scalar and fermionic contributions, the internal lines emerging
from each tree amplitude must have opposite helicities.  Therefore
the cuts in the $s_{12}$, $s_{34}$, and $s_{45}$ channels will vanish, because
the resulting four-point amplitudes will have the helicity structure
$(-{}+{}+{}+)$, and the corresponding tree amplitudes all vanish.  We are
left with contributions only in the $s_{23}$ and $s_{51}$ channels.
We will compute the coefficient of $\ln(-s_{23})$; the coefficient
of $\ln(-s_{51})$ is related by the reflection symmetry
$(12345)\leftrightarrow (21543)$.  In this particular
amplitude, the triangles generate no squared logarithms.

The scalar contribution to the $s_{23}$ cut is,
\begin{eqnarray}
&& \sum_{h=\pm} A_4^\tree(-\ell_{2s}^{h},2^-,3^+,\ell_{4s}^{-h})
A_5^\tree(-\ell_{4s}^h,4^+,5^+,1^-,\ell_{2s}^{-h})
\nonumber\\
&&= 
%%%%% begin : scalar1
- 2 \frac{\spa{(-\ell_2)}.2^2\spa{\ell_4}.2^2
              \spa{(-\ell_4)}.1^2\spa{\ell_2}.1^2
         }{\spa{(-\ell_2)}.2\spa2.3\spa3.{\ell_4}\spa{\ell_4}.{(-\ell_2)}
     \,\spa{\ell_2}.{(-\ell_4)}\spa{(-\ell_4)}.4\spa4.5\spa5.1\spa1.{\ell_2}}
%%%%% end : scalar1
%\nonumber\\
%&&= - 2 \frac{\spa{\ell_4}.1\spa{\ell_4}.2 (
%           \sandmp2.{\Ellin_2\Ellin_4}.1\sandmp2.{\Ellin_4\Ellin_2}.1 )
%         }{\spa2.3\spa3.{\ell_4} s_{23}^2\spa{\ell_4}.4\spa4.5\spa5.1}
\nonumber\\
&&=
%%%%% begin : scalar2
   2 \frac{\spa{\ell_4}.1\spa{\ell_4}.2 
          \sandmp2.{\Kinsl_{23}\Ellin_{4}}.1\sandmp2.{\Ellin_{4}\Kinsl_{23}}.1 
         }{\spa2.3\spa3.{\ell_4} s_{23}^2\spa{\ell_4}.4\spa4.5\spa5.1}\,.
%%%%% end : scalar2
\end{eqnarray}
(The `helicity' label for a scalar distinguishes particle and 
antiparticle, which give equal contributions.)

We can separate the $\ell_4$-dependent denominators, using
the Schouten identity,
\begin{equation}
\spa{\ell_4}.{1} \spa{3}.{4}\ =\ 
  \spa{\ell_4}.{3} \spa{1}.{4} + \spa{\ell_4}.{4} \spa{3}.{1} \,,
\end{equation}
as a kind of partial-fraction decomposition.  We get,
\begin{equation}
\frac{\spa{\ell_4}.1\spa{\ell_4}.2}{\spa3.{\ell_4} \spa{\ell_4}.4} =
\frac{1}{\spa3.4} \biggl(
\frac{\spa1.3\sand2.{\Ellin_4}.3}{(\ell_4+k_3)^2}
+\frac{\spa1.4\sand2.{\Ellin_4}.4}{(\ell_4-k_4)^2}
\biggr)\,.
\label{PartialFractioned}
\end{equation}
Each term now corresponds to a different triangle integral, the first
to a one-mass one and the second to a two-mass one.  Both still have
non-trivial numerators.  The one-mass triangle turns out to give
a vanishing contribution to our amplitude.

%%%%%% FIGURE %%%%%%%%%%%
\begin{figure}[t]
\centerline{\epsfxsize 1.6 truein\epsfbox{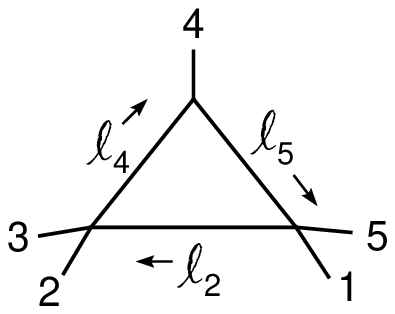}}
\caption{The two-mass triangle arising from the second term in
\eqn{PartialFractioned}. }
\label{Tri2mFigure}
\end{figure}
%%%%%%%%%%%%%%%%%

The two-mass triangle integral depicted in \fig{Tri2mFigure},
coming from the second term in \eqn{PartialFractioned}, 
has the form,
\begin{eqnarray}
&&2 {\spa1.4 \over \spa2.3\spa3.4\spa4.5\spa5.1\,s_{23}^2}
\nonumber\\
&&\hskip1cm \times 
\int {d^{4-2\e}\ell\over(2\pi)^{4-2\e}}
{ \sand2.{\Ellin_4}.4 \sandmp2.{K_{23}\Ellin_4}.1
                       \sandmp2.{\Ellin_4K_{23}}.1
 \over \ell_2^2 \ell_4^2 \ell_5^2 } \,.
\label{triexample}
\end{eqnarray}
Here, and below, the $\ell_i$ are taken to be off-shell loop momenta 
corresponding to the cut momenta denoted by the same variables above.
We can expand the loop momenta in the numerator of \eqn{triexample}
as discussed in \sect{UniStructureSubsection}, using
the basis in \eqn{TwoMassTriangleBasis}:
two real vectors, $v_1^\mu = K_{23}^\mu$ and $v_2^\mu = K_{51}^\mu$,
 as well as
two complex vectors, $v_3^\mu = \sandpm4.{\Kinsl_{23}\gamma^\mu}.4$
and $v_4^\mu = \sandmp4.{\Kinsl_{23}\gamma^\mu}.4$.
This choice is convenient, because powers of these two complex vectors
will give rise to vanishing integrals~\cite{OPP}.  It generalizes in
a straightforward way to processes beyond the five-point example
considered here.  In terms of the $v_i$ basis, the loop momentum
expansion is,
\begin{eqnarray}
\ell_4^\mu &=& 
%%%%% begin : ellc1
-\biggl[\frac{(\ell_4+K_{23})^2-s_{23}}{s_{51}-s_{23}} 
  +\frac{s_{51}+s_{23}}{(s_{51}-s_{23})^2} (\ell_4-k_4)^2\biggr]
%%%%% end : ellc1
 v_1^\mu
\nonumber\\
&&
%%%%% begin : ellc2
-\biggl[\frac{(\ell_4+K_{23})^2-s_{23}}{s_{51}-s_{23}} 
  +2\frac{s_{23}}{(s_{51}-s_{23})^2} (\ell_4-k_4)^2\biggr] 
%%%%% end : ellc2
v_2^\mu
\nonumber\\
&&
%%%%% begin : ellc3
+\frac{1}{2 (s_{51}-s_{23})^2}\sandmp4.{\Kinsl_{23}\Ellin_{4}}.4 
%%%%% end : ellc3
v_3^\mu
%%%%% begin : ellc4
+\frac{1}{2 (s_{51}-s_{23})^2}\sandpm4.{\Kinsl_{23}\Ellin_{4}}.4 
%%%%% end : ellc4
v_4^\mu \,.
\end{eqnarray}
Referring to \fig{Tri2mFigure}, we see that the factor 
$(\ell_4+K_{23})^2 = \ell_2^2$ will cancel a denominator, 
leaving a bubble integral with a massless leg, 
which vanishes in dimensional regularization.
We can therefore drop such terms.  A factor of $(\ell_4-k_4)^2 = \ell_5^2$ will
also cancel a propagator, leaving a bubble with external momentum $K_{23}$.
It may contribute to the coefficient of the logarithm we are computing.

After substituting this form for $\ell_4$ into the numerator
of the integrand, we will
obtain terms leading to integrals in the classes,
\begin{equation}
\I_3[(\ell_4\cdot v_3)^{a_3} (\ell_4\cdot v_4)^{a_4}]
\ \ {\rm\ and\ }\ \ 
\I_2[(\ell_4\cdot k_4)^{a_1} (\ell_4\cdot v_3)^{a_3} 
     (\ell_4\cdot v_4)^{a_4}]\,.
\end{equation}
The underlying tensor integrals can be expressed in terms of the
metric $g^{\mu\nu}$ and either $v_{1,2}^\mu$ in the three-point case,
or $v_1^\mu$ alone in the two-point case.  Because $v_3$ and $v_4$ are
null vectors, and moreover $v_1\cdot v_{3,4} = v_2\cdot v_{3,4} = 0$, 
pure powers of either $\ell_4\cdot v_3$ or $\ell_4\cdot v_4$
in the three-point case ($a_4 = 0$ or $a_3 = 0$), 
or pure powers of them with $\ell_4\cdot v_1$ (for any value of $a_1$),
will yield vanishing integrals.  Such numerator factors in a sense
produce total derivatives, and thereby give an explicit realization
of the vanishing integrals introduced in ref.~\cite{OPP}.
\def\indentA{\hskip 15mm}
We can use an identity,
\begin{eqnarray}
&&\sandpm4.{\Kinsl_{23}\Ellin_4}.4
\sandmp4.{\Kinsl_{23}\Ellin_4}.4 = (\ell_4-k_4)^2 
  \times {1\over2} \tr[\ksl_4\Ksl_{23}\ellsl_4\Ksl_{23}]\nonumber\\
&&\indentA =
%%%%% begin : mixed
(\ell_4-k_4)^2 \Bigl[(s_{51}-s_{23}) ((\ell_4+K_{23})^2 -s_{23})
                           - 2 s_{23} \ell_4\cdot k_4\Bigr]\,,
%%%%% end : mixed
\end{eqnarray}
to eliminate mixed powers of the numerator factors 
($a_3\neq 0\neq a_4$).  Only equal powers, $a_3=a_4$, will give 
a nonvanishing contribution.

We can thus derive the following replacement rules for the integrand,
\begin{eqnarray}
\ell_4^\mu &\rightarrow&
v_+^\mu \equiv 
%%%%% begin : ellrep1
\frac{s_{23}(v_1^\mu+v_2^\mu)}{s_{51}-s_{23}} 
  -\frac{s_{51}v_1^\mu+s_{23}(v_1^\mu+2v_2^\mu)}{(s_{51}-s_{23})^2} 
   (\ell_4-k_4)^2
%%%%% end : ellrep1
\nonumber\\
\ell_4^\mu\ell_4^\nu &\rightarrow& 
%%%%% begin : ellrep2
v_+^\mu v_+^\nu 
-\frac{s_{23} (\ell_4-k_4)^2}{4 (s_{51}-s_{23})^4}
 (s_{51}-s_{23}+2 \ell_4\cdot k_4)
  (v_3^\mu v_4^\nu +v_4^\mu v_3^\nu)
%%%%% end : ellrep2
\nonumber\\
\ell_4^\mu\ell_4^\nu\ell_4^\rho &\rightarrow& 
%%%%% begin : ellrep3
v_+^\mu v_+^\nu v_+^\rho
-\frac{s_{23} (\ell_4-k_4)^2}{4 (s_{51}-s_{23})^4}
(s_{51}-s_{23}+2 \ell_4\cdot k_4)
\nonumber\\
&&\hskip 20mm\times
  (v_3^\mu v_4^\nu v_+^\rho+{\rm permutations\ of\ }(\mu,\nu,\rho))
%%%%% end : ellrep3
\label{TensorReplacement}
\end{eqnarray}
Only terms without factors of $2\ell_4\cdot k_4 = -(\ell_4-k_4)^2$
can produce a two-mass triangle.  These terms come from the
replacement $\ell_4^\mu \rightarrow c (v_1+v_2)^\mu = -c k_4^\mu$, 
which makes the integrand in \eqn{triexample} vanish.  
We are left only with bubble integrals.  
We can think of these replacements as an explicit realization 
of the reductions introduced in ref.~\cite{AguilaPittau}.

To evaluate these terms, we can use the following
little table,
\begin{eqnarray}
\I_2[\ell_4^\mu](s_{23})\ &=&\
 -\frac{1}{2} K_{23}^\mu \, \I_2(s_{23})\,,\nonumber\\
\I_2[\ell_4^\mu\ell_4^\nu](s_{23})\ &=&\ 
  -\frac{1}{4 (D-1)} s_{23} g^{\mu\nu} \, \I_2(s_{23})
  +\frac{D}{4 (D-1)} K_{23}^\mu K_{23}^\nu \, \I_2(s_{23})
\,,
\label{BubbleTable}
\end{eqnarray}
which are the only tensor reductions we need to perform explicitly.
The bubble integral itself is, from \eqn{I2},
\begin{equation}
\I_2(s_{23}) = i \, \cg \biggl(\frac{1}{\e} 
        + \ln\biggl(\frac{\mu^2}{-s_{23}}\biggr)-2\biggr) + \Ord(\e)\,.
\end{equation}
The singular parts can also be extracted from the cuts.  We will
include them in the full cut part below.  The rational
part we will not include, because
as explained in prior subsections, additional information beyond 
the four-dimensional cuts is required to compute the finite rational
terms fully.
The metric term will drop out when evaluating integrals with numerator
powers of $\ell_4\cdot k_4$, so that in our computation we are left with,
\begin{eqnarray}
\I_2[2\ell_4\cdot k_4](s_{23})\ &=&\ 
\onehalf(s_{23}-s_{51}) \, \I_2(s_{23})\,,\nonumber\\
\I_2[(2\ell_4\cdot k_4)^2](s_{23})\ &=& \
\frac{D}{ 4(D-1)} (s_{23}-s_{51})^2 \, \I_2(s_{23})\,.
\end{eqnarray}
Using these values, the terms containing cuts in the $s_{23}$ channel 
are,
\begin{eqnarray}
&&
%%%%% begin : disc23
- \frac{i\cg}{6}{1\over\spa2.3 \spa3.4\spa4.5\spa5.1} 
   \ln\biggl(\frac{-s_{23}}{\mu^2}\biggr)
\nonumber\\
&&\indentA\times\biggl(
2
\frac{\spa2.3 \spb3.4\spa4.1 \spa2.4\spb4.5\spa5.1 B}{(s_{51}-s_{23})^3}
-\frac{\spa1.2^2 B}{s_{51}-s_{23}}
+\spa1.2^3
\biggr)\,.
%%%%% end : disc23
\label{S23coefficient}
\end{eqnarray}
in which $B = \spa2.3\spb3.4\spa4.1+\spa2.4\spb4.5\spa5.1$.

Most of the factors appearing in the denominators, along with the factors
in the denominator in \eqn{K12coefficient}, are nearest-neighbor 
spinor products.  These correspond to genuine physical singularities
of the one-loop amplitude: as at tree level, their vanishing corresponds
to a collinear limit.  The $s_{51}-s_{23}$ and $(s_{51}-s_{23})^3$
denominators do not correspond to such physical singularities, and
as such cannot be present in the amplitude.  The singularity in the
cut-containing terms is canceled by the contribution from the $s_{51}$
channel, which is related by the flip symmetry 
$(12345) \leftrightarrow (21543)$.  It combines with the expression
in \eqn{S23coefficient} via the replacement,
\begin{equation}
\ln\biggl(\frac{-s_{23}}{\mu^2}\biggr) \longrightarrow
\ln\biggl(\frac{-s_{23}}{-s_{51}}\biggr)\,.
\end{equation}
The logarithms in these contributions thus show up in the following
functional forms containing spurious singularities,
\begin{equation}
\frac{\ln r}{1-r}\qquad{\rm\ and\ }\qquad
\frac{\ln r}{(1-r)^3}\,,
\end{equation}
where $r$ is the ratio of two momentum invariants.
The singularity in these functions as $r\to1$ is purely rational,
\begin{eqnarray}
\frac{\ln r}{1-r}\ &\sim&\ -1 + \frac{r-1}{2} - \frac{(r-1)^2}{3} + \cdots\,,
\nonumber\\
\frac{\ln r}{(1-r)^3}\ &\sim&\ -\frac{1}{(r-1)^2} + \frac{1}{2 (r-1)}
  -\frac13 + \cdots\,.
\end{eqnarray}
The remaining singular terms must therefore be canceled by rational
terms.  We show how to compute these terms in the next section.  It
will be convenient, however, to anticipate the cancellation, and to define
new functions which are manifestly free of spurious singularities,
\begin{eqnarray}
\Ll_0(r)\ &=&\ \frac{\ln r}{1-r}\,,\nonumber\\
\Ll_1(r)\ &=&\ \frac{\ln r + 1-r}{(1-r)^2}\,, \label{Lldef}  \\
\Ll_2(r)\ &=&\ \frac{\ln r - (r-1/r)/2}{(1-r)^3}\,.\nonumber
\end{eqnarray}

We can now assemble the complete cut-containing terms for 
the internal-scalar contributions to $A_5^\oneloop(1^-,2^-,3^+,4^+,5^+)$,
\def\indentB{\hskip 5mm}
\begin{eqnarray}
&&A_5^{\rm scalar}(1^-,2^-,3^+,4^+,5^+)\big|_{\rm cut-containing} =
\nonumber\\
&&\indentB 
%%%%% begin : scalarResult
i\cg \Biggl\{
 \frac{\spa1.2^3}{6 \spa2.3\spa3.4\spa4.5\spa5.1} \biggl[
{2\over\e} - \ln\biggl(\frac{-s_{23}}{\mu^2}\biggr)
           - \ln\biggl(\frac{-s_{51}}{\mu^2}\biggr)\biggr]
\nonumber\\
&&\indentB-\frac{1}{3\spa2.3 \spa3.4\spa4.5\spa5.1} 
\nonumber\\
&&\indentB\indentB\times\biggl[
\frac{\spa2.3 \spb3.4\spa4.1 \spa2.4\spb4.5\spa5.1 B
   \Ll_2\Bigl(\frac{-s_{23}}{-s_{51}}\Bigr)}{s_{51}^3}
-\frac{\spa1.2^2 B \Ll_0\Bigl(\frac{-s_{23}}{-s_{51}}\Bigr)}{2 s_{51}}
\biggr] \Biggr\} \,.
%%%%% end : scalarResult
\label{Cutmmppp}
\end{eqnarray}
The simplifications due to the choice of basis used here in the
evaluation of two-mass triangle contributions should generalize to other
amplitudes.  Using the bases given in
\eqns{OneMassTriangleBasis}{TwoMassTriangleBasis}, we will obtain
replacements similar to those given in~\eqn{TensorReplacement}.

%%%%%%%%%%%%%%%%%%%%%%%%%%%%%

\subsection{A Spinorial Approach}

The above example illustrates how triangle and bubble integral
coefficients can be evaluated in an ordinary 
two-particle cut using projections
and vanishing integrals~\cite{AguilaPittau,OPP}.  For more complicated
cases, it is useful to have a systematic procedure that still keeps 
the analytic expressions relatively compact.  (Feynman parametrization
of cut loop integrals, for example, is a systematic method, 
but it tends to generate  an explosion of terms.)
In refs.~\cite{BBCF,BFM}, Britto  {\it et al.} used
spinorial variables in an efficient and systematic technique for 
evaluating generic one-loop unitarity cuts.
The cut integration is effectively performed by residue extraction. 
Applying these ideas, they computed the
cut-containing terms for the most complicated of the six-gluon helicity
amplitudes with a scalar or fermion circulating in the loop. 

This method makes use of an elegant decomposition of
phase-space integrals in terms of spinor variables~\cite{CSW},
\begin{equation}
\int d^4 \ell \, \delta^{(+)}(\ell^2) f(\ell) = \int_0^\infty t \, dt \int 
         \spa{\ell}.{d \ell} \spb{\ell}.{d \ell} f(\ell)\,,
\end{equation}
where
$
\spa{\ell}.{d \ell} \equiv \varepsilon^{\alpha \beta} (\lambda_\ell)_\alpha
(d\lambda_\ell)_\beta $, 
$\spb{\ell}.{d \ell} \equiv \varepsilon^{\dot \alpha \dot \beta}
     (\tlambda_\ell)_{\dot \alpha} (d\tlambda_\ell)_{\dot \beta}
$
and $\ell_{\mu} = t  \sand{\ell}.{\gamma_\mu}.{\ell}/2$.
On the contour of integration $\lambda_\ell$ and $\tlambda_\ell$ 
should be treated as complex conjugate variables, although elsewhere in
this review we treat them as independent.

In a two-particle cut, this decomposition gives,
\begin{eqnarray}
C_2 &=& \int d^4 \ell \, \delta^{(+)} (\ell^2) \, 
\delta^{(+)}((\ell - K)^2) \, f(\lambda_\ell,\tlambda_\ell,t )
 \nn \\
&= &
\int_0^\infty t \, dt \int \spa{\ell}.{d\ell} \spb{\ell}.{d\ell}
\delta^{(+)}(K^2- t \sand{\ell}.{K}.{\ell}) \, 
f(\lambda_\ell, \tlambda_\ell, t) \,.
\end{eqnarray}
In the kinematic region $K^2 > 0$, the delta function is non-vanishing
when the integration over $t$ is performed, 
\begin{equation}
C_2 = \int \spa{\ell}.{d\ell} \spb{\ell}.{d\ell}  
{K^2 \over \sand{\ell}.{K}.{\ell}^2 } \, 
f\Bigl(\lambda_\ell,\tlambda_\ell,
 t=\frac{K^2}{\sand{\ell}.{K}.{\ell}}\Bigr) \,.
\end{equation}
It is possible to split $f$ into functions whose numerators
depend only on $\lambda_\ell$, whose integration gives rise to
box integrals and three-mass triangle integrals, and
remaining terms whose numerators depend on both $\lambda_\ell$
and $\tlambda_\ell$.   For the latter terms,
the key idea is to rewrite the cut as a total derivative of the
form,
\begin{equation}
C_2  = \int \spa{\ell}.{d \ell} \spb{d \ell}.{\partial_\ell}\,
g(\lambda_\ell, \tlambda_\ell) \,,
\end{equation}
plus a set of residues.   The total derivative integrates to zero.
The residues arise from $1/\spa{\ell}.{a}$  
poles in $g(\lambda_\ell, \tlambda_\ell)$.
At these singularities, one obtains delta-function
contributions~\cite{HolomorphicAnomaly} using,
\begin{equation}
\spb{d \ell}.{\partial_\ell}
{1\over \spa{\ell}.a} = - 2 \pi \bar \delta(\spa{\ell}.a ) \,,
\label{deltafunction}
\end{equation}
where $\int \spa{\ell}.{\d \ell} \bar\delta(\spa{\ell}.a ) A(\lambda_\ell,
\tlambda_\ell) = -i A(\lambda_a, \tlambda_a)$. 
This localization by delta functions relies on treating
$\lambda_\ell$ and $\tlambda_\ell$ as complex conjugates~\cite{CSW}.
The delta functions allow us to replace the spinor integration variable
$\lambda_\ell$ with spinors of the external momenta,
since the delta-function condition in \eqn{deltafunction} is satisfied
whenever $\lambda_\ell \propto \lambda_a$ and $\tlambda_\ell \propto
\tlambda_a$.  (The constant of proportionality always drops out.)

In more complicated cases, where there are multiple poles in
$\tlambda_\ell$, one must first perform a partial-fraction decomposition
or a Feynman parametrization in order to rewrite the integrand as a
total derivative plus localized contributions. This
procedure has also been extended to the cases of
$D$-dimensional unitarity~\cite{ABFKM} and massive particles circulating
in the loop~\cite{BFMassive}.
           
%%%%%%%%%%%%%%%%%%%%%%%%%%%%%%%%%%%%%%%

\section{On-Shell Recursion at One Loop}
\label{LoopRecursionSection}

In this section, we discuss how to compute rational terms
in one-loop amplitudes.  We assume that the terms containing 
cuts have already been computed via four-dimensional
unitarity, as described in the previous section.
A number of analytic methods for determining the rational terms 
have been proposed in the literature.  A recent proposal is to 
make use of $D$-dimensional unitarity~\cite{BMST,ABFKM,BFMassive},
following the early work in refs.~\cite{DDimU,OneLoopReview}.
Another proposal relies on the observation that only a limited number
of loop-momentum integrals can contribute to
rational terms, making their evaluation simpler~\cite{XYZ,BinothRat}.
In this review, we focus on the on-shell recursive
approach~\cite{OnShellOneLoop,Qpap,Bootstrap,Genhel,LoopMHV}.
This approach is quite promising because it displays
only mild growth in computational complexity as the number
of external legs increases.
(In special cases, the pole structure of the rational coefficients in
front of integral functions is well-enough constrained that they
can be computed using this technique too, instead of using 
unitarity~\cite{SplitHelicityLoop}.)

At one loop as at tree level, on-shell recursion provides a 
systematic means of determining rational functions, using
knowledge of their poles and residues.  At loop level, however, 
there are a number of new issues that must be confronted.

The most obvious issue is the appearance of branch cuts
in the shifted amplitude.  We will find it helpful
to distinguish between terms that contain such cuts
and terms that do not.
Let us break up a one-loop amplitude, 
shifted by a $\Shift{j}{l}$ shift, as follows,
\begin{equation}
A_n^\oneloop(z) =  \cg {} \Bigl[\PureCut_n(z) + \Vertex_n(z) \Bigr] \,.
\label{PureCutDecomposition}
\end{equation}
where $\PureCut_n(z)$ denotes the pure cut-containing terms, 
and $\Vertex_n(z)$ gives the rational terms.
The rational parts $\Vertex_n$ are defined by setting all logarithms,
polylogarithms, and associated $\pi^2$ terms to zero,
\begin{equation}
\Vertex_n \equiv {1\over \cg} A_n^\oneloop\Bigr|_{\rm rat} \equiv
{1\over \cg} A_n^\oneloop\biggr|_{\ln, \Li, \pi^2 \rightarrow 0} \,.
\label{RationalDefinition}
\end{equation}
We assume that the cut-containing terms $\PureCut_n$ 
have already been computed.

As discussed in \sect{UniTriExampleSubsection}, loop amplitudes
contain spurious poles that cancel only
between the cut-containing and rational contributions.  Such spurious
poles would complicate the on-shell recursion relations because
they would require us to compute residues at poles corresponding
to unphysical singularities.  Such singularities
are not singularities of the total amplitude at all.
The simplest approach is to eliminate them completely.  
We can do this by adding in rational terms that manifestly
cancel all spurious singularities, 
before applying Cauchy's theorem to the other rational terms.
We gave examples of such `cut completions', associated with the structure
of two-mass triangle integrals, in \sect{UniTriExampleSubsection}.  
These completions introduce
functions $\Ll_i(r)$ defined in \eqn{Lldef}, which are nonsingular
as $r\to1$.  More intricate spurious singularities can also arise.
 One can make their absence manifest
using functions (based on higher-dimensional box and triangle
integrals) introduced in ref.~\cite{CampbellGloverMiller}.

In general, we denote the cut completion by $\Cuth_n$,
\begin{equation}
\Cuth_n(z) = C_n(z) + \widehat{CR}_n(z) \,,
\label{CuthCCR}
\end{equation}
where $\widehat{CR}_n(z)$ are the rational functions added in order to
cancel the unphysical spurious singularities in $z$.  For a given
shift we do not need to remove all spurious singularities, but only
those that depend on $z$.  Having added rational 
terms to the cuts, we must subtract them from the
rational part of the amplitude.  This procedure defines the `remaining'
rational terms,
\begin{equation}
\Remaining_n(z) =   \Vertex_n(z) - \widehat{CR}_n(z)\,.
\end{equation}
Instead of the pure-cut plus pure-rational
decomposition~(\ref{PureCutDecomposition}), we thus consider the 
completed-cut decomposition,
\begin{equation}
A_n^\oneloop(z) = \cg {}\Bigl[ \Cuth_n(z) + \Remaining_n(z) \Bigr] \,,
\label{CompletedCutDecomposition}
\end{equation}
and analyze the properties of $\Remaining_n(z)$ in the complex plane.
This decomposition is not unique; rational functions having no
spurious singularities can be moved between $\Cuth_n$ and $\Remaining_n$.

As a concrete example, consider the cut parts of the scalar-loop
contribution to the five-point amplitude
$A_5^{\rm scalar}(1^-,2^-,3^+,4^+,5^+)$  given in \eqn{Cutmmppp}.
The $\Ll_i$ functions defined in \eqn{Lldef} incorporate
rational terms and automatically provide a satisfactory cut completion 
for this amplitude.  The spurious poles located at $s_{23} - s_{51} = 0$ 
are manifestly absent from the cut parts.  Hence they will
also not appear in the remaining rational terms.

%%%%%% FIGURE %%%%%%%%%%%
\begin{figure}[t]
\centerline{\epsfxsize 5.0 truein\epsfbox{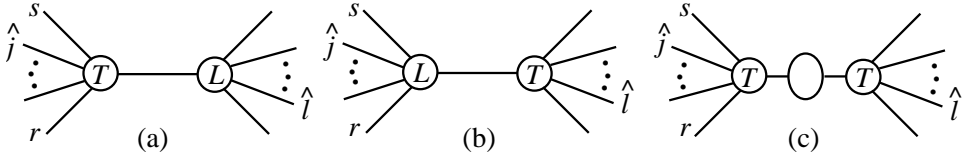}}
\caption{Diagrammatic representation of recursive contributions 
at one loop with a $\Shift{j}{l}$ shift.
The labels `$T$' and `$L$' on the vertices indicate tree and loop vertices,
respectively. The factorization-function contribution (c) does not appear
for MHV amplitudes.}
\label{LoopGenericFigure}
\end{figure}
%%%%%%%%%%%%%%%%%

In general, the complete amplitude cannot have any unphysical poles.
After we eliminate them from the completed-cut parts, they
cannot appear in the remaining rational terms $\Remaining_n(z)$ either.
We can then construct an on-shell recursion relation for $\Remaining_n(z)$,
summing only over residues at physical poles~\cite{Bootstrap,Genhel}.
The terms in the sum are in one-to-one
correspondence with the factorization channels, just as at tree level.
At one loop, though, there are typically more terms in each channel.
For each intermediate state helicity, there are generically three
contributions to one-loop factorization as
$K_{r\cdots s}^2 \rightarrow 0$, which are
depicted in \fig{LoopGenericFigure},
\begin{equation}
A_n^\oneloop\ \sim\ A_L^\tree \, {i\over K_{r\cdots s}^2 } \, A_R^\oneloop
+A_L^\oneloop \, {i\over K_{r\cdots s}^2  } \, A_R^\tree
+A_L^\tree \, {i \, \Fact^\oneloop\over K_{r\cdots s}^2} \, A_R^\tree\,.
\label{FullOneLoopFact}
\end{equation}
In the first two terms, one of the factorized amplitudes is a one-loop
amplitude and the other is a tree amplitude.  The last term
naively corresponds to a one-loop correction to the propagator.
However, massless theories contain infrared divergences from soft
and collinear virtual momenta.  These divergences do not commute
with the factorization limit $K_{r\cdots s}^2 \to 0$.
For this reason, the `factorization function' $\Fact^\oneloop$
is more subtle in massless theories; it can 
contain `pole-crossing' logarithms (logarithms of momentum invariants 
containing momenta from both sides of the pole)~\cite{BernChalmers}.
However, we are only interested in the rational terms.
These terms do have a simple interpretation in terms of 
propagator corrections.

Following similar logic as at tree level, and dropping the pure-cut
pieces in \eqn{FullOneLoopFact},
we obtain an on-shell recursion relation for the rational terms,
corresponding to the diagrams in \fig{LoopGenericFigure},
\def\indentA{\hskip 7mm}
\begin{eqnarray}
R^D_n &\equiv& \ 
 -\sum_{{\rm poles}\,\alpha} \Res_{z=z_\alpha} {\Vertex_n(z)\over z}
\nonumber\\
&=&\ 
\sum_{r,s,h} \Biggl\{
\Vertex_L(z = z_{rs}) \,
{i\over K_{r\cdots s}^2} \, 
A^\tree_R(z= z_{rs}) 
+ A^\tree_L(z= z_{rs})
\, {i\over K_{r\cdots s}^2} \,
\Vertex_R(z= z_{rs}) \nn\\
&& \null \hskip 1. cm
+ A_L^\tree(z= z_{rs})  
\, {i \Fact(K_{r\cdots s})\over K_{r\cdots s}^2} \,
A^\tree_R(z= z_{rs}) \Biggr\}
 \,.  \label{RationalRecursion}
\end{eqnarray}
The `vertices' $\Vertex_L$ and $\Vertex_R$ in this recursion relation 
are the pure rational parts (according to the 
definition~(\ref{RationalDefinition}) of the lower-point one-loop
amplitudes appearing on the left and right sides of \eqn{FullOneLoopFact}.
The factorization function ${\cal F}$ may be found in ref.~\cite{Genhel}. 
It only contributes in multi-particle channels, and only if the
tree amplitude contains a pole in that channel.
The superscript $D$ on $R^D_n$ indicates that this set of (recursive)
diagrammatic contributions to the remaining rational terms $\hat{R}_n$
is not the whole answer.

%%%%%% FIGURE %%%%%%%%%%%
\begin{figure}[t]
\centerline{\epsfxsize 2.0 truein\epsfbox{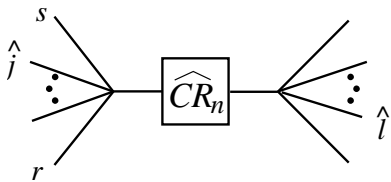}}
\caption{A generic overlap contribution with 
a $\Shift{j}{l}$ shift.  The diagram indicates the channel where
the residue is to be extracted.}
\label{OverlapGenericFigure}
\end{figure}

We have included some of the rational terms $\CuthRat_n$ in the 
completed cuts $\Cuth_n$ in shifting from \eqn{PureCutDecomposition} 
to~(\ref{CompletedCutDecomposition}).  Therefore we need to correct 
\eqn{RationalRecursion} 
for any residues that $\CuthRat_n(z)$ may have at the physical poles.  
We know $\CuthRat_n$ explicitly, so it is simple to compute the 
correction or `overlap' terms $\Overlap_n$ by performing 
the shift~(\ref{SpinorShift}) 
and extracting the residues of $\CuthRat_n(z)$ at each physical pole,
\begin{equation}
\Overlap_n \equiv \sum_{{\rm poles}\,\alpha} \Res_{z=z_\alpha}
                 {\CuthRat_n(z)\over z} \, .
\label{OverlapSum}
\end{equation}
These overlap contributions may be assigned a diagrammatic
interpretation, as depicted in \fig{OverlapGenericFigure}.
Each diagram corresponds to a different physical factorization channel.
Although the definition of the completed-cut terms $\Cuth_n$ is
not unique, the ambiguity cancels in the sum of $\Cuth_n(0)$ and the
overlap terms $\Overlap_n$.

We are not yet done, because we have not discussed the behavior 
of the shifted one-loop amplitude at large $z$.  
The straightforward application of Cauchy's theorem underlying 
our computation requires that the shift of the remaining rational
parts $\Remaining_n(z)$ fall off as $z\to\infty$.
At tree level, as discussed in \sect{TreeRecursionSection},
one can prove that $A_n^\oneloop(z)$ vanishes as $z \rightarrow \infty$
for various classes of shifts.  At one loop there are no such theorems,
and we are sometimes forced to use shifts 
for which $A_n^\oneloop(z)$ does not vanish.
Let us denote by $\Inf A_n^\oneloop$ a function which, when shifted, 
matches the behavior of $A_n^\oneloop(z)$ as $z \rightarrow \infty$, 
that is $A_n^\oneloop(z) - \Inf A_n^\oneloop(z) \to 0$ as $z \rightarrow \infty$.
(This function is rational in the existing examples.)  
A practical approach for determining $\Inf A_n^\oneloop$ is by
using an auxiliary on-shell recursion relation~\cite{Genhel}.
The completed-cut terms $\Cuth_n(z)$ may also be nonvanishing at
large $z$; we define an analogous function $\Inf \Cuth_n$ which 
matches their large $z$ behavior.  Neither $\Inf A_n^\oneloop$ nor
$\Inf \Cuth_n$ contains poles at finite $z$, so they do not affect
\eqns{RationalRecursion}{OverlapSum}.

Putting together all the different pieces, 
the full one-loop amplitude is~\cite{Genhel},
\begin{equation}
A_n^\oneloop(0)
= \Inf A_n^\oneloop +  \cg {} 
\Bigl[\Cuth_n(0) - \Inf\Cuth_n + R_n^D + O_n \Bigr]
\,.
\label{BasicFormula}
\end{equation}
Equivalently, the terms to be added to the completed-cut terms $\Cuth_n$
are,
\begin{equation}
\hat{R}_n
= {1\over \cg} \Inf A_n^\oneloop - \Inf\Cuth_n + R_n^D + O_n \,.
\label{BasicFormulaRhat}
\end{equation}
The role of $\Inf A_n^\oneloop$ and $(-\Inf\Cuth_n)$ is to ensure
the proper large-$z$ behavior under the shift.  The overlap 
terms $O_n$ remove double-counting of residues at physical poles
between the completed-cut terms $\Cuth_n(0)$ and the recursive diagrams
$R_n^D$.

One subtlety connected with the use of complex momenta
 is the appearance of physical poles at finite values of $z$
whose residues are not known {\it a priori}.
The factorization properties of amplitudes with real
external momenta are well studied at tree and loop 
level~\cite{TreeReview,Neq4Oneloop,BernChalmers,SplitUnitarity,TwoLoopSplit}.
In multi-particle factorization channels, this information is sufficient
to evaluate the residues at the complex poles corresponding to the 
recursive diagrams in \eqn{RationalRecursion}.
However, factorization in two-particle channels, such as in
\eqn{MHVBCFdiagram}, is more subtle in complex kinematics
than in real kinematics.  

The additional complexity in two-particle channels is due
to the fact that, in a real collinear limit, $k_i \parallel k_j$, 
the spinor products $\spa{i}.{j}$ and $\spb{i}.{j}$ vanish at the same rate,
whereas one of the two remains nonzero in complex on-shell three-point
kinematics.   
In a particular collinear limit where gluons $i$ and $j$ have positive
helicity, and the intermediate gluon with momentum $K_{ij}$ has negative
helicity, there is a potential violation of helicity conservation by 3 units.
Associated with this is a collinear behavior
\begin{equation}
A(\ldots,i^+,j^+,\ldots) 
\sim { \spb{i}.{j} \over \spa{i}.{j}^2 } \,.
\label{flip3units}
\end{equation}
Such behavior is absent at tree level, because as noted
in~\eqn{threeplusvanish}, $A_3^\tree(1^+,2^+,3^+)$ vanishes.  The
typical tree-level behavior in a collinear limit is $A \sim
1/\spa{i}.{j}$ or $A \sim 1/\spb{i}.{j}$.  At one loop, the behavior
(\ref{flip3units}) does arise.  For real kinematics, the behavior in
\eqn{flip3units} is equally singular in magnitude, $|A| \sim
1/\sqrt{\vph{}s_{ij}}$.  With complex momenta, however, one can
approach a kinematic point at which $\spa{i}.{j}$ vanishes but
$\spb{i}.{j}$ does not; the more singular behavior~(\ref{flip3units})
then leads to double poles in the $z$ plane.  The residues in such
channels are not yet fully understood, and may not even be universal,
because they peer deeper into the Taylor expansion around the pole.

One can also find examples of channels in which loop amplitudes
contain `unreal poles' $\spb{i}.{j}/\spa{i}.j$ which can be singular
for complex momenta but are finite for real momenta~\cite{Qpap}.  At
tree level this phenomenon also does not occur.  As the residues at
unreal poles are not yet fully understood either, the best strategy is
to choose shifts that avoid channels with unknown factorization
properties.  For $n$-gluon amplitudes the problematic channels always
have precisely two identical-helicity gluons on one side of the
factorization, as in \eqn{flip3units}.  A general strategy for
avoiding such channels has been given in ref.~\cite{Genhel}.

%%%%%%%%%%%%%%%%%%%%%%%%%%%

\subsection{One-loop five-point example}

We illustrate the on-shell recursive approach to determining loop
amplitudes by recomputing the rational part of a one-loop five-gluon
QCD amplitude first computed in ref.~\cite{GGGGG}.

Consider the amplitude $A_{5}^{\rm scalar}(1^-,2^-,3^+, 4^+, 5^+)$
for five external gluons, with a colored scalar circulating in the loop.
We begin with the cut-containing parts of this amplitude.  
They are given in \eqn{Cutmmppp}. For convenience, in defining the 
completed-cut terms $\Cuth_5$
we add in a rational term proportional to the tree amplitude.
(As explained in the last section, we are free to add any rational terms 
that do not introduce new spurious singularities in $z$.) 
We use,
\begin{eqnarray}
\Cuth_5 & = &
{1\over \cg} \, 
A_5^{\rm scalar}(1^-,2^-,3^+,4^+,5^+)\big|_{\rm cut-containing}\nn \\
&& \null
+ {8 \over 9} A_5^\tree(1^-,2^-,3^+,4^+,5^+) \,,
\label{CutCompletionmmppp}
\end{eqnarray}
where the tree amplitude is given in \eqn{MHVtree}, with $j=1$, $k=2$ and 
$n=5$. Taking the rational part of this expression gives
\begin{eqnarray}
\CuthRat_5\ &=&\     
%%%%% begin : Cuth5Rat
i  \biggl( {1\over 3 \eps} + {8 \over 9} \biggl) {\spa1.2^4 \over \spa1.2
    \spa2.3 \spa3.4 \spa4.5\spa5.1}
 -{i\over 6} {s_{51} + s_{23} \over s_{23} s_{51} (s_{51} -  s_{23})^2}
\nonumber\\
&&\hskip0cm \times 
   {\spb3.4\spa4.1\spa2.4\spb4.5
 \Bigl(\spa2.3\spb3.4\spa4.1+\spa2.4\spb4.5\spa5.1 \Bigr)\over\spa3.4\spa4.5}
%%%%% end : Cuth5Rat
    \,,
\label{CuthRatFive}
\end{eqnarray}
which is needed to determine the overlap terms.

We construct the rational parts using a $\Shift12$ shift,
\begin{equation}
\tlambda_1 \rightarrow \tlambda_1 - z \tlambda_2 \,, \hskip 2 cm
\lambda_2 \rightarrow \lambda_2 + z \lambda_1 \,.
\label{SpinorShift12}
\end{equation}
We will assume that the shifted amplitude vanishes at large $z$,
so that $\Inf A_5 = 0$.  We can verify the validity of this assumption
at the end of the calculation.
It is not difficult to check that the
rational part of the cut term $\CuthRat_5$, given in
\eqn{CuthRatFive}, vanishes at large $z$; thus $\Inf \Cuth_5 = 0$ as well
in this example.

%%%%%%%%%%%%%%%%%%%%
%FIGURE
%
\begin{figure}[t]
\centerline{\epsfxsize 3.9 truein \epsfbox{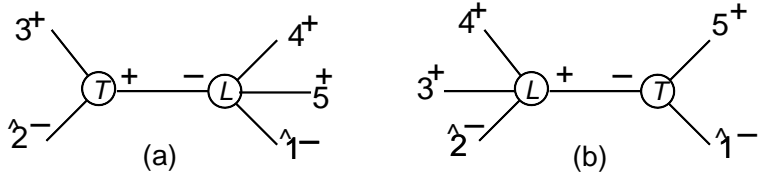}}
\caption{The recursive diagrams for computing the rational parts of
$A_{5}^{\rm scalar}(1^-, 2^-, 3^+, 4^+, 5^+)$ with the $\Shift12$ shift
given in \eqn{SpinorShift12}. `$T$' signifies a tree vertex and `$L$'
a loop vertex.}
\label{RecursmmpppFigure}
\end{figure}
%%%%%%%%%%%%%%%%%%%%%%

With the $\Shift{1}{2}$ shift, all diagrams (including loop diagrams)
with a $(-{}-{}+)$-type three-vertex containing leg 2 
vanish.  Similarly, all diagrams with a $(-{}+{}+)$-type three-vertex 
containing leg 1 also vanish.
One-loop three-vertices for which the external gluons have 
opposite helicity vanish as well.  Some of these vanishings are 
related to those of tree-level three-vertices with `wrong' kinematic
solutions, as explained in ref.~\cite{Genhel}.

We have just two nonvanishing recursive diagrams. 
Diagram (a) in \fig{RecursmmpppFigure} is given by 
\begin{eqnarray}
D_5^{\rm (a)} &=&  A_3^\tree(\hat 2^-, 3^+, -\Kh_{23}^+) 
          \,  {i\over s_{23}} \,
                 \Vertex_4 (\hat 1^-, \Kh_{23}^-, 4^+, 5^+)
\,.
\label{D5a1}
\end{eqnarray}
The required three-point tree amplitude is
given in \eqn{threeptfinalconj}.   The four-vertex is easily
constructed from the known four-point amplitude~\cite{FDH},
\begin{equation}
A_{4}^{\rm scalar}(1^-,2^-, 3^+, 4^+) =
\cg A_4^\tree(1^-,2^-, 3^+, 4^+) \biggl({1\over 3\eps}  -
 {1\over 3} \ln\biggl({ -s_{23} \over \mu^2} \biggr) + {8\over 9}
   \biggr) \,.
\end{equation}
Following \eqn{RationalDefinition} and setting the logarithm
to zero gives us the four-vertex,
\begin{equation}
R_4(1^-,2^-, 3^+, 4^+) =
\biggl({1\over 3\eps}  + {8\over 9} \biggr) A_4^\tree(1^-,2^-, 3^+, 4^+)\, .
\label{R4mmpp}
\end{equation}
Because this vertex is proportional to the four-point tree amplitude,
the evaluation of this diagram is identical to the (unique) recursive
diagram for the five-point tree amplitude with the same helicities
and shift.  The result is
\begin{equation}
D_5^{\rm (a)} = 
%%%%% begin : D5a
i \biggl( {1\over 3 \eps} + {8 \over 9} \biggl) {\spa1.2^4 \over \spa1.2
    \spa2.3 \spa3.4 \spa4.5\spa5.1}
%%%%% end : D5a
\label{D5a2}
 \,.
\end{equation}

The second diagram is,
\begin{equation}
D_5^{\rm (b)} 
 = A_3^\tree(5^+,\hat 1^-, -\Kh_{51}^-) \times {i\over s_{51}} \times
                 \Vertex_4 (\hat 2^-, 3^+, 4^+, \Kh_{51}^+) \,.
\end{equation}
Again the one-loop four-vertex comes directly from the 
known four-point amplitude~\cite{FDH}.  
For this helicity configuration, the amplitude is
purely rational so we have 
\begin{equation}
 \Vertex_4 (1^-, 2^+,3^+,4^+)  = {1\over \cg} 
A_4^{\rm scalar}(1^-,2^+, 3^+, 4^+) = 
   {i \over 3} {\spa2.4 \spb2.4^3 \over \spb1.2 \spa2.3 \spa3.4 \spb4.1} \,.
\end{equation}
The diagram evaluates to,
\begin{eqnarray}
D_5^{\rm (b)} 
 &=& - {i\over 3} {\spash{\hat 1}.{(-\Kh_{51})}^3 \over \spash{5}.{\hat 1} 
           \spash{(-\Kh_{51})}.5 \vphantom{\hat{K}}} \, {1\over s_{51}} \,
       {\spash3.{\Kh_{51}} \spbsh3.{\Kh_{51}}^3 \over 
       \spbsh{\hat 2}.3 \spa3.4 \spash4.{\Kh_{51}} \spbsh{\Kh_{51}}.{\hat 2}
       \vphantom{\hat{K}} }
                     \nn\\
 &=&
%%%%% begin : D5b1
  {i\over 3} {\sand{1}.{5}.2 ^3 \over \spa{5}.{1} 
           \sand5.{1}.2 }  \, {1\over \spa5.1 \spb1.5 \sand1.5.2^2} \,
       {\sand3.{4}.2 \sand1.{5}.3^3 \over 
       \spb2.3 \spa3.4 \sand4.{3}.2 \sand1.5.2} \nn \\
%%%%% end : D5b1
 & = & 
%%%%% begin : Df5
- {i \over 3} {\spb2.4\spb3.5^3\over \spa3.4\spb1.2\spb1.5\spb2.3^2} \,.
%%%%% end : Df5
\end{eqnarray}
%

%%%%%%%%%%%%%%%%%%%%
%FIGURE
%
\begin{figure}[t]
\centerline{\epsfxsize 4 truein \epsfbox{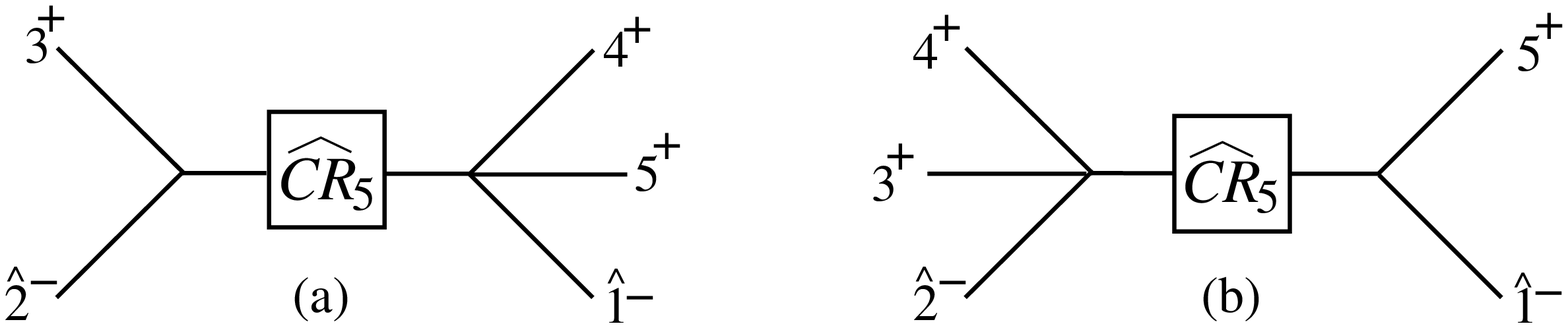}}
\caption{The five-point overlap diagrams using the $\Shift{1}{2}$
shift given in \eqn{SpinorShift12}.}
\label{Overlap5Figure}
\end{figure}
%%%%%%%%%%%%%%%%%%%%%%

The overlap contributions from \eqn{OverlapSum} are 
depicted in \fig{Overlap5Figure}.
The rational parts of the completed-cut terms, $\CuthRat_5$, are given 
in \eqn{CuthRatFive}.  
Applying the shift~(\ref{SpinorShift12}) to $\CuthRat_5$, we obtain,
\begin{eqnarray}
\CuthRat_5(z) & = &
%%%%% begin : Cuth5RatZ
i \biggl( {1\over 3 \eps} + {8 \over 9} \biggl) {\spa1.2^4 \over 
  \spa1.2  (\spa2.3 + z \spa1.3) \spa3.4 \spa4.5\spa5.1} \nn \\
&& \null \hskip .2cm 
 - {i\over 6} {\spb3.4\spa4.1(\spa2.4 + z \spa1.4) \spb4.5
    \over\spa3.4\spa4.5} \nn \\
&& \null \hskip .3cm \times
 { (\spa2.3+ z \spa1.3)  \spb3.4 \spa4.1+ (\spa2.4 + z \spa1.4)
            \spb4.5\spa5.1 
 \over (\spa2.3 + z \spa1.3) \spb3.2 \spa1.5 }
   \\
&& \null \hskip .3cm \times
    {s_{51} + s_{23}  -z \sand1.5.2  + z \sand1.3.2 
         \over (\spb5.1 -z \spb5.2)
           (s_{51} - s_{23} - z \sand1.{(5+3)}.2)^2 }
%%%%% end : Cuth5RatZ
 \,. \nn
\end{eqnarray}
The residues of $\CuthRat_5(z)/z$ have to be evaluated
at the following values of $z$,
\begin{equation}
z^{\rm (a)} =  - {\spa2.3 \over \spa1.3}\,,  \hskip 2 cm 
z^{\rm (b)} =    {\spb1.5 \over \spb2.5} \,, \hskip 2 cm 
\end{equation}
corresponding to the two overlap diagrams in \fig{Overlap5Figure}. 
The first residue is 
\begin{eqnarray}
O_5^{\rm (a)} &=&
%%%%% begin : O5a
-i \biggl( {1\over 3 \eps} + {8 \over 9} \biggl) {\spa1.2^4 \over \spa1.2
    \spa2.3 \spa3.4 \spa4.5\spa5.1}
- {i\over6} {\spa1.2^2\spa1.4\spb3.4
           \over \spa1.5\spa2.3\spa3.4\spa4.5\spb2.3} 
%%%%% end : O5a
 \,.
\label{OverLapDiagrama}
\end{eqnarray}
With our choice of cut completion (\ref{CutCompletionmmppp}), the singular
term and associated rational terms cancel between the recursive and overlap
contributions in \eqns{D5a2}{OverLapDiagrama}.
The second overlap diagram, shown in \fig{Overlap5Figure}(b), gives 
\begin{eqnarray}
O_5^{\rm (b)} &=& 
%%%%% begin : O6b
{i\over6} {\spa1.4\spb3.4\spb3.5 \bigl(\spa1.4\spb3.4-\spa1.5\spb3.5\bigr)
            \over\spa1.5\spa3.4\spa4.5\spb1.5\spb2.3^2} 
%%%%% end : O6b
\,.
\end{eqnarray}

Summing over the contributions to \eqn{BasicFormulaRhat}, 
with $\Inf A_5 = \Inf \Cuth_5 = 0$, we obtain,
\begin{eqnarray}
\Remaining_5 &=& D_5^{\rm (a)} + D_5^{\rm (b)} 
              + O_5^{\rm (a)} + O_5^{\rm (b)} \nn \\
& = &
%%%%% begin : DiagramSum5
 {i\over6} \Biggl(
 - {2 \spb2.4 \spb3.5^3\over \spa3.4 \spb1.2 \spb1.5 \spb2.3^2}
-{\spa1.2^2 \spa1.4 \spb3.4\over \spa1.5 \spa2.3 \spa3.4 \spa4.5 
    \spb2.3}
  \nn \\
&& \null \hskip .8 cm 
 + {\spa1.4^2 \spb3.4^2 \spb3.5\over\spa1.5 \spa3.4 \spa4.5 \spb1.5 \spb2.3^2}
 -{\spa1.4 \spb3.4 \spb3.5^2\over\spa3.4 \spa4.5 \spb1.5 
   \spb2.3^2}
\Biggr)  
%%%%% end : DiagramSum5
\,. \hskip 1.5 cm 
\end{eqnarray}
The full amplitude is the sum of $\hat{R}_5$ 
and the completed-cut terms $\Cuth_5$ from \eqn{CutCompletionmmppp}. 
The sum agrees with the result of ref.~\cite{GGGGG}. 

This procedure has been applied to generate a variety of amplitudes,
including the $n$-gluon amplitudes with two negative-helicity gluons, 
and the rest of positive helicity~\cite{Bootstrap,Genhel},
which are the one-loop analogs of the MHV tree amplitudes in \eqn{MHVtree}.

%%%%%%%%%%%%%%%%%%%%%%%%%%%%%%%%%%%%%%%%%%%%%%%%%%%%%%%
 
\section{Conclusions and Outlook}
\label{ConclusionsSection}

Next-to-leading order computations of QCD processes will play an
important role in understanding and interpreting results from the
forthcoming Large Hadron Collider.  The current state of the art, at
parton level, is for the production of three final-state objects,
requiring one-loop five-point amplitudes.  New approaches proposed for
one-loop amplitudes with six or more external legs fall into three
general categories: purely numerical~\cite{NSSubtract,LMPTriboson},
improved traditional (including semi-numerical)~\cite{Passarino,GRACEQCD,
DDReduction, GieleGloverNumerical,AguilaPittau,EGZH4p, BGHPS,
EGZSemiNum, EGZ6g,  XYZ, BinothRat, OPP}, and on-shell analytic
methods~\cite{Neq4Oneloop,Neq1Oneloop,DDimU, UnitarityMachinery,
NeqOneNMHVSixPt, BCFGeneralized, BBDPSQCD, BBCF, OnShellOneLoop, Qpap,
Bootstrap, Genhel,LoopMHV, BFM, ABFKM,MastroliaTriple,
BFMassive}.   The status of state-of-the-art calculations with these
methods has been summarized in the Introduction.  

This review has focused primarily on on-shell methods for loop
calculations.  An early version of on-shell methods has been used to
compute the one-loop amplitudes for $e^+e^- \rightarrow Z \rightarrow
4$ partons and (by crossing) for $pp \rightarrow W,Z$ + 2
jets~\cite{Zqqgg}.  The latter have been implemented in the MCFM
program~\cite{MCFM}. More recently, on-shell methods have been used to
obtain a variety of one-loop QCD amplitudes, both particular sequences
of helicity amplitudes containing an arbitrary number of external
gluons, as well as four of six helicity configurations for six
gluons~\cite{NeqOneNMHVSixPt, BBDPSQCD, BBCF, BFM, Genhel, LoopMHV}.
These computations make use of recent advances in the evaluation of
coefficients of loop integrals~\cite{BCFGeneralized,BBCF,BFM} within
the framework of the unitarity method~\cite{Neq4Oneloop,Neq1Oneloop},
as well as loop-level on-shell recursion for the rational
terms~\cite{OnShellOneLoop,Qpap,Bootstrap,Genhel}, based on the
BCFW~\cite{BCFRecursion,BCFW} tree-level on-shell recursion.  (The
cut-containing parts of the remaining two helicity configurations in
the six-gluon amplitude were also determined in this
way~\cite{BBCF,BFM}, while the rational parts were computed in
ref.~\cite{XYZ}.)  The analytical results agree numerically (where
they have been compared) with the semi-numerical results of
ref.~\cite{EGZ6g}.  Other loop-level developments include various
improvements~\cite{BMST,ABFKM} to the $D$-dimensional variant of the
the unitarity method~\cite{DDimU}.

The challenge now is to provide complete one-loop QCD results
for more complex processes,  for all possible partonic subprocesses 
and for all possible helicity configurations.  
Especially important phenomenologically are processes 
including electroweak vector bosons, heavy quarks and Higgs bosons 
in the final state.
The six-gluon computations have outlined effective analytical
techniques, but more automation will be required to handle all
the cases needed for complete NLO predictions.

On-shell methods offer the promise of a relatively modest
growth in complexity as the number of external legs increases,
due to the recursive structure.  The
unitarity method builds one-loop amplitudes directly from known
on-shell tree amplitudes.  In its four-dimensional variant, it
provides an efficient means for generating the coefficients of the 
basis integrals, while the $D$-dimensional variant can be used to
compute complete amplitudes.  Loop-level on-shell recursion
can be used to determine rational terms, 
using as input the cut-containing terms
and lower-point rational terms.  Together they yield complete amplitudes.

The one-loop multi-parton matrix elements are most often the only
missing ingredient needed to construct a numerical program for NLO
differential cross sections.  Writing such programs is a non-trivial
task, but general formalisms are available for doing
so~\cite{FKS,GGK,CS,CDST}. The implementation of the matrix elements in
these programs raises a number of practical issues, including the
speed of numerical evaluation and numerical stability.  One would
need operational NLO programs to study these issues more carefully,
but early indications are promising for the forms of the results
found using on-shell methods.  For example, speed comparisons
between the on-shell and semi-numerical approach for six-gluon
amplitudes~\cite{EGZ6g,LoopMHV} indicate a rather significant speed
advantage for the former method.  

Studies of numerical 
stability, due to round-off error near spurious singularities,
have not been performed for amplitudes recently constructed by
on-shell methods.  However, experience with the processes 
$pp \rightarrow W,Z\, +$ 2 jets~\cite{CampbellPrivate} and $e^+ e^-
\rightarrow 4$ jets~\cite{ZFourPartonsNLODS} indicates that these
methods produce analytic expressions for amplitudes~\cite{Zqqgg} 
with sufficiently mild singular behavior that numerical stability
is not an issue. As explained in ref.~\cite{CampbellGloverMiller}, the
widths of numerically unstable regions depend heavily on the powers to
which potentially-singular denominator factors are raised.
Singularities that cancel within the functions used in the cut
completion, such as the $\Ll_i$ functions, can easily be patched, using
Taylor expansions, in their numerical implementation.
The remaining spurious singularities appearing in the matrix elements
do not cause any difficulties~\cite{CampbellPrivate,ZFourPartonsNLODS}.
This evidence suggests that NLO programs using
higher-point amplitudes computed from on-shell methods
should also be free of significant round-off instabilities.
Of course, this suggestion will need to be confirmed with detailed studies.
Beyond the application to parton-level physics programs, it
will be important to interface new one-loop matrix elements to the new
generation of parton showering programs~\cite{MCNLO,NLOPartonShowers}.

The techniques described in this review can be applied directly to
processes involving quarks and external vector bosons.  Such processes
are of crucial importance for understanding backgrounds to new physics
such as supersymmetry or other extensions of the Standard Model.
Other processes of interest containing massive particles inside the
loop, such as top production, will require further development.
Recent progress on applying $D$-dimensional
unitarity~\cite{DDimU,BMST,ABFKM} to massive particles in the loops
may be found in ref.~\cite{BFMassive}.

The large number of subprocesses needed for new applications to
collider phenomenology makes it highly desirable to have an automated
program for evaluating the amplitudes.  The on-shell methods we have
discussed in this review are systematic and thus should lend
themselves to automation, although much more work is required
to substantiate this assertion.

Besides the remaining practical issues there are a number of open
theoretical issues. The analytic properties of loop amplitudes with
complex momenta are not as well understood as one might like.  In
particular, we have an incomplete understanding of the `unreal' poles
encountered in on-shell recursion at the loop level.  The best 
strategy for dealing with these poles at present is to 
set up recursion relations that avoid them.  
A first-principles understanding of properties of one-loop scattering 
amplitudes, under factorization with complex momenta,
and at large values of the complex shift parameter, would be very helpful.  
One possible avenue for investigating these properties
is the link between tree-level on-shell recursion relations 
and Feynman rules~\cite{VamanYao} in a variant of light-cone gauge called
space-cone gauge~\cite{SpaceCone}.
Unitarity in $D$ dimensions~\cite{DDimU} may also assist in this formal
understanding.

We anticipate that the on-shell methods described in this review will,
with further development, be widely applicable to the higher-multiplicity 
amplitudes required for next-to-leading order computations of 
phenomenological interest at the Large Hadron Collider.

\section*{Acknowledgments}

We thank Carola Berger and Darren Forde for collaboration on topics
presented in this review.  We also thank Bo Feng for helpful discussions. The
figures were generated using Jaxodraw~\cite{Jaxo}, based on
Axodraw~\cite{Axo}.

% \appendix

% \section{}
% \label{}


% notes:

% \bibitem{label2}
% If there is a note, it should come last:
% \bibitem{label3} \note
% \end{subbibitems}


\begin{thebibliography}{00}

% 117 distinct references found.
\bibitem{MADGRAPH}
T.~Stelzer and W.~F.~Long,
%``Automatic generation of tree level helicity amplitudes,''
Comput.\ Phys.\ Commun.\  {\bf 81}, 357 (1994)
[hep-ph/9401258].
%%CITATION = HEP-PH 9401258;%%

\bibitem{CompHEP}
A.~Pukhov {\it et al.},
%``CompHEP: A package for evaluation of Feynman diagrams and integration  over
%multi-particle phase space. User's manual for version 33,''
hep-ph/9908288.
%%CITATION = HEP-PH/9908288;%%

\bibitem{AMEGIC}
F.~Krauss, R.~Kuhn and G.~Soff,
%``AMEGIC++ 1.0: A matrix element generator in C++,''
JHEP {\bf 0202}, 044 (2002)
[hep-ph/0109036].
%%CITATION = JHEPA,0202,044;%%

\bibitem{ALPGEN}
M.~L.~Mangano, M.~Moretti, F.~Piccinini, R.~Pittau and A.~D.~Polosa,
%``ALPGEN, a generator for hard multiparton processes in hadronic
%collisions,''
JHEP {\bf 0307}, 001 (2003)
[hep-ph/0206293].
%%CITATION = JHEPA,0307,001;%%

\bibitem{HELAC}
A.~Kanaki and C.~G.~Papadopoulos,
%``HELAC: A package to compute electroweak helicity amplitudes,''
Comput.\ Phys.\ Commun.\  {\bf 132}, 306 (2000)
[hep-ph/0002082].
%%CITATION = CPHCB,132,306;%%

\bibitem{BGRecursion}
F.~A.~Berends and W.~T.~Giele,
%``Recursive Calculations For Processes With N Gluons,''
Nucl.\ Phys. B {\bf 306}, 759 (1988).
%%CITATION = NUPHA,B306,759;%%

\bibitem{LaterRecursive}
D.~A.~Kosower,
%``Light Cone Recurrence Relations For QCD Amplitudes,''
Nucl.\ Phys.\ B {\bf 335}, 23 (1990);\\
%%CITATION = NUPHA,B335,23;%%
%
F.~Caravaglios and M.~Moretti,
%``An algorithm to compute Born scattering amplitudes without Feynman graphs,''
Phys.\ Lett.\ B {\bf 358}, 332 (1995)
[hep-ph/9507237];\\
%%CITATION = HEP-PH 9507237;%%
%
P.~Draggiotis, R.~H.~P.~Kleiss and C.~G.~Papadopoulos,
%``On the computation of multigluon amplitudes,''
Phys.\ Lett.\ B {\bf 439}, 157 (1998)
[hep-ph/9807207];\\
%%CITATION = HEP-PH 9807207;%%
%
F.~Caravaglios, M.~L.~Mangano, M.~Moretti and R.~Pittau,
%``A new approach to multi-jet calculations in hadron collisions,''
Nucl.\ Phys.\ B {\bf 539}, 215 (1999)
[hep-ph/9807570].
%%CITATION = HEP-PH 9807570;%%

\bibitem{BCFRecursion}
R.~Britto, F.~Cachazo and B.~Feng,
%``New recursion relations for tree amplitudes of gluons,''
Nucl.\ Phys.\ B {\bf 715}, 499 (2005)
[hep-th/0412308].
%%CITATION = HEP-TH 0412308;%%

\bibitem{BCFW}
R.~Britto, F.~Cachazo, B.~Feng and E.~Witten,
%``Direct proof of tree-level recursion relation in Yang-Mills theory,''
Phys.\ Rev.\ Lett.\ {\bf 94}, 181602 (2005)
[hep-th/0501052].
%%CITATION = HEP-TH 0501052;%%

\bibitem{VECBOS}
F.~A.~Berends, W.~T.~Giele and H.~Kuijf,
%``Exact Expressions for Processes Involving a Vector Boson and Up to Five
%Partons,''
Nucl.\ Phys.\  B {\bf 321}, 39 (1989).
%%CITATION = NUPHA,B321,39;%%

\bibitem{PYTHIA}
H.~U.~Bengtsson and T.~Sj\"ostrand,
%``The Lund Monte Carlo for Hadronic Processes: Pythia Version 4.8,''
Comput.\ Phys.\ Commun.\  {\bf 46}, 43 (1987);\\
%%CITATION = CPHCB,46,43;%%
T.~Sj\"ostrand, P.~Eden, C.~Friberg, L.~L\"onnblad, G.~Miu, 
S.~Mrenna and E.~Norrbin,
%``High-energy-physics event generation with PYTHIA 6.1,''
Comput.\ Phys.\ Commun.\ {\bf 135}, 238 (2001)
[hep-ph/0010017].
%%CITATION = CPHCB,135,238;%%
% T.~Sj\"ostrand, L.~L\"onnblad, S.~Mrenna and P.~Skands,
%``PYTHIA 6.3: Physics and manual,''
% hep-ph/0308153.
%%CITATION = HEP-PH/0308153;%%

\bibitem{HERWIG}
G.~Marchesini and B.~R.~Webber,
%``HERWIG: A NEW MONTE CARLO EVENT GENERATOR FOR SIMULATING HADRON EMISSION
%REACTIONS WITH INTERFERING GLUONS,''
Cavendish-HEP-87/9;\\
%%CITATION = CAVENDISH-HEP-87/9;%%
G.~Marchesini, B.~R.~Webber, G.~Abbiendi, I.~G.~Knowles, M.~H.~Seymour 
and L.~Stanco,
%``HERWIG: A Monte Carlo event generator for simulating hadron emission
%reactions with interfering gluons. Version 5.1 - April 1991,''
Comput.\ Phys.\ Commun.\  {\bf 67}, 465 (1992);\\
%%CITATION = CPHCB,67,465;%%
G.~Corcella {\it et al.},
%``HERWIG 6.5 release note,''
hep-ph/0210213.
%%CITATION = HEP-PH/0210213;%%

\bibitem{Matching}
S.~Catani, F.~Krauss, R.~Kuhn and B.~R.~Webber,
%``QCD matrix elements + parton showers,''
JHEP {\bf 0111}, 063 (2001)
[hep-ph/0109231];\\
%%CITATION = JHEPA,0111,063;%%
S.~Mrenna and P.~Richardson,
%``Matching matrix elements and parton showers with HERWIG and PYTHIA,''
JHEP {\bf 0405}, 040 (2004)
[hep-ph/0312274];\\
%%CITATION = JHEPA,0405,040;%%
M.~Mangano, 
%``The so-called MLM prescription for ME/PS matching'' (2004),
presented at the Fermilab ME/MC Tuning Workshop, October 4,
2004;\\
S.~H\"oche, F.~Krauss, N.~Lavesson, L.~L\"onnblad, M.~Mangano, 
A.~Sch\"alicke and S.~Schumann,
%``Matching parton showers and matrix elements,''
hep-ph/0602031;\\
%%CITATION = HEP-PH 0602031;%%
M.~L.~Mangano, M.~Moretti, F.~Piccinini and M.~Treccani,
%``Matching matrix elements and shower evolution for top-quark production in
%hadronic collisions,''
JHEP {\bf 0701}, 013 (2007)
[hep-ph/0611129].
%%CITATION = JHEPA,0701,013;%%

\bibitem{NNLOHiggs}
S.~Catani, D.~de Florian and M.~Grazzini,
%``Higgs production in hadron collisions: Soft and virtual QCD 
%corrections at NNLO,''
JHEP {\bf 0105}, 025 (2001)
[hep-ph/0102227];\\
%%CITATION = JHEPA,0105,025;%%
R.~V.~Harlander and W.~B.~Kilgore,
%``Next-to-next-to-leading order Higgs production at hadron colliders,''
Phys.\ Rev.\ Lett.\  {\bf 88}, 201801 (2002)
[hep-ph/0201206];\\
%%CITATION = HEP-PH 0201206;%%
C.~Anastasiou and K.~Melnikov,
%``Higgs boson production at hadron colliders in NNLO QCD,''
Nucl.\ Phys.\  B {\bf 646}, 220 (2002)
[hep-ph/0207004];\\
%%CITATION = NUPHA,B646,220;%%
V.~Ravindran, J.~Smith and W.~L.~van Neerven,
%``NNLO corrections to the total cross section for Higgs boson production  in
%hadron hadron collisions,''
Nucl.\ Phys.\  B {\bf 665}, 325 (2003)
[hep-ph/0302135].
%%CITATION = NUPHA,B665,325;%%

\bibitem{MCNLO}
S.~Frixione and B.R.~Webber,
%``Matching NLO QCD computations and parton shower simulations,''
JHEP {\bf 0206}, 029 (2002)
[hep-ph/0204244]; \\
%%CITATION = HEP-PH 0204244;%%
S.~Frixione, P.~Nason and B.R.~Webber,
%``Matching NLO QCD and parton showers in heavy flavour production,''
JHEP {\bf 0308}, 007 (2003)
[hep-ph/0305252].
%%CITATION = HEP-PH 0305252;%%

\bibitem{NLOPartonShowers}
P.~Nason,
%``A new method for combining NLO QCD with shower Monte Carlo algorithms,''
JHEP {\bf 0411}, 040 (2004)
[hep-ph/0409146];\\
%%CITATION = JHEPA,0411,040;%%
%
Z.~Nagy and D.~E.~Soper,
%``Matching parton showers to NLO computations,''
JHEP {\bf 0510}, 024 (2005)
[hep-ph/0503053];\\
%%CITATION = JHEPA,0510,024;%%
%
M.~Kr\"amer, S.~Mrenna and D.~E.~Soper,
%``Next-to-leading order QCD jet production with parton showers and
%hadronization,''
Phys.\ Rev.\  D {\bf 73}, 014022 (2006)
[hep-ph/0509127];\\
%%CITATION = PHRVA,D73,014022;%%
P.~Nason and G.~Ridolfi,
%``A positive-weight next-to-leading-order Monte Carlo for Z pair
%hadroproduction,''
JHEP {\bf 0608}, 077 (2006)
[hep-ph/0606275];\\
%%CITATION = JHEPA,0608,077;%%
O.~Latunde-Dada, S.~Gieseke and B.~Webber,
%``A positive-weight next-to-leading-order Monte Carlo for e+ e- annihilation
%to hadrons,''
JHEP {\bf 0702}, 051 (2007)
[hep-ph/0612281].
%%CITATION = JHEPA,0702,051;%%

\bibitem{FKS}
Z.~Kunszt and D.~E.~Soper,
%``Calculation of jet cross-sections in hadron collisions at order
%alpha-s**3,''
Phys.\ Rev.\  D {\bf 46}, 192 (1992);\\
%%CITATION = PHRVA,D46,192;%%
%
S.~Frixione, Z.~Kunszt and A.~Signer,
%``Three-jet cross sections to next-to-leading order,''
Nucl.\ Phys.\  B {\bf 467}, 399 (1996)
[hep-ph/9512328].
%%CITATION = NUPHA,B467,399;%%

\bibitem{GGK}
W.~T.~Giele and E.~W.~N.~Glover,
%``Higher Order Corrections To Jet Cross-Sections In E+ E- Annihilation,''
Phys.\ Rev.\  D {\bf 46}, 1980 (1992);\\
%%CITATION = PHRVA,D46,1980;%%
%
W.~T.~Giele, E.~W.~N.~Glover and D.~A.~Kosower,
%``Higher Order Corrections To Jet Cross-Sections In Hadron Colliders,''
Nucl.\ Phys.\  B {\bf 403}, 633 (1993)
[hep-ph/9302225].
%%CITATION = NUPHA,B403,633;%%

\bibitem{CS}
S.~Catani and M.~H.~Seymour,
%``The Dipole Formalism for the Calculation of QCD Jet Cross Sections at
%Next-to-Leading Order,''
Phys.\ Lett.\  B {\bf 378}, 287 (1996)
[hep-ph/9602277];
%%CITATION = PHLTA,B378,287;%%
%
%``A general algorithm for calculating jet cross sections in NLO QCD,''
Nucl.\ Phys.\  B {\bf 485}, 291 (1997)
[Erratum-ibid.\  B {\bf 510}, 503 (1998)]
[hep-ph/9605323].
%%CITATION = NUPHA,B485,291;%%

\bibitem{CDST}
S.~Catani, S.~Dittmaier, M.~H.~Seymour and Z.~Tr\'ocs\'anyi,
%``The dipole formalism for next-to-leading order QCD calculations with
%massive partons,''
Nucl.\ Phys.\  B {\bf 627}, 189 (2002)
[hep-ph/0201036].
%%CITATION = NUPHA,B627,189;%%

\bibitem{Antenna}
D.~A.~Kosower,
%``Antenna factorization of gauge-theory amplitudes,''
Phys.\ Rev.\  D {\bf 57}, 5410 (1998)
[hep-ph/9710213];\\
%%CITATION = PHRVA,D57,5410;%%
J.~M.~Campbell, M.~A.~Cullen and E.~W.~N.~Glover,
%``Four jet event shapes in electron positron annihilation,''
Eur.\ Phys.\ J.\  C {\bf 9}, 245 (1999)
[hep-ph/9809429].
%%CITATION = EPHJA,C9,245;%%

\bibitem{Gehrmann}
A.~Gehrmann-De Ridder, T.~Gehrmann and E.~W.~N.~Glover,
%``Antenna subtraction at NNLO,''
JHEP {\bf 0509}, 056 (2005)
[hep-ph/0505111].
%%CITATION = JHEPA,0509,056;%%

\bibitem{ttH}
W.~Beenakker, S.~Dittmaier, M.~Kr\"amer, B.~Pl\"umper, M.~Spira
and P.~M.~Zerwas,
%``Higgs radiation off top quarks at the Tevatron and the LHC,''
Phys.\ Rev.\ Lett.\  {\bf 87}, 201805 (2001)
[hep-ph/0107081];
%%CITATION = PRLTA,87,201805;%%
%W.~Beenakker, S.~Dittmaier, M.~Kr\"amer, B.~Pl\"umper, 
%M.~Spira and P.~M.~Zerwas,
%``NLO QCD corrections to t anti-t H production in hadron collisions. ((U)),''
Nucl.\ Phys.\  B {\bf 653}, 151 (2003)
[hep-ph/0211352];\\
%%CITATION = NUPHA,B653,151;%%
S.~Dawson, L.~H.~Orr, L.~Reina and D.~Wackeroth,
%``Associated top quark Higgs boson production at the LHC,''
Phys.\ Rev.\  D {\bf 67}, 071503 (2003)
[hep-ph/0211438];\\
%%CITATION = PHRVA,D67,071503;%%
S.~Dawson, C.~Jackson, L.~H.~Orr, L.~Reina and D.~Wackeroth,
%``Associated Higgs production with top quarks at the Large Hadron  Collider:
%NLO QCD corrections,''
Phys.\ Rev.\  D {\bf 68}, 034022 (2003)
[hep-ph/0305087].
%%CITATION = PHRVA,D68,034022;%%

\bibitem{NLOThreeJet}
W.~B.~Kilgore and W.~T.~Giele,
%``Next-to-leading order gluonic three jet production at hadron colliders,''
Phys.\ Rev.\  D {\bf 55}, 7183 (1997)
[hep-ph/9610433].
%%CITATION = PHRVA,D55,7183;%%
%
Z.~Nagy,
%``Three-jet cross sections in hadron hadron collisions at next-to-leading
%order,''
Phys.\ Rev.\ Lett.\  {\bf 88}, 122003 (2002)
[hep-ph/0110315];
%%CITATION = PRLTA,88,122003;%%
%
%Z.~Nagy,
%``Next-to-leading order calculation of three-jet observables in hadron
%hadron collision,''
Phys.\ Rev.\  D {\bf 68}, 094002 (2003)
[hep-ph/0307268].
%%CITATION = PHRVA,D68,094002;%%

\bibitem{MCFM}
J.~Campbell and R.~K.~Ellis,
%``Next-to-leading order corrections to W + 2 jet and Z + 2 jet production  at
% hadron colliders,''
Phys.\ Rev.\ D {\bf 65}, 113007 (2002)
[hep-ph/0202176];\\
%%CITATION = HEP-PH 0202176;%%
%
J.~Campbell, R.~K.~Ellis and D.~L.~Rainwater,
%``Next-to-leading order QCD predictions for W + 2jet and Z + 2jet  production
%at the CERN LHC,''
Phys.\ Rev.\ D {\bf 68}, 094021 (2003)
[hep-ph/0308195].
%%CITATION = HEP-PH 0308195;%%

\bibitem{HbbWbb}
S.~Dawson, C.~B.~Jackson, L.~Reina and D.~Wackeroth,
%``Higgs production in association with bottom quarks at hadron colliders,''
Mod.\ Phys.\ Lett.\  A {\bf 21}, 89 (2006)
[hep-ph/0508293];\\
%%CITATION = MPLAE,A21,89;%%
%
F.~Febres Cordero, L.~Reina and D.~Wackeroth,
%``NLO QCD corrections to W boson production with a massive b-quark jet pair
%at the Tevatron p anti-p collider,''
Phys.\ Rev.\  D {\bf 74}, 034007 (2006)
[hep-ph/0606102].
%%CITATION = PHRVA,D74,034007;%%

\bibitem{EGZHjj}
J.~M.~Campbell, R.~K.~Ellis and G.~Zanderighi,
%``Next-to-leading order Higgs + 2 jet production via gluon fusion,''
JHEP {\bf 0610}, 028 (2006)
[hep-ph/0608194].
%%CITATION = JHEPA,0610,028;%%

\bibitem{NLOttjet}
A.~Brandenburg, S.~Dittmaier, P.~Uwer and S.~Weinzierl,
%``Top quark pair + jet production at next-to-leading order: NLO QCD
%corrections to g g -> t t-bar g,''
Nucl.\ Phys.\ Proc.\ Suppl.\  {\bf 135}, 71 (2004)
[hep-ph/0408137];\\
%%CITATION = NUPHZ,135,71;%%
S.~Dittmaier, P.~Uwer and S.~Weinzierl,
%``NLO QCD corrections to t anti-t + jet production at hadron colliders,''
hep-ph/0703120.
%%CITATION = HEP-PH/0703120;%%

\bibitem{LMPTriboson}
A.~Lazopoulos, K.~Melnikov and F.~Petriello,
%``QCD corrections to tri-boson production,''
hep-ph/0703273.
%%CITATION = HEP-PH/0703273;%%

\bibitem{Davydychev}
A.~I.~Davydychev,
%``A Simple formula for reducing Feynman diagrams to scalar integrals,''
Phys.\ Lett.\  B {\bf 263}, 107 (1991).
%%CITATION = PHLTA,B263,107;%%

\bibitem{Passarino}
A.~Ferroglia, M.~Passera, G.~Passarino and S.~Uccirati,
%``All-purpose numerical evaluation of one-loop multi-leg Feynman diagrams,''
Nucl.\ Phys.\  B {\bf 650}, 162 (2003)
[hep-ph/0209219].
%%CITATION = NUPHA,B650,162;%%

\bibitem{GRACEQCD}
Y.~Kurihara {\it et al.},
%``QCD event generators with next-to-leading order matrix-elements and parton
%showers,''
Nucl.\ Phys.\  B {\bf 654}, 301 (2003)
[hep-ph/0212216];
%%CITATION = NUPHA,B654,301;%%
%Y.~Kurihara {\it et al.},
%``NLO-QCD calculation in GRACE,''
Nucl.\ Phys.\ Proc.\ Suppl.\  {\bf 157}, 231 (2006).
%%CITATION = NUPHZ,157,231;%%

\bibitem{DDReduction}
A.~Denner and S.~Dittmaier,
%``Reduction of one-loop tensor 5-point integrals,''
Nucl.\ Phys.\  B {\bf 658}, 175 (2003)
[hep-ph/0212259];
%%CITATION = NUPHA,B658,175;%%
%A.~Denner and S.~Dittmaier,
%``Reduction schemes for one-loop tensor integrals,''
Nucl.\ Phys.\  B {\bf 734}, 62 (2006)
[hep-ph/0509141].
%%CITATION = NUPHA,B734,62;%%

\bibitem{GieleGloverNumerical}
W.~T.~Giele and E.~W.~N.~Glover,
%``A calculational formalism for one-loop integrals,''
JHEP {\bf 0404}, 029 (2004)
[hep-ph/0402152].
%%CITATION = JHEPA,0404,029;%%

\bibitem{AguilaPittau}
F.~del Aguila and R.~Pittau,
%``Recursive numerical calculus of one-loop tensor integrals,''
JHEP {\bf 0407}, 017 (2004)
[hep-ph/0404120].
%%CITATION = JHEPA,0407,017;%%

\bibitem{EGZH4p}
R.~K.~Ellis, W.~T.~Giele and G.~Zanderighi,
%``Virtual QCD corrections to Higgs boson plus four parton processes,''
Phys.\ Rev.\  D {\bf 72}, 054018 (2005)
[Erratum-ibid.\  D {\bf 74}, 079902 (2006)]
[hep-ph/0506196].
%%CITATION = PHRVA,D72,054018;%%

\bibitem{BGHPS}
T.~Binoth, J.~P.~Guillet, G.~Heinrich, E.~Pilon and C.~Schubert,
%``An algebraic / numerical formalism for one-loop multi-leg amplitudes,''
JHEP {\bf 0510}, 015 (2005)
[hep-ph/0504267].
%%CITATION = JHEPA,0510,015;%%

\bibitem{EGZSemiNum}
R.~K.~Ellis, W.~T.~Giele and G.~Zanderighi,
%``Semi-numerical evaluation of one-loop corrections,''
Phys.\ Rev.\  D {\bf 73}, 014027 (2006)
[hep-ph/0508308].
%%CITATION = PHRVA,D73,014027;%%

\bibitem{EGZ6g}
R.~K.~Ellis, W.~T.~Giele and G.~Zanderighi,
%``The one-loop amplitude for six-gluon scattering,''
JHEP {\bf 0605}, 027 (2006)
[hep-ph/0602185].
%%CITATION = JHEPA,0605,027;%%

\bibitem{XYZ}
Z.~Xiao, G.~Yang and C.~J.~Zhu,
%``The rational part of QCD amplitude. III: The six-gluon,''
Nucl.\ Phys.\  B {\bf 758}, 53 (2006)
[hep-ph/0607017].
%%CITATION = NUPHA,B758,53;%%

\bibitem{BinothRat}
T.~Binoth, J.~P.~Guillet and G.~Heinrich,
%``Algebraic evaluation of rational polynomials in one-loop amplitudes,''
JHEP {\bf 0702}, 013 (2007)
[hep-ph/0609054].
%%CITATION = JHEPA,0702,013;%%

\bibitem{OPP}
G.~Ossola, C.~G.~Papadopoulos and R.~Pittau,
%``Reducing full one-loop amplitudes to scalar integrals at the integrand
%level,''
Nucl.\ Phys.\  B {\bf 763}, 147 (2007)
hep-ph/0609007].
%%CITATION = NUPHA,B763,147;%%

\bibitem{NSSubtract}
M.~Kr\"amer and D.~E.~Soper,
%``Next-to-leading order numerical calculations in Coulomb gauge,''
Phys.\ Rev.\  D {\bf 66}, 054017 (2002)
[hep-ph/0204113];\\
%%CITATION = PHRVA,D66,054017;%%
%
Z.~Nagy and D.~E.~Soper,
%``General subtraction method for numerical calculation of one-loop QCD
%matrix elements,''
JHEP {\bf 0309}, 055 (2003)
[hep-ph/0308127].
%%CITATION = JHEPA,0309,055;%%

\bibitem{Neq4Oneloop}
Z.~Bern, L.~J.~Dixon, D.~C.~Dunbar and D.~A.~Kosower,
%``One-loop $n$-point gauge theory amplitudes, unitarity
% and collinear limits,''
Nucl.\ Phys.\ B {\bf 425}, 217 (1994)
[hep-ph/9403226].
%%CITATION = HEP-PH 9403226;%%

\bibitem{Neq1Oneloop}
Z.~Bern, L.~J.~Dixon, D.~C.~Dunbar and D.~A.~Kosower,
%``Fusing gauge theory tree amplitudes into loop amplitudes,''
Nucl.\ Phys.\ B {\bf 435}, 59 (1995)
[hep-ph/9409265].
%%CITATION = HEP-PH 9409265;%%

\bibitem{DDimU}
Z.\ Bern and A.\ G.\ Morgan,
%``Massive Loop Amplitudes from Unitarity,''
Nucl.\ Phys.\ B {\bf 467}, 479 (1996)
[hep-ph/9511336];\\
%%CITATION = HEP-PH 9511336;%%
%
Z.~Bern, L.~J.~Dixon, D.~C.~Dunbar and D.~A.~Kosower,
%``One-loop self-dual and N = 4 superYang-Mills,''
Phys.\ Lett.\ B {\bf 394}, 105 (1997)
[hep-th/9611127].
%%CITATION = HEP-TH 9611127;%%

\bibitem{UnitarityMachinery}
Z.\ Bern, L.\ J.\ Dixon and D.\ A.\ Kosower,
%``Unitarity-based techniques for one-loop calculations in QCD,''
Nucl.\ Phys.\ Proc.\ Suppl.\  {\bf 51C}, 243 (1996)
[hep-ph/9606378];
%%CITATION = HEP-PH 9606378;%%
%
% Z.~Bern, L.~J.~Dixon and D.~A.~Kosower,
%``A two-loop four-gluon helicity amplitude in QCD,''
JHEP {\bf 0001}, 027 (2000)
[hep-ph/0001001];\\
%%CITATION = HEP-PH 0001001;%%
%
Z.~Bern, A.~De Freitas and L.~J.~Dixon,
%``Two-loop helicity amplitudes for gluon gluon scattering in QCD and
%supersymmetric Yang-Mills theory,''
JHEP {\bf 0203}, 018 (2002)
[hep-ph/0201161].
%%CITATION = HEP-PH 0201161;%%

\bibitem{NeqOneNMHVSixPt}
S.~J.~Bidder, N.~E.~J.~Bjerrum-Bohr, L.~J.~Dixon and D.~C.~Dunbar,
%``N = 1 supersymmetric one-loop amplitudes and the holomorphic anomaly of
%unitarity cuts,''
Phys.\ Lett.\ B {\bf 606}, 189 (2005)
[hep-th/0410296].
%%CITATION = HEP-TH 0410296;%%

\bibitem{BCFGeneralized}
R.~Britto, F.~Cachazo and B.~Feng,
%``Generalized unitarity and one-loop amplitudes in N = 4  super-Yang-Mills,''
Nucl.\ Phys.\  B {\bf 725}, 275 (2005)
[hep-th/0412103].
%%CITATION = NUPHA,B725,275;%%

\bibitem{BBDPSQCD}
S.~J.~Bidder, N.~E.~J.~Bjerrum-Bohr, D.~C.~Dunbar and W.~B.~Perkins,
%``One-loop gluon scattering amplitudes in theories with N < 4
%supersymmetries,''
Phys.\ Lett.\ B {\bf 612}, 75 (2005)
[hep-th/0502028].
%%CITATION = HEP-TH 0502028;%%

\bibitem{BBCF}
R.~Britto, E.~Buchbinder, F.~Cachazo and B.~Feng,
%``One-loop amplitudes of gluons in SQCD,''
Phys.\ Rev.\  D {\bf 72}, 065012 (2005)
[hep-ph/0503132].
%%CITATION = PHRVA,D72,065012;%%

\bibitem{OnShellOneLoop}
Z.~Bern, L.~J.~Dixon and D.~A.~Kosower,
%``On-shell recurrence relations for one-loop QCD amplitudes,''
Phys.\ Rev.\  D {\bf 71}, 105013 (2005)
[hep-th/0501240].
%%CITATION = PHRVA,D71,105013;%%

\bibitem{Qpap}
Z.~Bern, L.~J.~Dixon and D.~A.~Kosower,
%``The last of the finite loop amplitudes in QCD,''
Phys.\ Rev.\  D {\bf 72}, 125003 (2005)
[hep-ph/0505055].
%%CITATION = PHRVA,D72,125003;%%

\bibitem{Bootstrap}
Z.~Bern, L.~J.~Dixon and D.~A.~Kosower,
%``Bootstrapping multi-parton loop amplitudes in QCD,''
Phys.\ Rev.\  D {\bf 73}, 065013 (2006)
[hep-ph/0507005];\\
%%CITATION = PHRVA,D73,065013;%%
%
D.~Forde and D.~A.~Kosower,
%``All-multiplicity one-loop corrections to MHV amplitudes in QCD,''
Phys.\ Rev.\  D {\bf 73}, 061701 (2006)
[hep-ph/0509358].
%%CITATION = PHRVA,D73,061701;%%

\bibitem{Genhel}
C.~F.~Berger, Z.~Bern, L.~J.~Dixon, D.~Forde and D.~A.~Kosower,
%``Bootstrapping one-loop QCD amplitudes with general helicities,''
Phys.\ Rev.\ D {\bf 74}, 036009 (2006)
[hep-ph/0604195].
%%CITATION = HEP-PH 0604195;%%

\bibitem{LoopMHV}
C.~F.~Berger, Z.~Bern, L.~J.~Dixon, D.~Forde and D.~A.~Kosower,
%``All one-loop maximally helicity violating gluonic amplitudes in QCD,''
Phys.\ Rev.\  D {\bf 75}, 016006 (2007)
[hep-ph/0607014].
%%CITATION = PHRVA,D75,016006;%%

\bibitem{BFM}
R.~Britto, B.~Feng and P.~Mastrolia,
%``The cut-constructible part of QCD amplitudes,''
Phys.\ Rev.\  D {\bf 73}, 105004 (2006)
[hep-ph/0602178].
%%CITATION = PHRVA,D73,105004;%%

\bibitem{ABFKM}
C.~Anastasiou, R.~Britto, B.~Feng, Z.~Kunszt and P.~Mastrolia,
%``D-dimensional unitarity cut method,''
Phys.\ Lett.\  B {\bf 645}, 213 (2007)
[hep-ph/0609191];
%
%%CITATION = PHLTA,B645,213;%%
%C.~Anastasiou, R.~Britto, B.~Feng, Z.~Kunszt and P.~Mastrolia,
%``Unitarity cuts and reduction to master integrals in d dimensions for
%one-loop amplitudes,''
JHEP {\bf 0703}, 111 (2007)
[hep-ph/0612277].
%%CITATION = JHEPA,0703,111;%%

\bibitem{MastroliaTriple}
P.~Mastrolia,
%``On triple-cut of scattering amplitudes,''
Phys.\ Lett.\  B {\bf 644}, 272 (2007)
[hep-th/0611091].
%%CITATION = PHLTA,B644,272;%%

\bibitem{BFMassive}
R.~Britto and B.~Feng,
%``Unitarity cuts with massive propagators and algebraic expressions for
%coefficients,''
hep-ph/0612089.
%%CITATION = HEP-PH/0612089;%%

\bibitem{NS6ph}
Z.~Nagy and D.~E.~Soper,
%``Numerical integration of one-loop Feynman diagrams for N-photon
%amplitudes,''
Phys.\ Rev.\  D {\bf 74}, 093006 (2006)
[hep-ph/0610028].
%%CITATION = PHRVA,D74,093006;%%

\bibitem{BinothPhotons}
T.~Binoth, T.~Gehrmann, G.~Heinrich and P.~Mastrolia,
%``Six-Photon Amplitudes,''
hep-ph/0703311.
%%CITATION = HEP-PH/0703311;%%

\bibitem{OPPPhotons}
G.~Ossola, C.~G.~Papadopoulos and R.~Pittau,
%``Numerical Evaluation of Six-Photon Amplitudes,''
0704.1271 [hep-ph].
%%CITATION = ARXIV:0704.1271;%%

\bibitem{Zqqgg}
Z.~Bern, L.~J.~Dixon and D.~A.~Kosower,
%``One-loop amplitudes for e+ e- to four partons,''
Nucl.\ Phys.\  B {\bf 513}, 3 (1998)
[hep-ph/9708239].
%%CITATION = NUPHA,B513,3;%%

\bibitem{OldUnitarity}
L.~D.~Landau,
%``On analytic properties of vertex parts in quantum field theory,''
Nucl.\ Phys.\  {\bf 13}, 181 (1959);\\
%%CITATION = NUPHA,13,181;%%
S.~Mandelstam,
%``Determination Of The Pion - Nucleon Scattering Amplitude From Dispersion
%Relations And Unitarity. General Theory,''
Phys.\ Rev.\  {\bf 112}, 1344 (1958);
%%CITATION = PHRVA,112,1344;%%
%S.~Mandelstam,
%``Analytic Properties Of Transition Amplitudes In Perturbation Theory,''
Phys.\ Rev.\  {\bf 115}, 1741 (1959).
%%CITATION = PHRVA,115,1741;%%

\bibitem{VanNeervenUnitarity}
W.~L.~van Neerven,
%``Dimensional Regularization Of Mass And Infrared Singularities In Two Loop
%On-Shell Vertex Functions,''
Nucl.\ Phys.\  B {\bf 268}, 453 (1986).
%%CITATION = NUPHA,B268,453;%%

\bibitem{TwoLoopSplit}
Z.~Bern, L.~J.~Dixon and D.~A.~Kosower,
%``Two-loop g $\to$ g g splitting amplitudes in QCD,''
JHEP {\bf 0408}, 012 (2004)
[hep-ph/0404293].
%%CITATION = HEP-PH 0404293;%%

\bibitem{Neq47pt}
Z.~Bern, V.~Del Duca, L.~J.~Dixon and D.~A.~Kosower,
%``All non-maximally-helicity-violating one-loop seven-gluon amplitudes in  N
%= 4 super-Yang-Mills theory,''
Phys.\ Rev.\  D {\bf 71}, 045006 (2005)
[hep-th/0410224].
%%CITATION = PHRVA,D71,045006;%%

\bibitem{FordeTriBub}
D.~Forde,
%``Direct extraction of one-loop integral coefficients,''
0704.1835 [hep-ph].
%%CITATION = ARXIV:0704.1835;%%

\bibitem{BMST}
A.~Brandhuber, S.~McNamara, B.~J.~Spence and G.~Travaglini,
%``Loop amplitudes in pure Yang-Mills from generalised unitarity,''
JHEP {\bf 0510}, 011 (2005)
[hep-th/0506068].
%%CITATION = JHEPA,0510,011;%%

\bibitem{OneLoopReview}
Z.\ Bern, L.\ J.\ Dixon and D.\ A.\ Kosower,
%``Progress in one-loop QCD computations,''
Ann.\ Rev.\ Nucl.\ Part.\ Sci.\  {\bf 46}, 109 (1996)
[hep-ph/9602280].
%%CITATION = HEP-PH 9602280;%%

\bibitem{BSTZ}
A.~Brandhuber, B.~Spence, G.~Travaglini and K.~Zoubos,
%``One-loop MHV Rules and Pure Yang-Mills,''
0704.0245 [hep-th].
%%CITATION = ARXIV:0704.0245;%%

\bibitem{CSLectures}
F.~Cachazo and P.~Svr\v{c}ek,
%``Lectures on twistor strings and perturbative Yang-Mills theory,''
PoS {\bf RTN2005}, 004 (2005)
[hep-th/0504194].
%%CITATION = POSCI,RTN2005,004;%%

\bibitem{TwistorReview}
L.~J.~Dixon,
%``Twistor string theory and QCD,''
PoS {\bf HEP2005}, 405 (2006)
[hep-ph/0512111].
%%CITATION = POSCI,HEP2005,405;%%

\bibitem{SW}
C.~Schwinn and S.~Weinzierl,
%``On-shell recursion relations for all Born QCD amplitudes,''
JHEP {\bf 0704}, 072 (2007)
[hep-ph/0703021].
%%CITATION = JHEPA,0704,072;%%

\bibitem{TheoreticalAdvances}
Z.~Bern, M.~Czakon, L.~J.~Dixon, D.~A.~Kosower and V.~A.~Smirnov,
%``The four-loop planar amplitude and cusp anomalous dimension in maximally
%supersymmetric Yang-Mills theory,''
Phys.\ Rev.\  D {\bf 75}, 085010 (2007)
[hep-th/0610248];\\
%%CITATION = PHRVA,D75,085010;%%
%
Z.~Bern, L.~J.~Dixon and R.~Roiban,
%``Is N = 8 supergravity ultraviolet finite?,''
Phys.\ Lett.\  B {\bf 644}, 265 (2007)
[hep-th/0611086];\\
%%CITATION = PHLTA,B644,265;%%
%
Z.~Bern, J.~J.~Carrasco, L.~J.~Dixon, H.~Johansson, D.~A.~Kosower 
and R.~Roiban,
%``Three-loop superfiniteness of N = 8 supergravity,''
Phys.\ Rev.\ Lett.\  {\bf 98}, 161303 (2007)
[hep-th/0702112].
%%CITATION = HEP-TH/0702112;%%

\bibitem{SpinorHelicity}
F.~A.~Berends, R.~Kleiss, P.~De Causmaecker, R.~Gastmans and T.~T.~Wu,
%``Single Bremsstrahlung Processes In Gauge Theories,''
Phys.\ Lett.\ B {\bf 103}, 124 (1981);\\
%%CITATION = PHLTA,B103,124;%%
%
P.~De Causmaecker, R.~Gastmans, W.~Troost and T.~T.~Wu,
%``Multiple Bremsstrahlung In Gauge Theories At High-Energies. 1. General
%Formalism For Quantum Electrodynamics,''
Nucl.\ Phys.\ B {\bf 206}, 53 (1982);\\
%%CITATION = NUPHA,B206,53;%%
%
Z.~Xu, D.~H.~Zhang and L.~Chang,
%``Helicity Amplitudes For Multiple Bremsstrahlung In Massless Nonabelian
% Gauge Theory. 1. New Definition Of Polarization Vector And Formulation Of
% Amplitudes In Grassmann Algebra,''
TUTP-84/3-TSINGHUA;\\
%\href{http://www.slac.stanford.edu/spires/find/hep/www?r=tutp-84\
%2F3-tsinghua}{SPIRES entry}
%
R.~Kleiss and W.~J.~Stirling,
%``Spinor Techniques For Calculating P Anti-P $\to$ W+- / Z0 + Jets,''
Nucl.\ Phys.\ B {\bf 262}, 235 (1985);\\
%%CITATION = NUPHA,B262,235;%%
%
J.~F.~Gunion and Z.~Kunszt,
% ``Improved Analytic Techniques For Tree Graph Calculations And The G G Q
% Anti-Q Lepton Anti-Lepton Subprocess,''
Phys.\ Lett.\ B {\bf 161}, 333 (1985);\\
%%CITATION = PHLTA,B161,333;%%
%
Z.~Xu, D.~H.~Zhang and L.~Chang,
%``Helicity Amplitudes For Multiple Bremsstrahlung In Massless Nonabelian
%Gauge Theories,''
Nucl.\ Phys.\ B {\bf 291}, 392 (1987).
%%CITATION = NUPHA,B291,392;%%

\bibitem{TreeReview}
M.~L.~Mangano and S.~J.~Parke,
%``Multiparton Amplitudes In Gauge Theories,''
Phys.\ Rept.\ {\bf 200}, 301 (1991);\\
%%CITATION = PRPLC,200,301;%%
%
L.~J.~Dixon,
%``Calculating scattering amplitudes efficiently,''
in {\it QCD \& Beyond: Proceedings of TASI '95},
ed. D.\ E.\ Soper (World Scientific, 1996)
[hep-ph/9601359].
%%CITATION = HEP-PH 9601359;%%

\bibitem{TreeColor}
J.~E.~Paton and H.~M.~Chan,
%``Generalized Veneziano Model With Isospin,''
Nucl.\ Phys.\ B {\bf 10}, 516 (1969);\\
%%CITATION = NUPHA,B10,516;%%
%
P.~Cvitanovi\'c, P.~G.~Lauwers and P.~N.~Scharbach,
%``Gauge Invariance Structure Of Quantum Chromodynamics,''
Nucl.\ Phys.\ B {\bf 186}, 165 (1981);\\
%%CITATION = NUPHA,B186,165;%%
%
D.~Kosower, B.~H.~Lee and V.~P.~Nair,
%``Multi Gluon Scattering: A String Based Calculation,''
Phys.\ Lett.\ B {\bf 201}, 85 (1988).
%%CITATION = PHLTA,B201,85;%%

\bibitem{BKColor}
Z.~Bern and D.~A.~Kosower,
%``Color Decomposition Of One Loop Amplitudes In Gauge Theories,''
Nucl.\ Phys.\ B {\bf 362}, 389 (1991).
%%CITATION = NUPHA,B362,389;%%

\bibitem{TwoQuarkThreeGluon}
Z.~Bern, L.~J.~Dixon and D.~A.~Kosower,
%``One Loop Corrections To Two Quark Three Gluon Amplitudes,''
Nucl.\ Phys.\  B {\bf 437}, 259 (1995)
[hep-ph/9409393].
%%CITATION = NUPHA,B437,259;%%

\bibitem{WittenTopologicalString}
E.~Witten,
%``Perturbative gauge theory as a string theory in twistor space,''
Commun.\ Math.\ Phys.\  {\bf 252}, 189 (2004)
[hep-th/0312171].
%%CITATION = HEP-TH 0312171;%%

\bibitem{Penrose}
R.~Penrose,
%``Twistor Algebra,''
J.\ Math.\ Phys.\  {\bf 8}, 345 (1967).
%%CITATION = JMAPA,8,345;%%

\bibitem{GoroffSagnotti}
M.~H.~Goroff and A.~Sagnotti,
%``Quantum Gravity At Two Loops,''
Phys.\ Lett.\  B {\bf 160}, 81 (1985);\\
%%CITATION = PHLTA,B160,81;%%
% M.~H.~Goroff and A.~Sagnotti,
%``The Ultraviolet Behavior Of Einstein Gravity,''
Nucl.\ Phys.\  B {\bf 266}, 709 (1986).
%%CITATION = NUPHA,B266,709;%%

\bibitem{ParkeTaylor}
S.~J.~Parke and T.~R.~Taylor,
%``An Amplitude For N Gluon Scattering,''
Phys.\ Rev.\ Lett.\  {\bf 56}, 2459 (1986).
%%CITATION = PRLTA,56,2459;%%

\bibitem{BGSix}
F.~A.~Berends and W.~Giele,
%``The Six Gluon Process As An Example Of Weyl-Van Der Waerden Spinor
%Calculus,''
Nucl.\ Phys.\ B {\bf 294}, 700 (1987).
%%CITATION = NUPHA,B294,700;%%

\bibitem{MPX}
M.~L.~Mangano, S.~J.~Parke and Z.~Xu,
%``Duality And Multi - Gluon Scattering,''
Nucl.\ Phys.\ B {\bf 298}, 653 (1988).
%%CITATION = NUPHA,B298,653;%%

\bibitem{BGKS}
S.~D.~Badger, E.~W.~N.~Glover, V.~V.~Khoze and P.~Svr\v{c}ek,
%``Recursion relations for gauge theory amplitudes with massive particles,''
JHEP {\bf 0507}, 025 (2005)
[hep-th/0504159].
%%CITATION = JHEPA,0507,025;%%

\bibitem{RSVTree}
R.~Roiban, M.~Spradlin and A.~Volovich,
%``Dissolving N = 4 loop amplitudes into QCD tree amplitudes,''
Phys.\ Rev.\ Lett.\  {\bf 94}, 102002 (2005)
[hep-th/0412265].
%%CITATION = PRLTA,94,102002;%%

\bibitem{SplitHelicityTree}
R.~Britto, B.~Feng, R.~Roiban, M.~Spradlin and A.~Volovich,
%``All split helicity tree-level gluon amplitudes,''
Phys.\ Rev.\  D {\bf 71}, 105017 (2005)
[hep-th/0503198].
%%CITATION = PHRVA,D71,105017;%%

\bibitem{DTW}
M.~Dinsdale, M.~Ternick and S.~Weinzierl,
%``A comparison of efficient methods for the computation of Born gluon
%amplitudes,''
JHEP {\bf 0603}, 056 (2006)
[hep-ph/0602204].
%%CITATION = JHEPA,0603,056;%%

\bibitem{Risager}
K.~Risager,
%``A direct proof of the CSW rules,''
JHEP {\bf 0512}, 003 (2005)
[hep-th/0508206].
%%CITATION = HEP-TH 0508206;%%

\bibitem{CSW}
F.~Cachazo, P.~Svr\v{c}ek and E.~Witten,
%``MHV vertices and tree amplitudes in gauge theory,''
JHEP {\bf 0409}, 006 (2004)
[hep-th/0403047].
%%CITATION = JHEPA,0409,006;%%

\bibitem{LWdeFZ}
M.~Luo and C.~Wen,
%``Recursion relations for tree amplitudes in super gauge theories,''
JHEP {\bf 0503}, 004 (2005)
[hep-th/0501121];
%%CITATION = HEP-TH 0501121;%%
%M.-x.~Luo and C.-k.~Wen,
%``Compact formulas for all tree amplitudes of six partons,''
Phys.\ Rev.\ D {\bf 71}, 091501 (2005)
[hep-th/0502009];\\
%%CITATION = HEP-TH 0502009;%%
%
D.~de Florian and J.~Zurita,
%``The last of the seven-parton tree amplitudes,''
JHEP {\bf 0611}, 080 (2006)
[hep-ph/0609099];
%%CITATION = JHEPA,0611,080;%%
%
%D.~de Florian and J.~Zurita,
%``Seven parton amplitudes from recursion relations,''
JHEP {\bf 0605}, 073 (2006)
[hep-ph/0605291].
%%CITATION = JHEPA,0605,073;%%

\bibitem{MassiveRecursion}
S.~D.~Badger, E.~W.~N.~Glover and V.~V.~Khoze,
%``Recursion relations for gauge theory amplitudes with massive vector bosons
%and fermions,''
JHEP {\bf 0601}, 066 (2006)
[hep-th/0507161];\\
%%CITATION = JHEPA,0601,066;%%
%
D.~Forde and D.~A.~Kosower,
%``All-multiplicity amplitudes with massive scalars,''
Phys.\ Rev.\  D {\bf 73}, 065007 (2006)
[hep-th/0507292];\\
%%CITATION = PHRVA,D73,065007;%%
%
K.~J.~Ozeren and W.~J.~Stirling,
%``Scattering amplitudes with massive fermions using BCFW recursion,''
Eur.\ Phys.\ J.\  C {\bf 48}, 159 (2006)
[hep-ph/0603071];\\
%%CITATION = EPHJA,C48,159;%%
%
C.~F.~Berger, V.~Del Duca and L.~J.~Dixon,
%``Recursive construction of Higgs+multiparton loop amplitudes: The last of
%the phi-nite loop amplitudes,''
Phys.\ Rev.\  D {\bf 74}, 094021 (2006)
[hep-ph/0608180].
%%CITATION = PHRVA,D74,094021;%%

\bibitem{OSQED}
K.~J.~Ozeren and W.~J.~Stirling,
%``MHV techniques for QED processes,''
JHEP {\bf 0511}, 016 (2005)
[hep-th/0509063].
%%CITATION = JHEPA,0511,016;%%

\bibitem{BrownFeynman}
L.~M.~Brown and R.~P.~Feynman,
%``Radiative corrections to Compton scattering,''
Phys.\ Rev.\  {\bf 85} (1952) 231;\\
%%CITATION = PHRVA,85,231;%%
L.~M.~Brown, Nuovo Cim.\ {\bf 21}, 3878 (1961).

\bibitem{PassarinoVeltman}
G.~Passarino and M.~J.~G.~Veltman,
%``One Loop Corrections For E+ E- Annihilation Into Mu+ Mu- 
%In The Weinberg Model,''
Nucl.\ Phys.\  B {\bf 160}, 151 (1979).
%%CITATION = NUPHA,B160,151;%%

\bibitem{Vermaseren}
W.~L.~van Neerven and J.~A.~M.~Vermaseren,
%``Large Loop Integrals,''
Phys.\ Lett.\  B {\bf 137}, 241 (1984);\\
%%CITATION = PHLTA,B137,241;%%
G.~J.~van Oldenborgh and J.~A.~M.~Vermaseren,
%``New Algorithms for One Loop Integrals,''
Z.\ Phys.\  C {\bf 46}, 425 (1990).
%%CITATION = ZEPYA,C46,425;%%

\bibitem{Pittau}
R.~Pittau,
%``A simple method for multi-leg loop calculations,''
Comput.\ Phys.\ Commun.\  {\bf 104}, 23 (1997)
[hep-ph/9607309];
%%CITATION = CPHCB,104,23;%%
%R.~Pittau,
%``A simple method for multi-leg loop calculations. II: A general
%algorithm,''
Comput.\ Phys.\ Commun.\  {\bf 111}, 48 (1998)
[hep-ph/9712418].
%%CITATION = CPHCB,111,48;%%

\bibitem{Weinzierl}
S.~Weinzierl,
%``Reduction of multi-leg loop integrals,''
Phys.\ Lett.\  B {\bf 450}, 234 (1999)
[hep-ph/9811365].
%%CITATION = PHLTA,B450,234;%%

\bibitem{FDH}
Z.~Bern and D.~A.~Kosower,
%``The Computation of loop amplitudes in gauge theories,''
Nucl.\ Phys.\  B {\bf 379}, 451 (1992).
%%CITATION = NUPHA,B379,451;%%

\bibitem{FDH2}
Z.~Bern, A.~De Freitas, L.~J.~Dixon and H.~L.~Wong,
%``Supersymmetric regularization, two-loop QCD amplitudes and coupling
%shifts,''
Phys.\ Rev.\  D {\bf 66}, 085002 (2002)
[hep-ph/0202271].
%%CITATION = PHRVA,D66,085002;%%

\bibitem{Melrose}
D.~B.~Melrose,
%``Reduction Of Feynman Diagrams,''
Nuovo Cim.\  {\bf 40}, 181 (1965).
%%CITATION = NUCIA,40,181;%%

\bibitem{BDKPentagon}
Z.~Bern, L.~J.~Dixon and D.~A.~Kosower,
%``Dimensionally Regulated One Loop Integrals,''
Phys.\ Lett.\  B {\bf 302}, 299 (1993)
[Erratum-ibid.\  B {\bf 318}, 649 (1993)]
[hep-ph/9212308];\\
%%CITATION = PHLTA,B302,299;%%
Z.~Bern, L.~J.~Dixon and D.~A.~Kosower,
%``Dimensionally regulated pentagon integrals,''
Nucl.\ Phys.\  B {\bf 412}, 751 (1994)
[hep-ph/9306240].
%%CITATION = NUPHA,B412,751;%%

\bibitem{OtherPentagons}
J.~Fleischer, F.~Jegerlehner and O.~V.~Tarasov,
%``Algebraic reduction of one-loop Feynman graph amplitudes,''
Nucl.\ Phys.\  B {\bf 566}, 423 (2000)
[hep-ph/9907327];\\
%%CITATION = NUPHA,B566,423;%%
T.~Binoth, J.~P.~Guillet and G.~Heinrich,
%``Reduction formalism for dimensionally regulated one-loop N-point
%integrals,''
Nucl.\ Phys.\  B {\bf 572}, 361 (2000)
[hep-ph/9911342];\\
%%CITATION = NUPHA,B572,361;%%
G.~Duplan\v{c}i\'c and B.~Ni\v{z}i\'c,
%``Reduction method for dimensionally regulated one-loop N-point Feynman
%integrals,''
Eur.\ Phys.\ J.\  C {\bf 35}, 105 (2004)
[hep-ph/0303184].
%%CITATION = EPHJA,C35,105;%%

\bibitem{Cutkosky}
R.~E.~Cutkosky,
%``Singularities and discontinuities of Feynman amplitudes,''
J.\ Math.\ Phys.\  {\bf 1}, 429 (1960).
%%CITATION = JMAPA,1,429;%%

\bibitem{ELOP}
R.~J.~Eden, P.~V.~Landshoff, D.~I.~Olive, J.~C.~Polkinghorne,
{\it The Analytic S Matrix} (Cambridge University Press, 1966).

\bibitem{HolomorphicAnomaly}
F.~Cachazo, P.~Svr\v{c}ek and E.~Witten,
%``Gauge theory amplitudes in twistor space and holomorphic anomaly,''
JHEP {\bf 0410}, 077 (2004)
[hep-th/0409245].
%%CITATION = JHEPA,0410,077;%%

\bibitem{SplitHelicityLoop}
Z.~Bern, N.~E.~J.~Bjerrum-Bohr, D.~C.~Dunbar and H.~Ita,
%``Recursive calculation of one-loop QCD integral coefficients,''
JHEP {\bf 0511}, 027 (2005)
[hep-ph/0507019].
%%CITATION = JHEPA,0511,027;%%

\bibitem{CampbellGloverMiller}
J.~M.~Campbell, E.~W.~N.~Glover and D.~J.~Miller,
%``One-loop tensor integrals in dimensional regularisation,''
Nucl.\ Phys.\  B {\bf 498}, 397 (1997)
[hep-ph/9612413].
%%CITATION = NUPHA,B498,397;%%

\bibitem{BernChalmers}
Z.~Bern and G.~Chalmers,
%``Factorization in one loop gauge theory,''
Nucl.\ Phys.\ B {\bf 447}, 465 (1995)
[hep-ph/9503236].
%%CITATION = HEP-PH 9503236;%%

\bibitem{SplitUnitarity}
Z.~Bern, V.~Del Duca and C.~R.~Schmidt,
%``The infrared behavior of one-loop gluon amplitudes at
%next-to-next-to-leading order,''
Phys.\ Lett.\  B {\bf 445}, 168 (1998)
[hep-ph/9810409];\\
%%CITATION = PHLTA,B445,168;%%
%
Z.~Bern, V.~Del Duca, W.~B.~Kilgore and C.~R.~Schmidt,
%``The infrared behavior of one-loop {QCD} amplitudes at
%next-to-next-to-leading order,''
Phys.\ Rev.\  D {\bf 60}, 116001 (1999)
[hep-ph/9903516];\\
%%CITATION = PHRVA,D60,116001;%%
D.~A.~Kosower and P.~Uwer,
%``One-loop splitting amplitudes in gauge theory,''
Nucl.\ Phys.\  B {\bf 563}, 477 (1999)
[hep-ph/9903515];\\
%%CITATION = NUPHA,B563,477;%%
%
S.~D.~Badger and E.~W.~N.~Glover,
%``Two-loop splitting functions in QCD,''
JHEP {\bf 0407}, 040 (2004)
[hep-ph/0405236].
%%CITATION = JHEPA,0407,040;%%

\bibitem{GGGGG}
Z.~Bern, L.~J.~Dixon and D.~A.~Kosower,
%``One loop corrections to five gluon amplitudes,''
Phys.\ Rev.\ Lett.\  {\bf 70}, 2677 (1993)
[hep-ph/9302280].
%%CITATION = PRLTA,70,2677;%%

\bibitem{CampbellPrivate}
J.~M.~Campbell, private communication.

\bibitem{ZFourPartonsNLODS}
L.~J.~Dixon and A.~Signer,
%``Complete O(alpha(s)**3) results for e+ e- $\to$ (gamma,Z) $\to$ four
%jets,''
Phys.\ Rev.\ D {\bf 56}, 4031 (1997)
[hep-ph/9706285].
%%CITATION = HEP-PH 9706285;%%

\bibitem{VamanYao}
D.~Vaman and Y.~P.~Yao,
%``QCD recursion relations from the largest time equation,''
JHEP {\bf 0604}, 030 (2006)
[hep-th/0512031].
%%CITATION = HEP-TH 0512031;%%

\bibitem{SpaceCone}
G.~Chalmers and W.~Siegel,
%``Simplifying algebra in Feynman graphs. II: Spinor helicity from the
%spacecone,''
Phys.\ Rev.\  D {\bf 59}, 045013 (1999)
[hep-ph/9801220].
%%CITATION = PHRVA,D59,045013;%%

\bibitem{Jaxo}
D.~Binosi and L.~Theussl,
%``JaxoDraw: A graphical user interface for drawing Feynman diagrams,''
Comput.\ Phys.\ Commun.\ {\bf 161}, 76 (2004)
[hep-ph/0309015].
%%CITATION = HEP-PH 0309015;%%

\bibitem{Axo}
J.~A.~M.~Vermaseren,
%``Axodraw,''
Comput.\ Phys.\ Commun.\ {\bf 83}, 45 (1994).
%%CITATION = CPHCB,83,45;%%


\end{thebibliography}
\end{document}